\documentclass[fleqn,usenatbib]{mnras}

\usepackage{newtxtext,newtxmath}
\usepackage[T1]{fontenc}

\DeclareRobustCommand{\VAN}[3]{#2}
\let\VANthebibliography\thebibliography
\def\thebibliography{\DeclareRobustCommand{\VAN}[3]{##3}\VANthebibliography}

\usepackage{graphicx}	% Including figure files
\newcommand\ddfrac[2]{\frac{\displaystyle #1}{\displaystyle #2}}

\title[Finding Orphan Afterglows with DWF]{A Fast-cadenced Search for Gamma-Ray Burst Orphan Afterglows with the Deeper, Wider, Faster Programme}

\author[Freeburn et al.]{
James Freeburn,$^{1,2}$\thanks{E-mail: jfreeburn@swin.edu.au}
Jeff Cooke,$^{1,2}$
Anais M\"oller,$^{1,2}$
Dougal Dobie,$^{1,2}$
Jielai Zhang,$^{1,2}$
\newauthor
{Om Sharan} Salafia,$^{3,4}$
Karelle Siellez,$^{2,5}$
Katie Auchettl,$^{2,6,7}$
Simon Goode,$^{2,8}$
Timothy M. C. Abbott,$^{9}$
\newauthor
Igor Andreoni,$^{10,11,12}$
Rebecca Allen,$^{1}$
Natasha Van Bemmel$^{1,2}$ and 
Sara Webb$^{1}$
\\
$^{1}$Centre for Astrophysics and Supercomputing, Swinburne University of Technology, John St, Hawthorn, VIC 3122, Australia\\
$^{2}$ARC Centre of Excellence for Gravitational Wave Discovery (OzGrav), John St, Hawthorn, VIC 3122, Australia\\
$^{3}$INAF -- Osservatorio Astronomico di Brera, via Brera 28, I-20121 Milan (MI), Italy\\
$^{4}$INFN -- Sezione di Milano-Bicocca, piazza della Scienza 3, I-20126 Milan (MI), Italy\\
$^{5}$School of Mathematics \& Physics, University of Tasmania, Churchill Ave, Hobart, TAS 7005, Australia\\
$^{6}$School of Physics, The University of Melbourne, Parkville, VIC 3010, Australia \\ 
$^{7}$Department of Astronomy and Astrophysics, University of California, Santa Cruz, CA 95064, USA \\
$^{8}$School of Physics and Astronomy, Monash University, VIC 3800, Australia\\
$^{9}$NOIRLab/MSO/CTIO Casilla 603, La Serena, Chile\\
$^{10}$Joint Space-Science Institute, University of Maryland, College Park, MD 20742, USA \\
$^{11}$Department of Astronomy, University of Maryland, College Park, MD 20742, USA \\
$^{12}$Astrophysics Science Division, NASA Goddard Space Flight Center, Mail Code 661, Greenbelt, MD 20771, USA
}
\date{Accepted XXX. Received YYY; in original form ZZZ}

\pubyear{2024}

\begin{document}
\label{firstpage}
\pagerange{\pageref{firstpage}--\pageref{lastpage}}
\maketitle

\begin{abstract}

The relativistic outflows that produce Long GRBs (LGRBs) can be described by a structured jet model where prompt $\gamma$-ray emission is restricted to a narrow region in the jet's core.  Viewing the jet off-axis from the core, a population of afterglows without an associated GRB detection can be predicted.  In this work, we conduct an archival search for these `orphan' afterglows (OAs) with minute-cadence, deep ($g\sim23$) data from the Dark Energy Camera (DECam) taken as part of the Deeper, Wider, Faster programme (DWF).  We introduce a method to select fast-evolving OA candidates within DWF data that comprises a machine learning model, based on a realistic synthetic population of OAs. Using this classifier, we recover 51 OA candidates.  Of these candidates, 42 are likely flare events from M-class stars.  The remaining nine possess quiescent, coincident sources in archival data with angular profiles consistent with a star and are inconsistent with the expected population of LGRB host galaxies.  We therefore conclude that these are likely Galactic events.  We calculate an upper limit on the rate of OAs down to $g<22$ AB mag of 7.46\,deg$^{-2}$yr$^{-1}$ using our criteria and constrain possible jet structures.  We also place an upper limit of the characteristic angle between the $\gamma$-ray emitting region and the jet's half opening angle. For a smooth power-law and a power-law with core jet model respectively, these values are $58.3^{\circ}$ and $56.6^{\circ}$, for a power-law index of 0.8 and $75.3^{\circ}$ and $76.8^{\circ}$ for a power-law index of 1.2.

\end{abstract}


%%%%%%%%%%%%%%%%%%%%%%%%%%%%%%%%%%%%%%%%%%%%%%%%%%

%%%%%%%%%%%%%%%%% BODY OF PAPER %%%%%%%%%%%%%%%%%%

\section{Introduction}\label{sec:intro}

Gamma-ray Bursts (GRBs) are understood to occur as a result of a relativistic outflow of material (or jet) projected from a central engine.  Understanding the GRB population requires measuring properties such as their intrinsic energy release, event rates and circumburst medium densities.  These properties are degenerate with the jet collimation \citep{afterglow_obs}.  Additionally, constraints on jet geometry also allows us to probe the jet launching mechanism and its interaction with the surrounding medium \citep{structure_review}.

There are two prominent progenitor scenarios for GRBs; binary neutron star mergers and collapsars \citep{fireball}.  In this work, we consider only the more abundant observed population of long duration GRBs (LGRBs) with durations longer than 2\,s, they are commonly associated with collapsars \citep{lgrbsne,lgrb_sne1,lgrb_sne2}.  There are, however, notable counterexamples such as GRB\,211211A \citep{211211A_KN,LGRB_KN_Troja,Yang211211A,Gompertz211211A,Mei211211A} and GRB\,230307A \citep{230307A_KN,Sun230307A,Yang230307A,Gillanders230307A,Dichiara230307A} which were LGRBs associated with the merger of two compact objects.

The traditional, `top-hat', model for relativistic outflows comprises a jet that is constant in Lorentz factor ($\Gamma$) and energy throughout the jet's collimation angle and sharply goes to zero at its edges \citep{jet1,jet2,jet3}.  More complex angular jet profiles were initially proposed to explain the variety of observed energies associated with GRBs, arguing that they possessed a standard energy reservoir  \citep{structured1,structured2,structured3}.  Numerical simulations have provided a mechanism for how these jets would arise.  They show that mixing between the jetted material and the cocoon creates an interface layer which results in an angular jet profile exhibiting a gradual decay in energy at the edges \citep{jetstructure_sim,jetstructure_sim2,structure_review}.

The extremely bright GRB\,221009A showed evidence of a structured jet \citep{221009A_williams,Lesage221009A,Frederiks221009A,Kann221009A,LHAASO221009A}.  The shallow evolution of the post-break X-ray afterglow can be explained by emitting material outside the core of the jet.  The collimation angle of the core was determined to be many times smaller than angle swept out by the entire jet half-opening angle. Additionally, an unusually small viewing angle was determined for GRB\,221009A \citep{LHAASO221009A}.  The small viewing angle and large jet half-opening angle presents a contradiction in the rareity of events similar to GRB\,221009A, as one would expect a comparatively large population of similar events observed at larger viewing angles.  One possible way of explaining this contradiction is by restricting prompt $\gamma$-ray emission to the core of the jet \citep{oconnor_221009A}.  

With a viewing angle outside the $\gamma$-ray emission angle, an observer would be unable to detect a GRB.  An afterglow, however, would be detectable.  These are referred to as `orphan' afterglows (OAs) \citep{oa_detectability,DalalOAs,dirty_fireball,RhoadsOAs}.    If a large fraction of the GRB population posses a shallow jet structure, similar to GRB\,221009A, with $\gamma$-ray emission restricted to a small region in the jet's core, we would expect to see a significant population of OAs \citep{oconnor_221009A}.

\citet{221009A_shallow} found that the multi-wavelength observations of GRB\,221009A were consistent with a shallow jet structure, characteristic of a weakly magnetised jet.  With such a structure, a prominent jet-break should not be observed at late times.  It is rare that GRB afterglows are observable long enough for the jet-break to be observed, and among the afterglows with late-time detections, there is a sub-population that do not exhibit a jet-break.  This could be a result of a shallow angular jet profile similar to GRB\,221009A \citep{oconnor_221009A}.

\citet{structure_constraints} provide constraints on jet structure using the observed population of prompt GRBs and their counterparts.  These constraints, however, break down in a shallow structured jet scenario where prompt $\gamma$-ray emission is restricted to the core, similar to that observed with GRB\,221009A. Searches for OAs, therefore, provide an avenue to further constrain GRB jet structure by observing GRBs at viewing angles larger than where prompt $\gamma$-ray emission is detectable \citep{on-axis_orphans}.

For this work, we only consider OAs emitted from misaligned structured jets.  However, there are a number of scenarios in which OAs are produced.  One of which occurs when observing a GRB entirely off-axis from the jet.  These are difficult to observe originating from collapsars in optical wavelengths as they are expected to be substantially fainter than the accompanying supernova emission \citep{OAs_SNe}.  

OAs can also emitted from a collapsar if the jet is initially loaded with baryons.  The Lorentz factor of such a jet is insufficient to produce a GRB, but peaks at lower energies, resulting in an OA which follows a similar evolution to an on-axis afterglow \citep{dirty_fireball}.  These are commonly referred to as `dirty fireballs' or `failed GRBs.'  Authors  \citet{orphan_afterglows}, \citet{OA_search_Rau} and \citet{ZTF_OA_search} conducted searches for OAs in this scenario.

These searches, along with other surveys, have yielded a number of suspected OAs.  The earliest discovery of such an event achieved the the Palomar Transient Factory \citep{PTF} was PTF11agg \citep{OA_cand}.  In recent years, the Zwicky Transient Facility \citep{ZTF} has dominated in this effort with events such as AT\,2019pim \citep{ZTF_OA_search,at2019pim}, AT\,2020blt \citep{ZTF_OA}, AT\,2021any \citep{ztfrest_afterglows} and AT 2021lfa \citep{ZTF_OA_search}.

Once an afterglow detection is made without an associated GRB, it becomes important to rule out the existence of prompt $\gamma$-ray emission that was simply not detected by GRB monitors like \textit{Swift} Burst Alert Telescope (BAT) and \textit{Fermi} GBM \citep{swift,fermi6yr}.  The best constraints to-date on GRB emission accompanying an OA candidate was AT2019pim, reported by \citet{at2019pim}.  They find that prompt, $\gamma$-ray emission accompanying AT2019pim is disfavoured. This presents evidence for the existence of a population of OAs, highlighting the opportunity for the discovery of further OAs.

However, OAs observed on-axis are fast evolving and can be observable for mere minutes \citep{GROND}.  Therefore, existing surveys with day cadences may be unable to detect a significant fraction of the OA population.  The highest cadence data used for an OA search was in \citet{ZTF_OA_search} with up to three visits per night.  No OA search to-date has been conducted with a cadence on minute timescales.

The Deeper, Wider, Faster Programme (DWF) involves observations with the Dark Energy Camera (DECam) mounted on the CTIO 4m telescope.  DWF's observing strategy involves minute cadence observations while reaching deeper ($g\sim23$) than other transient surveys such as ZTF ($\sim21$).  DWF utilises near real-time data reduction and analysis, designed to identify transients for rapid, spectroscopic follow-up \citep[e.g, ][]{DWF1}.  Although much of the data has been analysed in real-time, conducting late-time analysis allows for a comprehensive search with rigorous rate constraints.

This work comprises a search for OAs in 100 nights of archival DECam data across 18 fields, each one covering $\sim2.1$\,deg$^2$ of effective sky area.  With this search, we constrain GRB jet structure.  Other prominent transient surveys lack either the cadence or depth necessary to probe the population of OAs occurring from misaligned structured jets.

The DWF dataset has been previously used for other works studying minute timescale transients.  For example, \citet{DWF2} searched for extragalactic fast transients broadly and provided rate constraints but only analysed 25 nights across five fields, about a quarter of the data that is available as of the writing of this work.  \citet{sara_flares} searched a similarly large subset of the data but was targeted towards stellar flares within 500\,pc.  No work to-date has conducted a search tuned specifically for GRB afterglows on all of the applicable data.

This paper is organised as follows: In Section \ref{sec:data}, we describe the DWF DECam data used for this work.  Section \ref{sec:method} outlines our synthetic population of OAs, used for our search methodology, efficiency calculations and rate estimates.  We then describe the machine learning algorithm used to identify OAs in the data and their expected rates in Section \ref{sec:extracting}.  In Section \ref{sec:results}, we analyse the OA candidates found in the data.  A discussion on the implications of a non-detection in the data is detailed in Section \ref{sec:discussion}.  This includes constraints on GRB jet structure and prospects for future work.  We then conclude in Section \ref{sec:conclusion}.

\section{The Deeper Wider Faster Programme}\label{sec:data}

To-date, there have been 13 DWF coordinated operational runs (which we denote O1 through O13) spanning from December 2015 to January 2024.  Each of these runs lasted for six nights, observing 3--5 fields.  Typically, these fields are observed by taking continuous minute-cadenced imaging, 20\,s exposures with $\sim$30\,s readout and CCD clear, without dithering and for 1-3\,hr per night.  We call these `observing windows.'  Our classifier, described in Section \ref{sec:classifier}, was designed to ingest well sampled light curves, which comprise the majority of the DWF data, and its efficiency drops for light curves with a small number of exposures. As a result, we exclude observing windows where less than 20 exposures ($<$ 20\,min) were taken.  

We also exclude four DWF observing runs; O8, O9, O11 and O13.  This work focuses on data taken with DECam and DWF O8 and O11 used Subaru Hyper Suprime-Cam and KMTNet, respectively. DWF O9 experimented in using an alternate strategy with dithered exposures and a larger number of fields tailored to detect kilonovae and had fewer exposures in each field per night.  DWF O13 primarily observed the Large Magellanic Cloud, with crowded fields, less well suited for the discovery of extragalactic transients.  Finally, there were two DWF pilot runs that employed a dithering strategy, not compatible with the photometry pipeline used in this search.  We use only data from the remaining nine DWF runs (see table \ref{tab:dwf_fields}) which comprise 9033 images and 145 hours of observing time.

\begin{table}
 \caption{Fields and night coverage for this search.}
 \label{tab:dwf_fields}
 \begin{tabular*}{\columnwidth}{@{}l@{\hspace*{10pt}}l@{\hspace*{10pt}}l@{\hspace*{10pt}}l@{\hspace*{10pt}}l@{\hspace*{30pt}}}
  \hline
  Field & Coordinates & Gal. Latitude & Runs & Nights\\
  \hline
  FRB\,010724 & 01:18:06 -75:12:19 & -41.80 & Dec 2015 & 4\\
  3hr & 03:00:00 -55:25:00 & -53.43 & Dec 2015 & 5\\
  CDFS & 03:30:24 -28:06:00 & -54.93 & Dec 2015, & 9\\
    & & & Dec 2019, & \\
    & & & Sep 2021, & \\
    & & & Sep 2022 & \\
  4hr & 04:10:00 -55:00:00 & -44.76 & Dec 2015 & 5\\
  Prime & 05:55:07 -61:21:00 & -30.26 & Feb 2017 & 5\\
  FRB\,131104 & 06:44:00 -51:16:00 & -21.95 & Feb 2017 & 5\\
  8hr & 08:16:00 -78:45:00 & -22.62 & Jun 2018 & 2\\
  Dusty 10 & 10:12:00 -80:50:00 & -19.96 & Jun 2018 & 1\\
  Antlia & 10:30:00 -35:20:00 & 19.17 & Feb 2017, & 4\\ 
   & & &  Jun 2018 & \\
  Dusty 12 & 11:46:00 -84:33:00 & -21.89 & Jul 2016 & 3\\ 
  14hr & 14:34:00 -78:06:00 & -16.30 & Jul 2016 & 3\\
  NGC 6101 & 16:26:00 -73:00:00 & -16.37 & Jul 2016,  & 10\\
   & & & Aug 2016, & \\
   & & & Jun 2019 & \\
  NGC 6744 & 19:09:46 -63:51:27 & -26.15 & Jul 2016,  & 22\\
   & & & Aug 2016, & \\
   & & & Jun 2019, & \\
   & & & Sep 2021, & \\
   & & & Sep 2022 & \\
  Field 3 & 21:00:00 -42:48:00 & -49.41 & Jun 2019 & 3\\
  NSF2 & 21:28:00 -66:48:00 & -39.82 & Jul 2016 & 2\\
  FRB\,190711 & 21:57:41 -80:21:29 & -33.90 & Sep 2021, & 10\\
   & & & Sep 2022 & \\
  FRB\,171019 & 22:17:31 -08:39:32 & -49.24 & Dec 2019, & 4\\
   & & & Sep 2021 & \\
  HDFS & 22:33:26 -60:38:09 & -49.22 & Sep 2021 & 3\\  
  \hline
  Total &  &  & & 100\\
  \hline
 \end{tabular*}
\end{table}

\subsection{Photometric Pipeline}\label{sec:phot}

For this search, we use a data processing pipeline that takes in DECam images, calibrated with the NOIRLab community pipeline \citep{DECam_pipeline}, and outputs light curves for each source in the field.  The full details of this pipeline will be presented in a future publication (Freeburn et al. in prep.).

The source extraction software, \textsc{sextractor} is utilised for our photometric pipeline \citep{sextractor}.  We run adaptive aperture photometry (known as `MAG\_AUTO') in double image mode.  Double image mode requires a separate, detection image to identify sources.  It then measures photometry, in the science image, at the location of the sources found in the detection image.

The image with the largest value of the full width at half maximum (FWHM) is chosen as the detection image for a given observing window.  This corresponds to the image with the most unfavourable seeing conditions and typically shallowest depth.  An initial \textsc{sextractor} run is conducted on the science image that comprises a given observing window.  Any new sources found during this run that are not present in the detection image, are then injected into the detection image.

We then match the PSF of the detection image to the rest of the observing window's images with \textsc{hotpants}, using Gaussian convolution.  Light curves are then obtained by conducting a second \textsc{sextractor} run on the convolved images, using the detection image.  The instrumental photometry is then calibrated with photometric catalogs from \textit{SkyMapper} \citep{skymapper_dr2} or \textit{Pan-STARRS} \citep{panstarrs}.

\section{A Synthetic Population of Orphan Afterglows}\label{sec:method}
\begin{figure}
    \includegraphics[width=\columnwidth]{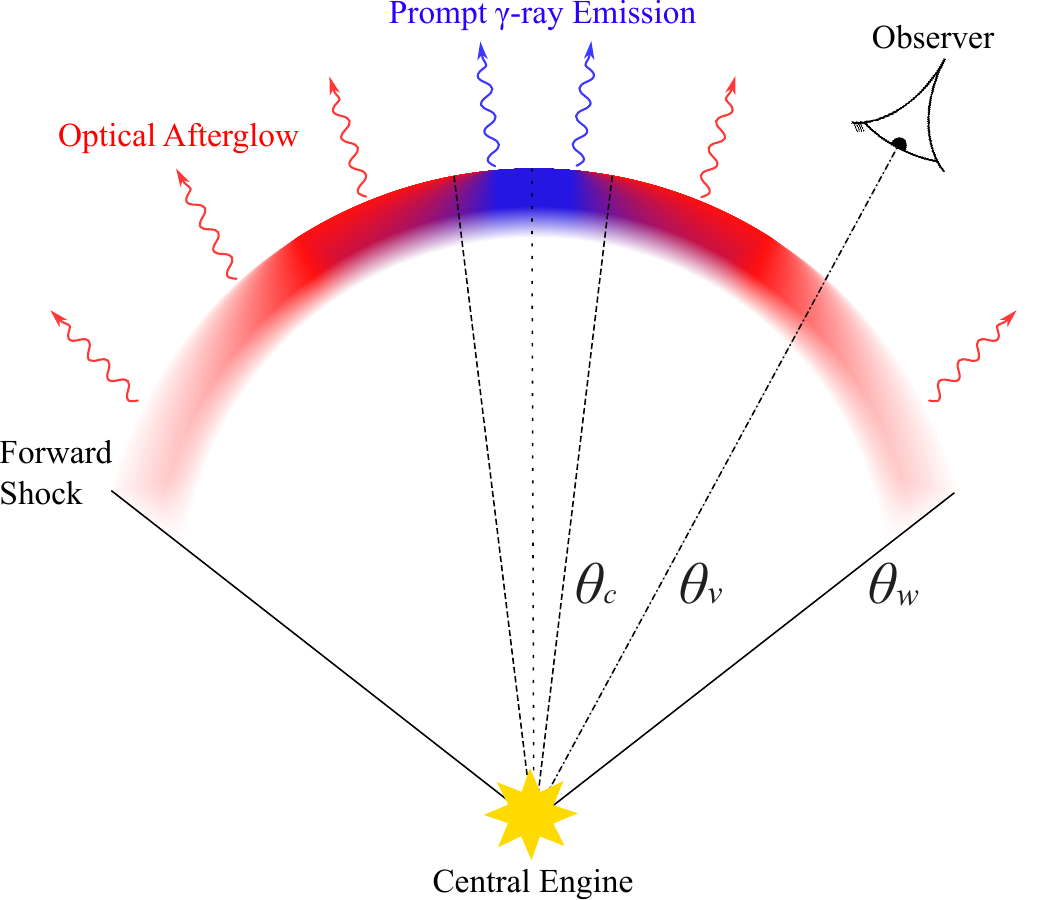}
    \caption{Diagram of the structured jet model considered in this work, adapted from \citet{oconnor_221009A}.  In this model, $\gamma$-ray emission is beamed at an angle of $\theta_c$.  The energy of the jet decays as a power-law out to an angle of $\theta_w$ according to Eq. \ref{eq:pl_jet} and Eq. \ref{eq:plcore_jet}.  OAs are therefore detectable with viewing angles, that satisfy $\theta_c <\theta_v < \theta_w$.}
    \label{fig:structure_diagram}    
\end{figure}

We consider GRB jets to be described by three collimation angles shown in Figure \ref{fig:structure_diagram}.  The angle of the core, $\theta_c$ describes the inner region of the jet that is approximately constant in $\Gamma$ and energy.  We restrict the angle at which a GRB is produced to $\theta_c$.  $\theta_w$ defines the entire angular extent of the jet.  The angle at which we view the jet is denoted by $\theta_v$. Viewing the jet outside the core, $\theta_v > \theta_c$,  will result in an observable OA.  With a viewing angle outside the jet's half-opening angle, $\theta_c > \theta_w$, OA detection becomes very difficult, as explained in Section \ref{sec:intro}.  In this work, we explore only the scenario where we observe the jet inside its half-opening angle but outside the core, $\theta_c > \theta_v > \theta_w$.  This provides a more luminous and fast-evolving population of OAs \citep{KumarGranotStructured,BeniaminiStructured,BeniaminiGillStructured}.

GRB afterglow light curves vary significantly in their evolution.  The rise time, peak luminosity, fade rate and the time of the jet break all depend on a number of free parameters.  In order to accurately calculate the rates of orphan afterglows we expect to detect for a given jet structure and effectively identify them, it is necessary to use a representative synthetic dataset of afterglows tailored to the DWF data.

In Section \ref{sec:synth_data}, we describe how we generate a synthetic population of GRBs.  For this population, we then model their corresponding afterglows in Section \ref{sec:afterglow_sample}.  In Section \ref{sec:fakes}, we inject these afterglows into DWF images.

\subsection{GRB Population Synthesis} \label{sec:synth_data}

The \textit{Swift} BAT6 complete sample comprises \textit{Swift}/BAT LGRB detections with fluxes >\,$2.6$\,ph\,cm$^{-2}$\,s$^{-1}$.  These LGRBs have a 90\,\% completeness in $z$ and possess well-defined detection rate in \textit{Swift} BAT's half-coded region of $\mathcal{R}_{\mathrm{\textit{Swift}}} \sim 15$ events $\mathrm{sr}^{-1} \mathrm{yr}^{-1}$ \citep{BAT6_sample}. 

\citet[G13]{pop_synth} use the BAT6 sample to generate a synthetic population that reproduces the properties and rates of observed LGRBs.  We use the results from G13, to produce a population of LGRBs that are representative of observed LGRB and afterglow fluxes and rates.

G13 assume a standard rest frame energy reservoir of $1.5\times10^{48}$\,erg and peak energy of $1.5$\,keV with a GRB formation rate based on \citet{GRBFR} and \citet{HandB}. T90 values are generated in a log-normal distribution centred at 27.5\,s with a dispersion of 0.35, truncated at 2\,s.  We assume an isotropic distribution of viewing angles which corresponds to a probability density that scales with $\sin\theta_v$.  Using this formalism, distributions of the jet half-opening angle, $\theta_j$ and the initial Lorentz factor of the jet, $\Gamma_0$ were fit to the \textit{Swift} BAT6 complete sample.  The relation in Equation \ref{eq:lognorm_dist} was derived using this methodology.
\begin{equation}
    \log\theta_{*,\mathrm{jet}} = -\frac{1}{m}\log\Gamma_{*} + q
    \label{eq:lognorm_dist}
\end{equation}
$\log\Gamma_{*}$ is the central value of the $\log\Gamma_0$ distribution.  For each value of $\Gamma_0$ generated, a value of $\log\theta_{\mathrm{jet}}$ is generated with a central value, $\log\theta{*,\mathrm{jet}}$.  The best-fit parameters calculated in G13 are denoted $m=2.5$ and $q=1.45$.  The log-normal distributions have dispersions of $\sigma_{\log\Gamma_0}=0.65$\,dex and $\sigma_{\log\theta{\mathrm{jet}}}=0.3$\,dex.

The isotropic equivalent energy release, $E_{\mathrm{iso}}$ can then be calculated from $\Gamma_0$ and $\theta_j$ using
\begin{equation}
    E_{\mathrm{iso}} = 
    \begin{cases}
        E_{\gamma}/(1-\cos\theta_j), & \text{ if } 1/\Gamma_0 \leq \sin\theta_j\\
        E_{\gamma}/(1+\beta_0)\Gamma_0^2, & \text{ if } 1/\Gamma_0 > \sin\theta_j
    \end{cases}
    \label{eq:eiso}
\end{equation}
where $E_\gamma$ is the total energy release in the observer frame and $\Gamma_0 = 1/(1 - \beta_0)^{1/2}$. The resultant synthetic population successfully reproduces the distribution of fluences and $E_{\mathrm{peak}}$ values of \textit{Swift} BAT6 LGRBs.  Additionally, modelling the afterglows of these bursts in R-band reproduce the observed flux distribution of \textit{Swift} BAT6 LGRB afterglows 11hrs post burst as shown in \citet{unveiling_population}.

\subsection{Generating a Sample of Synthetic Afterglows}\label{sec:afterglow_sample}

Afterglows are characterised by a further four parameters: The fraction of the forward shock's thermal energy in electrons and the magnetic field are described by $\epsilon_e$ and $\epsilon_B$ respectively, the electron distribution power-law index, $p$ and the circumburst number density, $n$.  These parameters are currently poorly constrained due to degeneracy in predicting afterglow light curves.

\citet{radio_popsynth,unveiling_population} used the same synthetic population to reproduce observed properties of afterglows in radio, optical and X-ray wavelengths.  We adopt the same ensemble of parameters in those works (see Table~\ref{tab:oa_params}), as they are consistent with observed GRB afterglows and are values derived from first-principles simulations of particle acceleration in relativistic shocks \citep{shock_sims}. \citet{unveiling_population} also show that the population synthesis models are able to recover the distribution of observed afterglow R-band fluxes for the \textit{Swift} BAT6 complete sample.

We use {\scshape afterglowpy}, a \textsc{python} package for generating afterglow light curves \citep{afterglowpy}.  Power-law jet models have been found to be consistent with hydrodynamical simulations \citep{jetstructure_sim2} and observations of GRB\,221009A's afterglow \citep{221009A_shallow}. We therefore investigate both a smooth power-law jet and a power-law with a uniform core in this work.  The energy distributions for these jet models are given in Equations \ref{eq:pl_jet} and \ref{eq:plcore_jet} respectively:
\begin{equation}
    E(\theta) =  
    \begin{cases}
        E_0 \left( 1 + \frac{\theta^2}{b\theta_c^2}\right)^{-b/2}, & \text{if } \theta < \theta_w\\
        0, & \text{if } \theta > \theta_w
        \label{eq:pl_jet}
    \end{cases}
\end{equation}

\begin{equation}
    E(\theta) = 
    \begin{cases}
        E_0, & \text{if } \theta < \theta_c\\
        E_0 \left(\frac{\theta}{\theta_c}\right)^{-b}, & \text{if } \theta_c < \theta < \theta_w\\
        0, & \text{if } \theta > \theta_w\\
    \end{cases}
    \label{eq:plcore_jet}
\end{equation}

Where the model is normalised with $E_0$, $\theta$ is the angle from the centre of the jet, $\theta_c$ is the characteristic width of the distribution and $b$ is the power-law index. We set $E_0$ from Equations \ref{eq:pl_jet} and \ref{eq:plcore_jet} to $E_{\mathrm{iso}}$ from Equation \ref{eq:eiso}.  

Eq. \ref{eq:pl_jet} applies to the jet out to an angle of $\theta_w$, where $E(\theta)$ drops to zero.  We assume that, for angles larger than $\theta_c$, prompt $\gamma$-ray emission is not detectable.  G13 assumed that prompt $\gamma$-ray emission is produced for $\theta_v < \theta_j$.  Therefore, we take the $\theta_j$ values generated in our population synthesis model to be $\theta_c$ for generating afterglows.  We adopt a uniform distribution of $\theta_c < \theta_w < 90^{\circ}$.

We draw values $b$ uniformly such that $0<b<3$, as with the data used for this work, we are insensitive to OAs originating from structured jets with indices larger than $b=3$.  This is highlighted in Section \ref{sec:detectability}.  Additionally, simulations favour angular jet structures with $0.7 \lessapprox b \lessapprox 2.8$  \citep{jetstructure_sim,jetstructure_sim2} and GRB\,221009A has a measured value of $b=0.8$, well within the range of $b$ explored in this work \citep{221009A_shallow}.

Each combination of $b$ and $\theta_c - \theta_w$ can be considered a distinct jet structure, for which an OA rate can be calculated.

We note that, by default, \textsc{afterglowpy} does not model the jet's deceleration phase, which affects the afterglow's rise.  It also means that $\Gamma_0$ is not taken into account directly \citep{afterglowpy}.  However, due to \textsc{afterglowpy}'s flexibility and low computational cost, we find that it is optimal for this work.  Low $\Gamma_0$ events may have an early evolution that departs from our modelling.  This may slightly affect our total efficiency in finding these events but we leave this to future work.

\begin{table}
 \caption{Afterglow parameters used for generating our synthetic population.  The physical meaning of these parameters is explained in Section \ref{sec:afterglow_sample}.}
 \label{tab:oa_params}
 \begin{tabular*}{\columnwidth}{@{}l@{\hspace*{12pt}}l@{\hspace*{12pt}}l@{\hspace*{12pt}}l@{\hspace*{12pt}}l@{\hspace*{12pt}}l@{\hspace*{12pt}}}
  \hline
  $p$ & $n$ & $\epsilon_e$ & $\epsilon_B$ & $b$ & $\theta_w$\\
  \hline
  $2.3$ & $0.1 > n > 30$ & $0.02$ & $0.008$ & $0 < b < 3$ & $\theta_c < \theta_w < 90^{\circ}$\\
  \hline
 \end{tabular*}
\end{table}

\subsection{Injecting Synthetic Afterglows into DWF Images} \label{sec:fakes}

Realistic model afterglow light curves, as they would be detected with DWF, are important for both training data and the calculation of detection and classification efficiencies.  We inject a population of point sources in consecutive images forming light curves (fakes) from the synthetic sample of afterglows described in Section \ref{sec:afterglow_sample} into a representative subset of the DWF images.  Our photometry pipeline, described in Section \ref{sec:phot} is then run on these images to recover the injected source light curves.  

\begin{figure*}
    \includegraphics[width=0.49\textwidth]{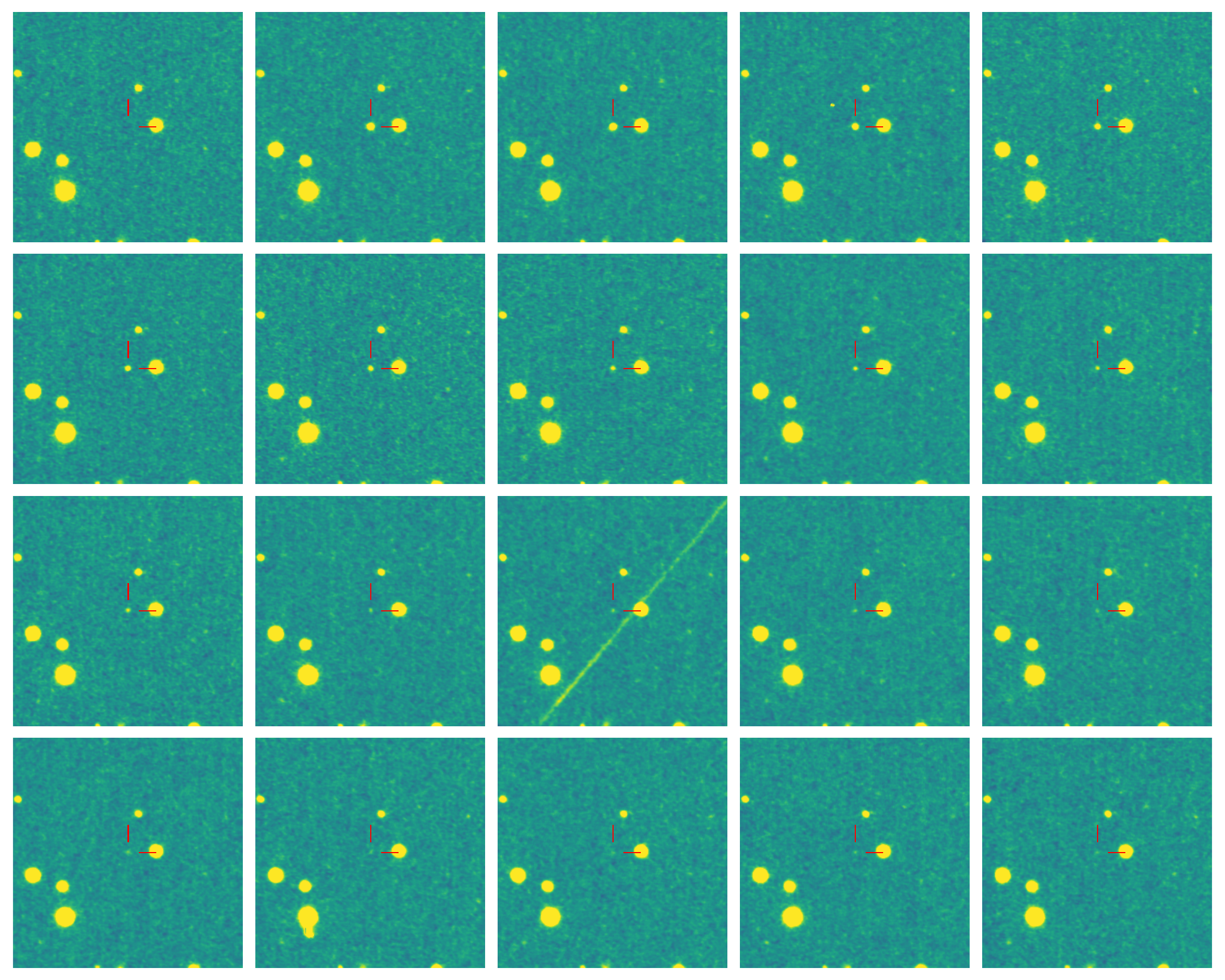}
    \includegraphics[width=0.49\textwidth]{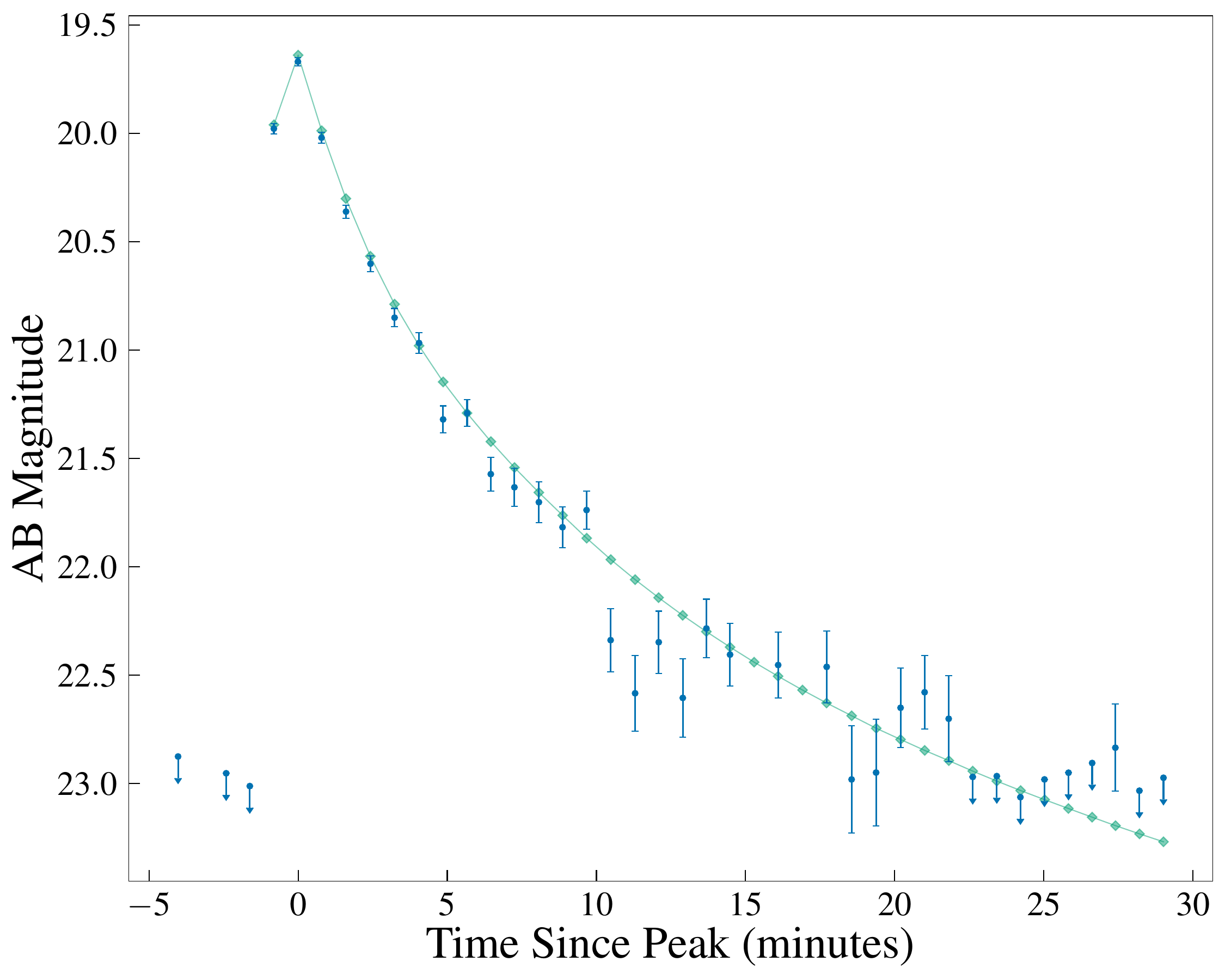}
    \caption{Example of an afterglow model injected into DWF images.  The left-hand side shows cutouts at the location of the injected afterglow (centred in the images).  The right-hand side shows the resultant light curve from the pipeline described in Section \ref{sec:phot}.  The blue points are the extracted photometry of the injected source and the green points show the injected magnitude of the afterglow.  The parameters for this afterglow are $\theta_v=72.8^{\circ}$, $\theta_w=79.1^{\circ}$, $\theta_c=12.0^{\circ}$, $b=0.30$, $z=2.95$, $n=1.8\times10^2$\,$\mathrm{cm}^{-3}$, $E_{\mathrm{iso}}=2.73\times10^{51}$.}
    \label{fig:fake_example}    
\end{figure*}

\begin{figure}
    \includegraphics[width=\columnwidth]{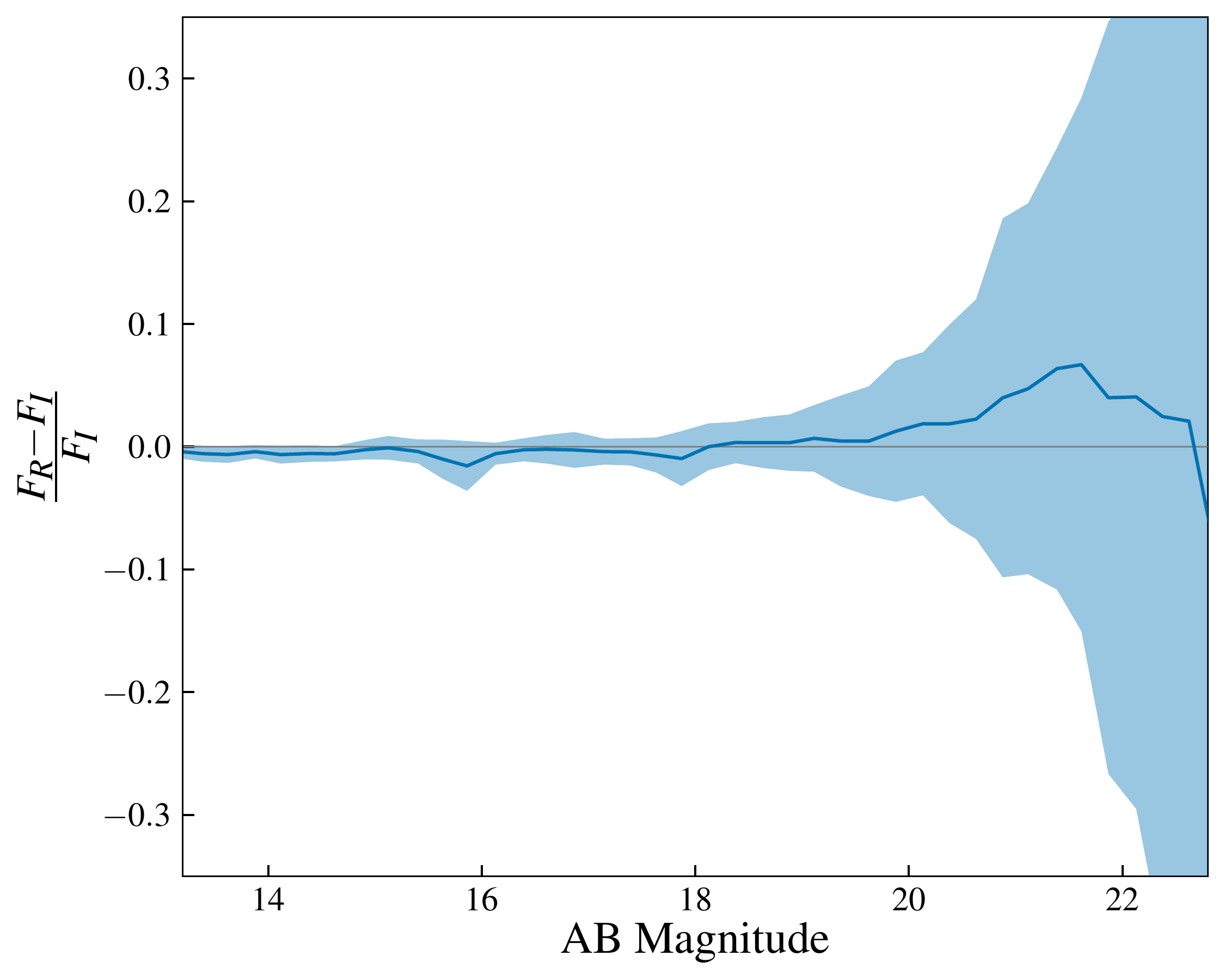}
    \caption{Relative difference in recovered and injected fluxes plotted against the magnitude of the injected source.  The blue line shows the median flux difference in each magnitude bin and the shaded blue region shows the 1-$\sigma$ variance in flux difference at each magnitude bin.  The difference in recovered versus injected flux scales with the expected error, reflecting the limiting magnitudes of our observations.}
    \label{fig:fakes_photometry}
\end{figure}

The fake sources are modelled using a Moffat profile \citep{moffat} with a FWHM matching the exposure and stellar point sources in the charge-coupled device (CCD) image in which the fakes are injected.  Injected sources are initially calibrated to each image by injecting and recovering five sources with varying instrumental flux values.  We present an example of one of these injected afterglows in Figure \ref{fig:fake_example}. Figure \ref{fig:fakes_photometry} shows the difference in injected versus recovered flux with magnitude.  We see broad agreement between injected and recovered flux up to a magnitude of $g \sim 22$.

We inject the fakes both randomly throughout the field and coincident with visible galaxies. \citet{GRB_hosts} analyse a sample of LGRB host galaxies from The Optically Unbiased GRB Host (TOUGH) survey \citep{TOUGH}, BAT6, the Gamma-Ray Burst Optical and Near-Infrared Detector (GROND) \citep{GROND,dusty_hosts} and the Swift Gamma-Ray Burst Host Galaxy Legacy Survey (SHOALS) \citep{SHOALS}. The vast majority of LGRB hosts in this sample are fainter than 23 AB magnitude in $R$-band.  Using our synthetic population of afterglows, we find that DWF's depth and cadence results in a higher median redshift than the sample used in \citet{GRB_hosts}.  We therefore assume that the OAs found in our search will have hosts fainter than 23 AB mag in $g$-band and only consider the randomly distributed fakes in our efficiency calculations.  The sources that were injected onto galaxies will be used as part of the training data for completeness and to prevent a bias towards transients without a visible coincident galaxy.

\section{Investigating Orphan Afterglow Candidates}\label{sec:extracting}

We use the synthetic dataset described in Section \ref{sec:method} to train a machine learning classifier that extracts OA candidates from the DWF data.  This is outlined in Section \ref{sec:classifier}. In Section \ref{sec:detectability}, we then discuss the efficiency of this classifier in probing this theoretical population of OAs and predict how many OAs we expect to find.

\subsection{Light Curve Classifier}\label{sec:classifier}

Our photometry pipeline outputs approximately $\sim10^5$ light curves per night per field, with an average of 90 measurements in each light curve. We obtain $1.63\times10^7$ light curves total.  To search this large dataset, we use a series of cuts and a machine learning classifier.

The first cut we make on this pool of light curves is based on their variability.  We only consider light curves that exhibit significant variability during a single observing window. We use the von Neumann statistic \citep{VN_statistic} 
\begin{equation}
    \eta = \ddfrac{\sum_{i=1}^{N-1} (m_{i+1}-m_i)^2/(N-1)}{\sum_{i=1}^{N}(m_{i+1}-\Bar{m})^2/(N-1)}
    \label{eq:vn}
\end{equation}
to measure variability. Here, $m_i$ describes a series of measurements, equally spaced in time.  \citet{vn_comparison} find $\eta^{-1}$ is an effective variability indicator for photometric time-series data. A cut of $\eta_{\mathrm{lc}}^{-1} > 0.6$ removes 86.8\,\% of all DWF light curves while retaining 64.7\,\% of injected afterglows brighter than $g =$ 23 AB mag. Injected afterglows that exhibit low variability, with $\eta_{\mathrm{lc}}^{-1} < 0.6$, predominantly fade below our detection threshold before the next observing window, $\sim$24 hours later.  The lack of vital evolutionary information, in this case, would make classification almost impossible.  We, therefore, remove light curves from our search if they satisfy $\eta_{\mathrm{lc}}^{-1} < 0.6$.

For the light curves satisfying $\eta_{\mathrm{lc}}^{-1} > 0.6$, we use a sliding window of three detections and use the peak detection from the window with the largest median value to identify the peak, $t_{\mathrm{peak}}$. For light curves with quiescent emission, the quiescent magnitude is identified as the median of all the detections before the peak.  If the peak is identified in one of the first 5 exposures of the night, we take the quiescent magnitude to be the median of the final five detections.  After two consecutive detections after $t_{\mathrm{peak}}$ to below the quiescent magnitude, we consider this the end of the event, denoted by $t_{\mathrm{max}}$.  For light curves without quiescent emission, the event is defined as the detections between the peak and the second non-detection.

We measure the variability of the detections that comprise the event with $\eta^{-1}_{\mathrm{event}}$. Only 0.02\,\% of all light curves in the DWF data and 30.6\,\% of injected afterglows brighter than 23 AB mag satisfy a cut of $\eta^{-1}_{\mathrm{event}} > 4$.  We therefore separate all light curves satisfying this cut for human inspection.

For light curves with $\eta^{-1}_{\mathrm{event}} < 4$ and $\eta_{\mathrm{lc}}^{-1} > 0.6$, we use a XGBoost binary classifier which has been shown to be robust in light-curve classification tasks \citep{Moller:2016}.  The training data for our classifier comprises 848 each of a curated set of injected afterglow light curves (described in Section \ref{sec:fakes}) and a curated set of randomly selected light curves from the DWF data that do not exhibit afterglow-like variability.  While afterglow light curves vary in morphology, their fade is modelled well as a power-law decay.  We fit a power-law decay to all DWF light curves that satisfy $\eta_{\mathrm{lc}}^{-1} > 0.6$.

We do not model the rise phase of the light curve for two reasons.  First, it allows us to treat events that peak before our observing window and those that peak during in the same manner.   Second, our training data, generated using \textsc{afterglowpy} does not accurately model the rise (see Section \ref{sec:afterglow_sample}).

A power-law of the form shown in Equation \ref{eq:pl} is then fit to the detections between the $t_{\mathrm{peak}}$ and $t_{\mathrm{max}}$.
\begin{equation}
    m_{\mathrm{norm}} = a(t-c)^{-b} + d =\frac{m-\Bar{m}}{\sigma_{lc}}
    \label{eq:pl}
\end{equation}
$t$ is the time in hours after the peak of the light curve, $m$ is the apparent magnitude of a given detection, $\Bar{m}$ and $\sigma_{lc}$ is the median and standard deviation of the magnitudes that comprise the light curve. $a$,$b$,$c$ and $d$ are the parameters we fit for.  The following features comprise those used in our classifier which achieves the performance metrics in table \ref{tab:classifier_performance}:
\begin{enumerate}
    \item $a,b,c,d$, best fit parameters from equation \ref{eq:pl}
    \item $\nabla_{\mathrm{event}}$, average gradient of the event, scaled by the error.
    \item $\nabla_{\mathrm{q}}$, average gradient of the quiescent measurements, scaled by the error.
    \item $\mathrm{med}_5$, the median time of the five brightest detections subtracted by $t_{\mathrm{peak}}$.
    \item $N_{\mathrm{event}}$, number of detections between the $t_{\mathrm{peak}}$ and $t_{\mathrm{max}}$.
    \item $t_{\mathrm{peak}} - t_{\mathrm{max}}$, difference between the $t_{\mathrm{peak}}$ and $t_{\mathrm{max}}$.
    \item $dm_{\mathrm{event}}/dm_{\mathrm{lc}}$, the median magnitude error of the event scaled by the median error of the light curve.
    \item $\eta^{-1}_{\mathrm{event}}$, the variability between $t_{\mathrm{peak}}$ and $t_{\mathrm{max}}$.
    \item $\eta^{-1}_{\mathrm{resid}}$, the variability of the residuals of the power-law fit.
    \item $\eta^{-1}_{\mathrm{lc}}$, the variability of the entire light curve.
    \item $\chi_{\mathrm{red,event}}^2$, the chi-squared statistic of the power-law fit.
    \item $\sigma_{\mathrm{peak,2}}$, the significance of first detection after the peak.
    \item $\Delta m/\Bar{dm}$, difference between maximum and minimum magnitude, scaled with the median error of the detections that comprise the light curve.
\end{enumerate}

\begin{table}
 \caption{Performance of light curve classifier on the test set. Uncertainties are given by Poisson statistics. AUC denotes the Area Under the Curve statistic for classifier performance which is given by the integral of the curve tracing the true positive rate versus the false positive rate.  Efficiency, purity and accuracy are given by ${\rm TP}/({\rm TP} + {\rm FP})$, ${\rm TP}/{\rm P}$ and accuracy is the balanced accuracy score metric in \textsc{scikit-learn} \citep{sklearn}.  TP is the number of true positives, FP is the number of false positives and P is the number of positives in the test set.}
 \label{tab:classifier_performance}
 \begin{tabular*}{\columnwidth}{@{}l@{\hspace*{25pt}}l@{\hspace*{25pt}}l@{\hspace*{25pt}}l@{\hspace*{25pt}}}
  \hline
  Accuracy & AUC & Purity & Efficiency\\
  \hline
  $94.4\pm1.2$\,\% & $0.98$ & $94.67\pm1.7$\,\% & $94.12\pm1.8$\,\%\\
  \hline
 \end{tabular*}
\end{table}

\begin{figure}
    \includegraphics[width=\columnwidth]{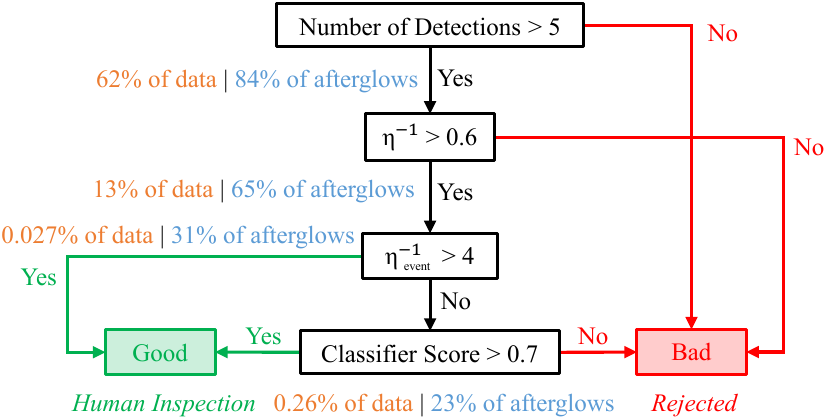}
    \caption{Steps involved in extracting afterglow light curves in the DWF data.} 
    \label{fig:classifier}    
\end{figure}

\begin{figure}
    \includegraphics[width=\columnwidth]{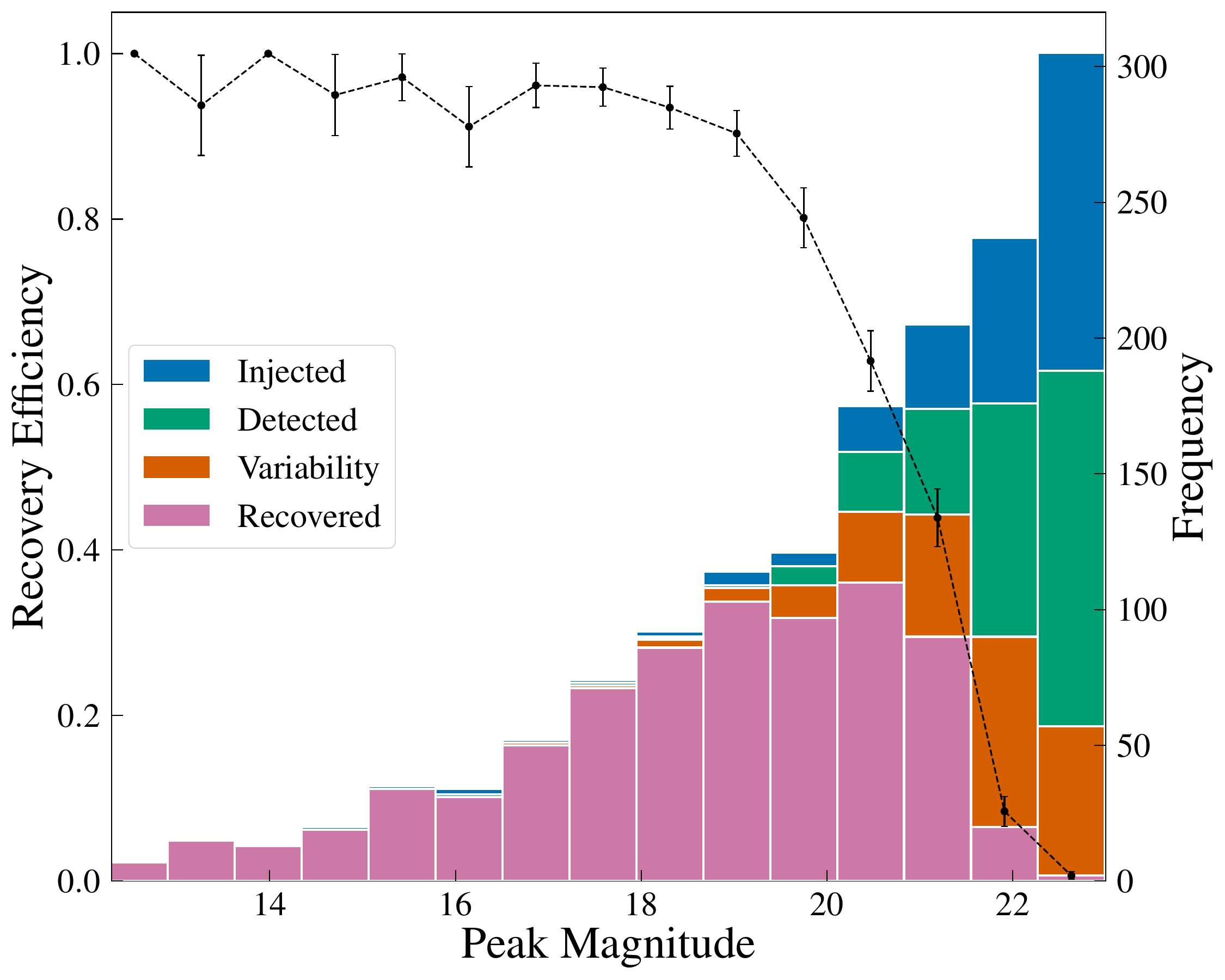}
    \caption{Classification efficiency of injected afterglows with magnitude and the distribution of a sample of injected afterglows.  The left-hand axis corresponds to the histogram of injected afterglows, the blue denotes the total afterglows that were  injected, green denotes those that had at least five detections, orange denotes those that had $\eta_{\mathrm{event}}^{-1} > 0.6$ and purple denotes those that are recovered by our classifier.  The right-hand axis and black line and points show the recovery efficiency of our classifier at each magnitude bin.  The total classification efficiency for events with peak magnitudes brighter than where our classification efficiency drops below 50\,\%, 21 AB mag, is $82.86 \pm 1.32$\,\%.  Light curves that are not recovered are primarily afterglows with slower fade rates in our observations and are outside the scope of this work.  The efficiency curve from this plot has been propagated into our rate calculations and jet constraints shown in Figures \ref{fig:expected_afterglows} and \ref{fig:probabilities}.} 
    \label{fig:efficiency}    
\end{figure}

\subsection{Expected Rates of Orphan Afterglows in DWF}\label{sec:detectability}

We use the sample of afterglows generated with the formalism described in Section \ref{sec:afterglow_sample} to evaluate the expected rate of OAs extracted from the DWF data \ref{sec:classifier}. We simulate DWF's observing strategy when generating light curves, using a distribution of minute-cadence observing windows based on the distribution of those in the data itself.  The values of time length of each observing window, $t_{\mathrm{window}}$, were binned for the data used in this search to generate typical $t_{\mathrm{window}}$ values for the synthetic light curves.  The simulated afterglow start times, defined by the time the prompt GRB is emitted, are uniformly distributed between 24\,hr before the beginning of the observing window and the end of the observing window.

We inject {\scshape afterglowpy} light curves into DWF images (as discussed in Section \ref{sec:fakes}) to determine classification efficiencies as a function of the magnitude of the light curve's peak magnitude, shown in Figure \ref{fig:efficiency}.  These values are propagated into our rate estimates.  For each simulated afterglow, we predict an efficiency based on its peak magnitude by interpolating the efficiency bins shown in Figure \ref{fig:efficiency}.  Each afterglow with a detection with $g<23$ AB mag is weighted based on this predicted efficiency.

\begin{figure*}
    \includegraphics[width=\textwidth]{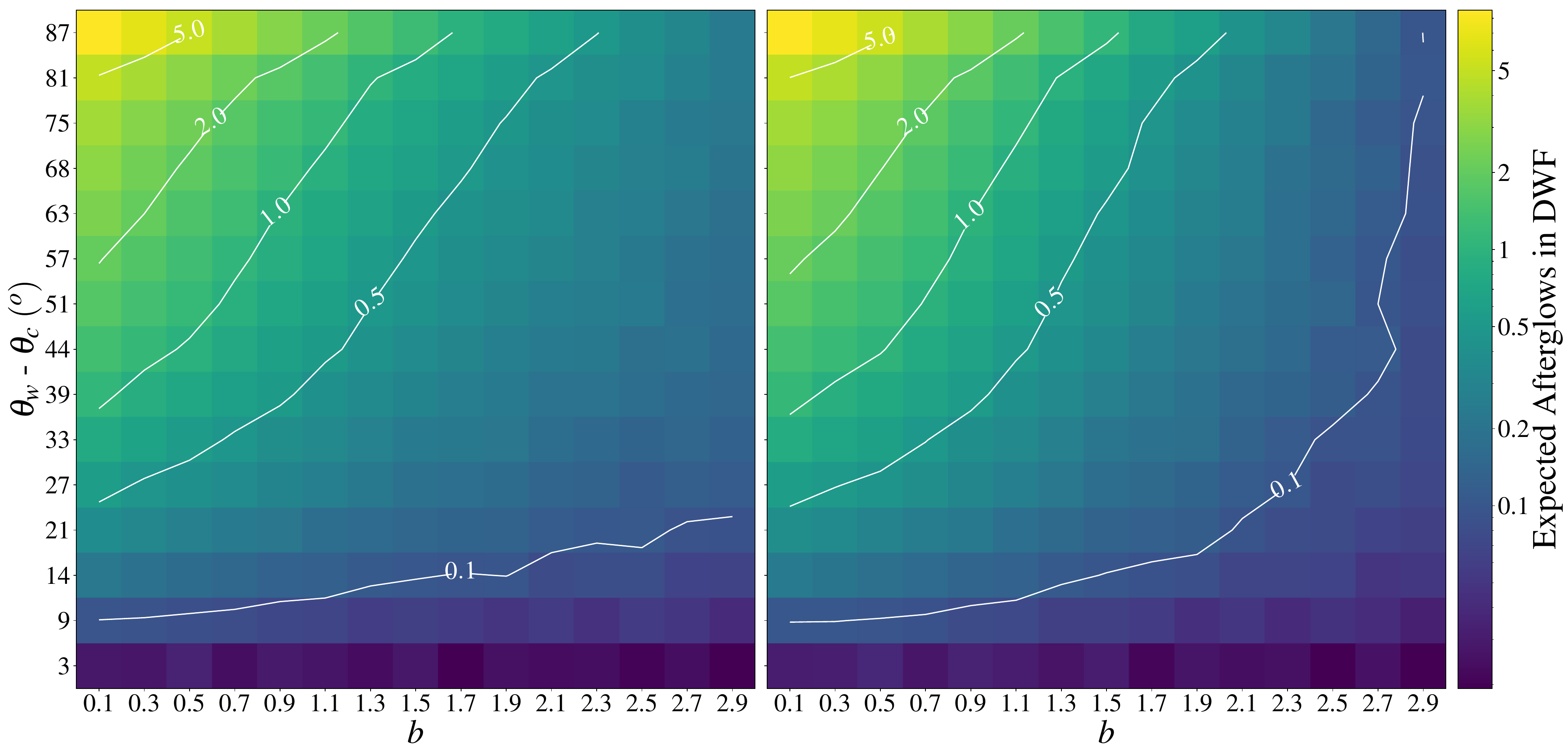}
    \caption{Expected value for the number of detectable afterglows in the DWF data for a range of jet profiles.  The left-hand plot shows the results for a smooth power-law model and the right-hand plot shows the results for a power-law with core model.  A jet with $b=0$ and $\theta_w-\theta_c\rightarrow90^{\circ}$ has isotropic afterglow emission whilst keeping the prompt GRB emission restricted to the jet's core.  Conversely, a jet with $b\rightarrow\infty$ and $\theta_w-\theta_c\sim0^{\circ}$ describes a top-hat model, where an afterglow is only detectable where the prompt GRB is also detectable.}
    \label{fig:expected_afterglows}
\end{figure*}

The OA detection rate varies between different values of $b$ and $\theta_w-\theta_c$.  For each of combination of $b$ and $\theta_w-\theta_c$ we can calculate a ratio of DWF-detected afterglows to \textit{Swift}-detected GRBs and an absolute rate with $\mathcal{R}_{\mathrm{\textit{Swift}}}$.  Figure \ref{fig:expected_afterglows} shows the resultant distribution of expected values for OA detections in the DWF data analysed in this search, according to the assumptions made in Section \ref{sec:afterglow_sample}.  Higher rates of OA detection are expected for shallower (small values of $b$), wider (large values of $\theta_w-\theta_c$) structures outside the core of the jet.

\section{Analysing Orphan Afterglow Candidates}\label{sec:results}
We present the candidates found in the DWF data using our classifier (Section \ref{sec:classifier}) in Section \ref{sec:candidates} and analyse the nature of their coincident sources in Section \ref{sec:coincidentsources}.

\subsection{Extracting Candidates}\label{sec:candidates}

The total number of light curves available in the processed data amounts to $2.6\times10^7$.  Requiring $\eta^{-1}_{\mathrm{lc}} > 0.6$ reduces the sample to $3.6\times10^6$ light curves.  Of these, 45961 were extracted from the data to be inspected based on a classification score $>0.7$ or $\eta_{\mathrm{event}}^{-1} > 4$.

We crossmatch our sample to \textit{Gaia} DR3 \citep{Gaia_DR3}, the ASAS-SN catalogue of variable stars \citep{ASAS-SN_Variables} and the catalog for RR Lyrae variable stars in DES Y6 \citep{DES_variables} to determine which sources in our sample are known variable stars.  We also crossmatched to the \textit{Gaia} DR3 to search for sources associated with known stars. We require a match within one arcsecond and a parallax measurement with >$3\sigma$ statistical significance.  Table \ref{tab:candidates} shows the numbers of candidates removed from these cuts, resulting in 33,123 candidates left for human inspection.

The largest contaminant of these candidates were artefacts including edge detections, cosmic rays, CCD pixel faults, crosstalk and saturated sources \citep{unsup_ml,ROBOT}.  The criteria for a candidate passing visual inspection is that no artefacts were evident in the candidate thumbnails during the event.  Once a candidate has passed the visual inspection, a candidate would then be rejected if there was rebrightening on subsequent nights of observation or had brightened and faded on previous nights.  It would also be rejected if a source was not detected with difference imaging of the candidate with stacked science and template images using \textsc{hotpants} \citep{hotpants}.

\begin{table*}
 \centering
 \caption{Evaluation of candidates extracted by our search methodology presented in Section~\ref{sec:candidates}.}
 \label{tab:candidates}
 \begin{tabular*}{1.\textwidth}{@{}l@{\hspace*{30pt}}l@{\hspace*{30pt}}l@{\hspace*{30pt}}l@{\hspace*{30pt}}l@{\hspace*{30pt}}l@{\hspace*{30pt}}l@{\hspace*{15pt}}}
  \hline
  Cut & Total & Known     & 3-$\sigma$ & Inspected & Rejected & Remaining \\
      &       & Variables & Parallax   &           &          &           \\
  \hline
  - & $1.62\times10^7$ & 39193 & $1.16\times10^6$ & - & - & $1.51\times10^7$\\
  No. Detections > 5 & $1.00\times10^7$ & 38130 & $1.13\times10^6$ & - & - & $8.85\times10^6$\\
  $\eta_{\mathrm{lc}}^{-1}>0.6$ & $2.15\times10^6$ & 12774 & $2.49\times10^5$ & - & - & $1.90\times10^6$\\
  $\eta_{\mathrm{event}}^{-1}>4$ & 4325 & 1081 & 1168 & 2743 & 2725 & 17\\
  $\eta_{\mathrm{event}}^{-1}>4$ or Classifier Score > 0.7 & 45960 & 2721 & 11631 & 33123 & 33072 & 51\\
  \hline
 \end{tabular*}
\end{table*}

\begin{table*}
    \centering
    \caption{List of coincident sources associated with OA candidates found in 100 nights of DWF data.  Colours are calculated from sources detected in \textit{DELVE} DR2.  A 3-$\sigma$ association with M-stars calculated from these colours using spectroscopically classified M-stars from \citet{SDSS_Mdwarfs}.}
\begin{tabular*}{\textwidth}{@{}l@{\hspace*{20pt}}l@{\hspace*{20pt}}l@{\hspace*{20pt}}l@{\hspace*{30pt}}l@{\hspace*{30pt}}l@{\hspace*{30pt}}l@{\hspace*{30pt}}}
\hline
Candidate No. &     Field &              Coordinates &  M-star Colours &               g-r &               r-i &               i-z \\
\hline
             0 &   Dusty10 & 10:11:22.85 -79:57:47.28 &     Yes &  $1.59 \pm 0.047$ &  $1.43 \pm 0.014$ & $0.63 \pm 0.0073$ \\
             1 &     Prime & 05:53:45.12 -60:54:55.01 &     Yes &  $1.42 \pm 0.023$ & $1.07 \pm 0.0085$ & $0.46 \pm 0.0036$ \\
             2 &       4hr & 04:08:24.64 -55:31:23.46 &     Yes &  $1.52 \pm 0.072$ &  $1.80 \pm 0.025$ &  $0.71 \pm 0.013$ \\
             3 & FRB010724 & 01:18:40.55 -74:24:40.25 &     Yes &   $1.78 \pm 0.11$ &  $1.64 \pm 0.027$ &  $0.74 \pm 0.013$ \\
             4 & FRB010724 & 01:13:53.74 -74:16:06.34 &     Yes & $1.64 \pm 0.0048$ & $0.55 \pm 0.0023$ & $0.33 \pm 0.0016$ \\
             5 &      NSF2 & 21:27:48.29 -67:35:20.72 &     Yes &  $1.48 \pm 0.015$ & $1.57 \pm 0.0056$ & $0.68 \pm 0.0023$ \\
             6 & FRB131104 & 06:45:04.60 -51:38:18.19 & Unknown &   - &   $>1.07$ &  $0.70 \pm 0.095$ \\
             7 & FRB171019 & 22:21:58.53 -09:33:58.02 &     Yes &  $1.43 \pm 0.012$ & $1.06 \pm 0.0065$ & $0.47 \pm 0.0046$ \\
             8 &   NGC\,6101 & 16:30:00.58 -72:54:01.27 & Unknown &  - &   - &  $0.44 \pm 0.064$ \\
             9 &   NGC\,6101 & 16:14:04.77 -73:25:10.68 &     Yes &  $1.49 \pm 0.017$ & $0.80 \pm 0.0063$ &  $0.42 \pm 0.004$ \\
            10 &   NGC\,6101 & 16:22:30.52 -73:27:12.53 &      No &  $0.61 \pm 0.012$ &   $1.28 \pm 0.01$ &  $0.65 \pm 0.013$ \\
            11 &   NGC\,6101 & 16:34:53.41 -72:35:28.25 &     Yes &  $1.26 \pm 0.045$ &  $0.75 \pm 0.022$ &  $0.37 \pm 0.011$ \\
            12 &   NGC\,6101 & 16:26:28.02 -72:25:06.15 &     Yes &  $1.51 \pm 0.034$ &  $1.49 \pm 0.014$ & $0.69 \pm 0.0038$ \\
            13 &   NGC\,6101 & 16:28:14.03 -72:33:11.39 &     Yes &   $1.71 \pm 0.04$ &  $0.53 \pm 0.029$ & $0.38 \pm 0.0074$ \\
            14 &   NGC\,6101 & 16:25:50.05 -72:47:42.86 &     Yes &  $1.38 \pm 0.043$ &   $0.94 \pm 0.02$ & $0.47 \pm 0.0079$ \\
            15 &   NGC\,6101 & 16:25:55.02 -73:05:09.30 &     Yes &  $1.42 \pm 0.074$ &  $1.55 \pm 0.029$ & $0.66 \pm 0.0068$ \\
            16 &   NGC\,6101 & 16:21:25.44 -72:06:14.24 &     Yes &  $1.43 \pm 0.096$ &  $1.41 \pm 0.018$ &  $0.61 \pm 0.011$ \\
            17 &   NGC\,6101 & 16:23:20.68 -72:23:12.22 &     Yes &  $1.39 \pm 0.097$ &  $1.96 \pm 0.018$ & $0.87 \pm 0.0065$ \\
            18 &   NGC\,6101 & 16:26:14.27 -72:28:12.30 &     Yes &  $1.52 \pm 0.087$ &  $1.40 \pm 0.036$ & $0.58 \pm 0.0096$ \\
            19 &   NGC\,6101 & 16:29:13.58 -72:47:20.82 &      No &  $0.79 \pm 0.051$ &  $1.09 \pm 0.033$ &  $0.57 \pm 0.011$ \\
            20 &   NGC\,6101 & 16:30:32.43 -72:52:17.17 &     Yes &   $>1.37$ &   $1.46 \pm 0.14$ &  $0.80 \pm 0.038$ \\
            21 &   NGC\,6101 & 16:30:38.65 -73:03:49.42 &      No &  $0.74 \pm 0.029$ &   $0.26 \pm 0.02$ &  $0.13 \pm 0.016$ \\
            22 &   NGC\,6101 & 16:28:51.93 -73:18:13.93 &     Yes &  $1.42 \pm 0.081$ &  $0.78 \pm 0.032$ &  $0.39 \pm 0.014$ \\
            23 &   NGC\,6101 & 16:17:15.82 -73:35:36.97 &     Yes &  $1.58 \pm 0.087$ &  $0.86 \pm 0.025$ &  $0.40 \pm 0.016$ \\
            24 &   NGC\,6101 & 16:27:13.37 -73:42:12.44 &     Yes &   $1.69 \pm 0.12$ &  $1.52 \pm 0.029$ &  $0.66 \pm 0.011$ \\
            25 &   NGC\,6101 & 16:14:36.96 -72:35:02.35 &     Yes &  $1.00 \pm 0.014$ & $0.46 \pm 0.0048$ & $0.30 \pm 0.0049$ \\
            26 &   NGC\,6101 & 16:36:52.53 -72:34:12.37 & Unknown & - & - &  $0.52 \pm 0.047$ \\
            27 &   NGC\,6101 & 16:30:37.04 -72:47:21.09 &     Yes &  $1.44 \pm 0.028$ &  $1.29 \pm 0.012$ & $0.57 \pm 0.0039$ \\
            28 &   NGC\,6101 & 16:34:11.76 -72:42:39.52 &     Yes &   $1.36 \pm 0.16$ &  $1.84 \pm 0.067$ &  $0.69 \pm 0.012$ \\
            29 &   NGC\,6101 & 16:34:14.93 -72:51:33.48 &     Yes &    $1.10 \pm 0.1$ &  $0.42 \pm 0.045$ &  $0.23 \pm 0.039$ \\
            30 & FRB190711 & 21:54:39.94 -80:38:04.39 &     Yes &   $>0.88$ &   $1.85 \pm 0.13$ &  $0.75 \pm 0.046$ \\
            31 & FRB190711 & 22:07:46.09 -79:59:53.17 &     Yes &  $1.67 \pm 0.067$ &  $1.33 \pm 0.013$ & $0.68 \pm 0.0096$ \\
            32 &      14hr & 14:41:39.11 -77:51:37.23 &     Yes &  $1.56 \pm 0.052$ &  $0.86 \pm 0.017$ &   - \\
            33 &      14hr & 14:37:36.02 -78:00:35.74 &     Yes &  $1.02 \pm 0.025$ &  $0.40 \pm 0.014$ &   - \\
            34 &      14hr & 14:43:56.32 -77:32:37.47 &     Yes &  $1.56 \pm 0.064$ &  $1.52 \pm 0.018$ &   - \\
            35 &      14hr & 14:37:42.80 -77:25:06.19 &     Yes &   $1.08 \pm 0.18$ &  $2.02 \pm 0.049$ &   - \\
            36 &      14hr & 14:39:34.38 -78:21:58.62 &     Yes &  $1.65 \pm 0.059$ &  $1.59 \pm 0.013$ &   - \\
            37 &   Dusty12 & 11:44:55.09 -84:11:32.95 &     Yes &   $>1.24$ &  $1.88 \pm 0.064$ &  $0.78 \pm 0.029$ \\
            38 &   NGC\,6744 & 19:08:00.50 -64:27:09.92 &     Yes &  $1.33 \pm 0.011$ & $0.72 \pm 0.0076$ & $0.36 \pm 0.0055$ \\
            39 &   NGC\,6744 & 19:07:37.88 -64:35:56.49 &     Yes &   $1.66 \pm 0.03$ &  $1.01 \pm 0.017$ &  $0.42 \pm 0.011$ \\
            40 &   NGC\,6744 & 19:02:04.71 -65:00:56.39 &     Yes &  $1.52 \pm 0.029$ &  $1.46 \pm 0.015$ & $0.62 \pm 0.0067$ \\
            41 &   NGC\,6744 & 18:59:10.89 -64:24:41.32 &      No &  $0.38 \pm 0.024$ &  $0.13 \pm 0.027$ &   $0.05 \pm 0.05$ \\
            42 &   NGC\,6744 & 19:07:10.75 -64:45:15.79 &     Yes &  $1.42 \pm 0.022$ &  $1.39 \pm 0.013$ &  $0.66 \pm 0.006$ \\
            43 &   NGC\,6744 & 19:00:03.66 -64:52:11.64 &     Yes &  $1.42 \pm 0.071$ &  $1.53 \pm 0.042$ &  $0.61 \pm 0.017$ \\
            44 &   NGC\,6744 & 19:02:43.34 -65:01:23.42 &     Yes &  $1.42 \pm 0.029$ &  $0.79 \pm 0.012$ & $0.41 \pm 0.0089$ \\
            45 &   NGC\,6744 & 19:02:09.04 -64:02:51.72 &     Yes &  $1.61 \pm 0.018$ & $1.44 \pm 0.0089$ & $0.65 \pm 0.0045$ \\
            46 &   NGC\,6744 & 19:01:44.94 -64:21:14.40 &     Yes &  $1.46 \pm 0.022$ &  $1.14 \pm 0.012$ & $0.49 \pm 0.0072$ \\
            47 &   NGC\,6744 & 19:12:34.76 -63:52:33.60 &      No &   $1.53 \pm 0.13$ &   $0.59 \pm 0.15$ &  $0.67 \pm 0.097$ \\
            48 &   NGC\,6744 & 19:08:41.89 -64:36:38.50 &      No &  $0.72 \pm 0.053$ &  $1.07 \pm 0.044$ &  $0.52 \pm 0.027$ \\
            49 &   NGC\,6744 & 19:02:00.78 -64:30:14.56 &     Yes &  $1.64 \pm 0.062$ &  $1.62 \pm 0.029$ &  $0.72 \pm 0.011$ \\
            50 &   NGC\,6744 & 19:03:07.89 -63:54:25.46 &     Yes &  $1.44 \pm 0.018$ &   $1.45 \pm 0.01$ & $0.64 \pm 0.0051$ \\
\hline
\end{tabular*}
\end{table*}

\subsection{Analysing Candidates' Coincident Sources}\label{sec:coincidentsources}

\begin{figure}
    \includegraphics[width=\columnwidth]{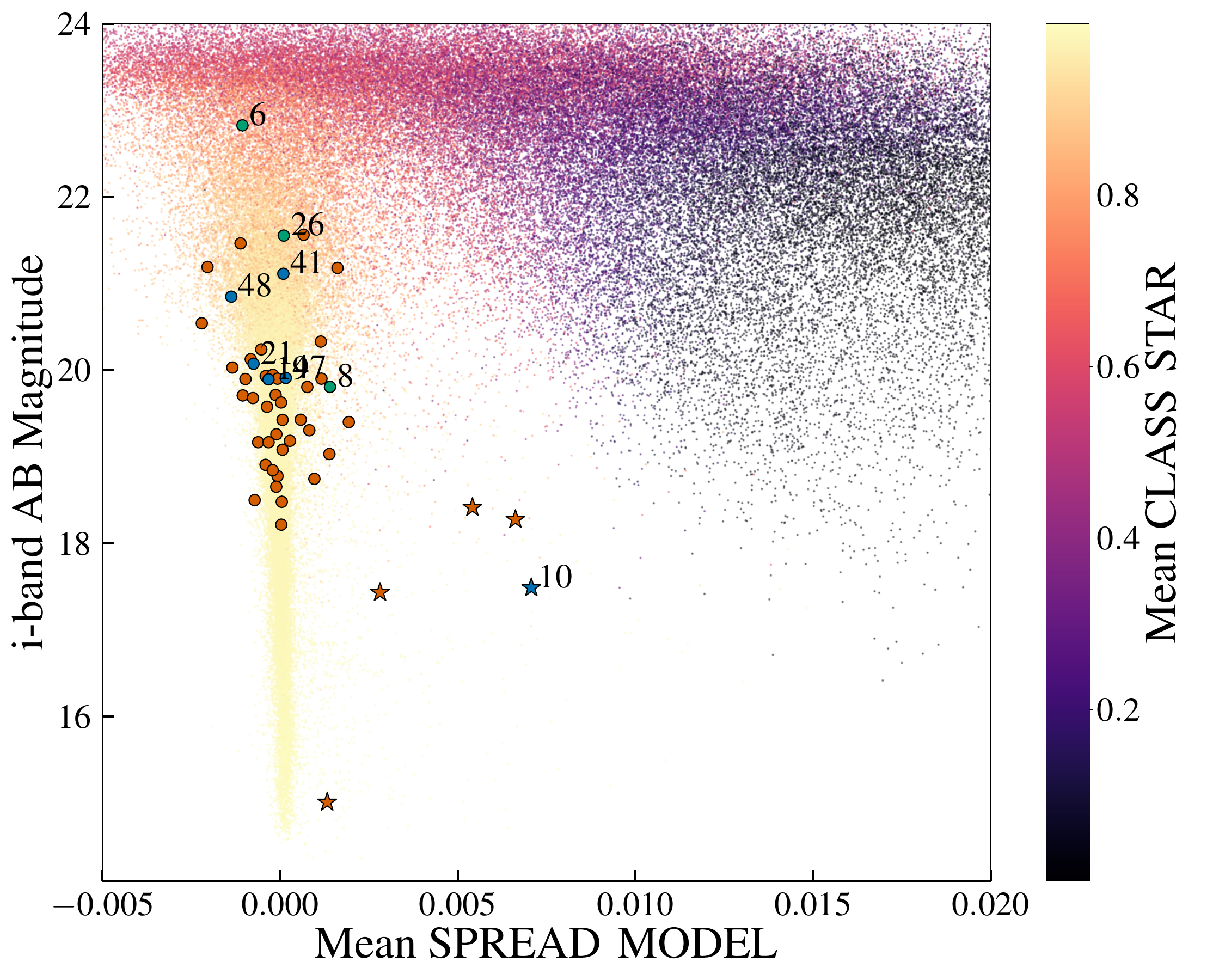}
    \caption{The mean of \textsc{sextractor}'s SPREAD\_MODEL parameter across $g$, $r$, $i$ and $z$-bands plotted against the $i$-band AB magnitude for the coincident sources of the candidates found in this work and a sample of sources detected in \textit{DELVE}.  The \textit{DELVE} sample is coloured with their the mean of the CLASS\_STAR parameter across $g$, $r$, $i$ and $z$-bands.  CLASS\_STAR is the output to a star/galaxy classifier that is run with \textsc{sextractor} where a source that appears star-like is near a value of 1.0 and galaxies typically range from 0.0 to $\gtrsim$ 0.9.  The candidates' coincident sources are plotted with the same colours as in Figure \ref{fig:colour-colour}.  They are plotted as stars if they are detected as two distinct sources in, within one arcsecond, in \textit{Gaia} DR3.  Two blended point sources within one arcsecond of each other would be detected as a single extended source with typical atmospheric seeing conditions.  We therefore conclude that all of the coincident sources are amongst the distribution of stars in the \textit{DELVE} sample.}
    \label{fig:spread_model}    
\end{figure}

After this process, 51 candidates remained.  All of these candidates possess coincident sources detected in \textit{DELVE}'s second data release \citep{DELVE}.  We use \textsc{sextractor} SPREAD\_MODEL parameter, a star/galaxy classifier based on PSF models, for these candidates and plot the results in Figure \ref{fig:spread_model}.  We find that all but four of our candidates are consistent with a single point-source.  The four sources which have large SPREAD\_MODEL values, uncharacteristic of point sources, are resolved as two distinct sources in \textit{Gaia} DR3. Therefore, we cannot conclude that any of our 51 candidates have extended, galaxy-like hosts.

\begin{figure*}
    \includegraphics[width=0.49\textwidth]{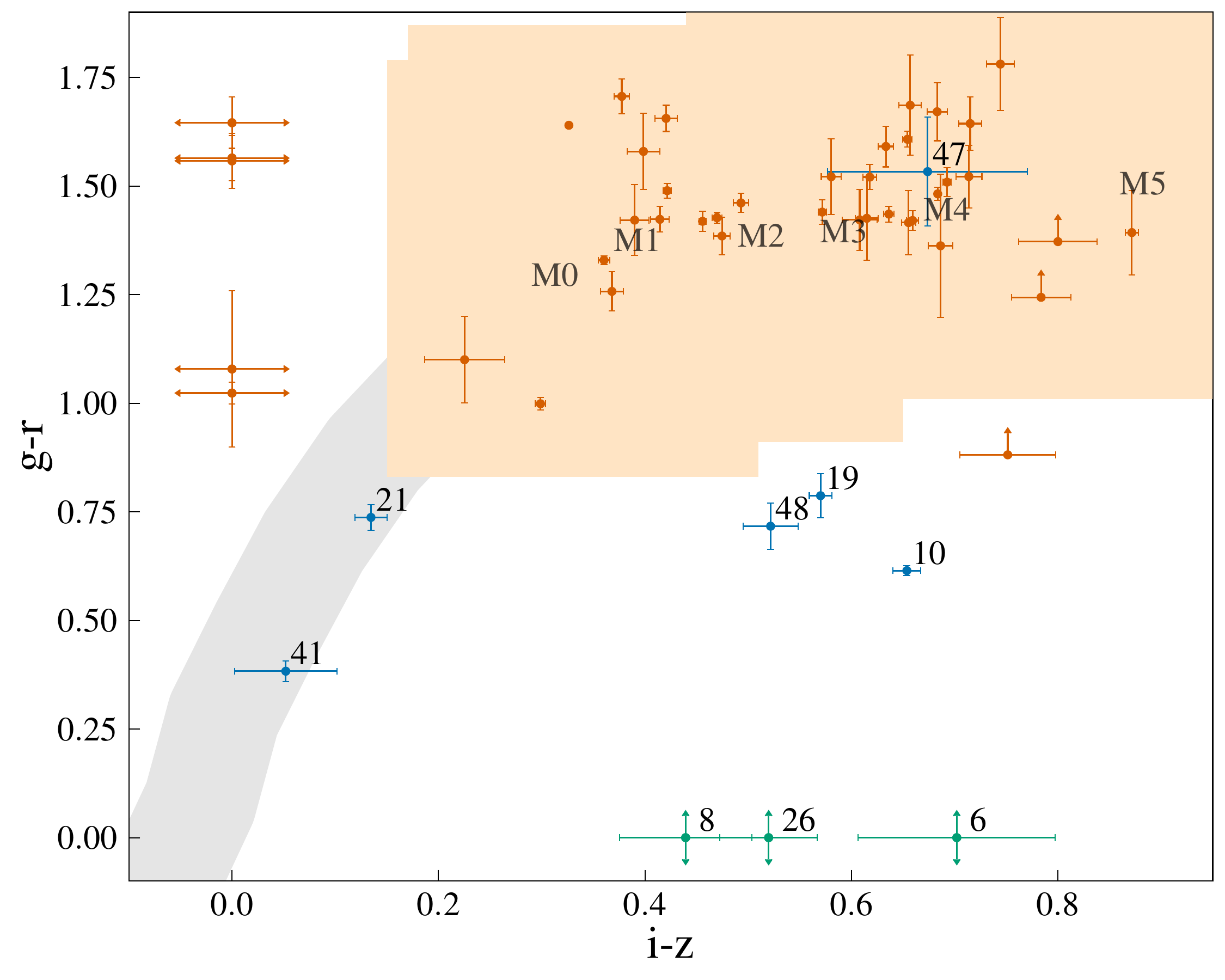}
    \includegraphics[width=0.49\textwidth]{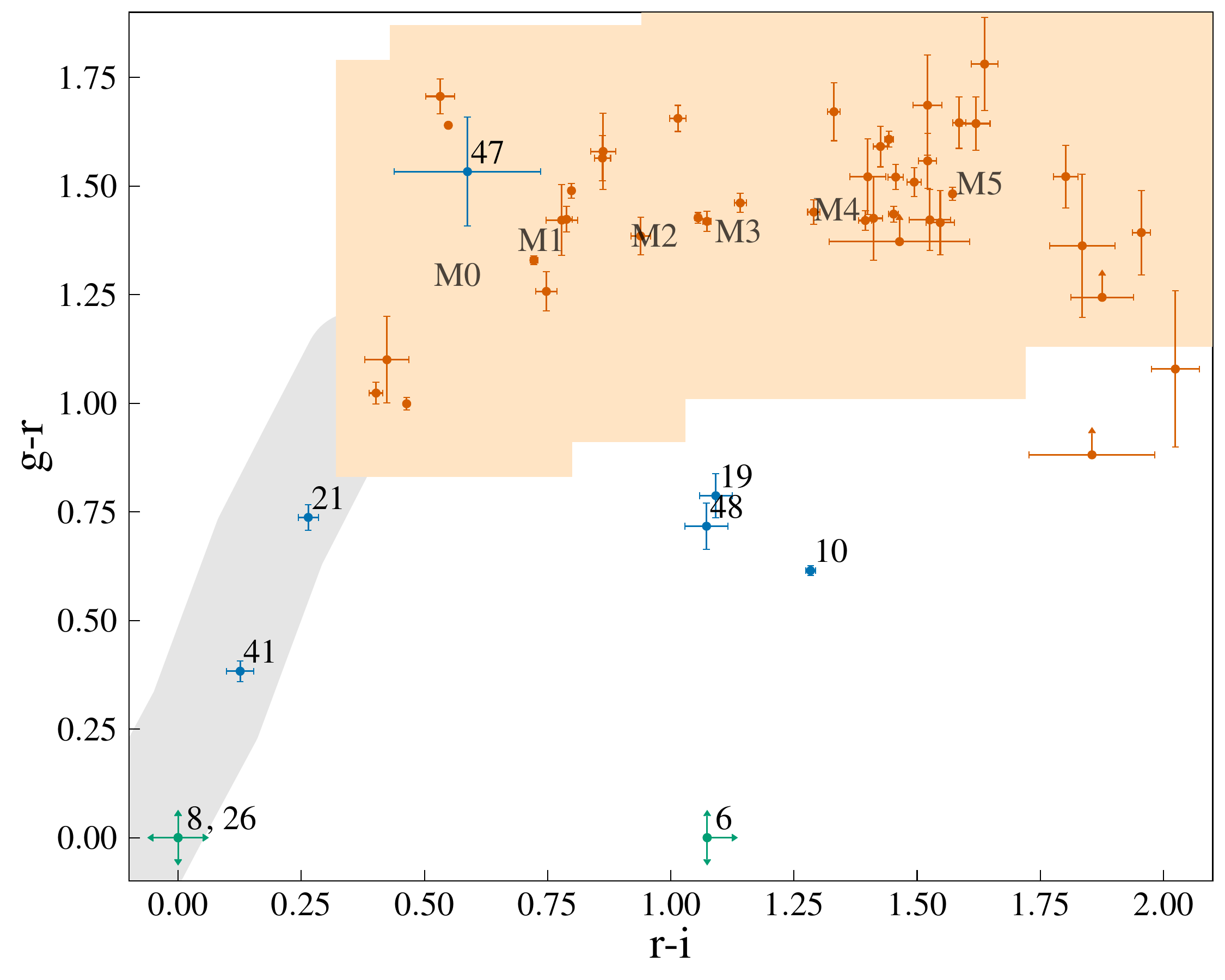}
    \includegraphics[width=0.49\textwidth]{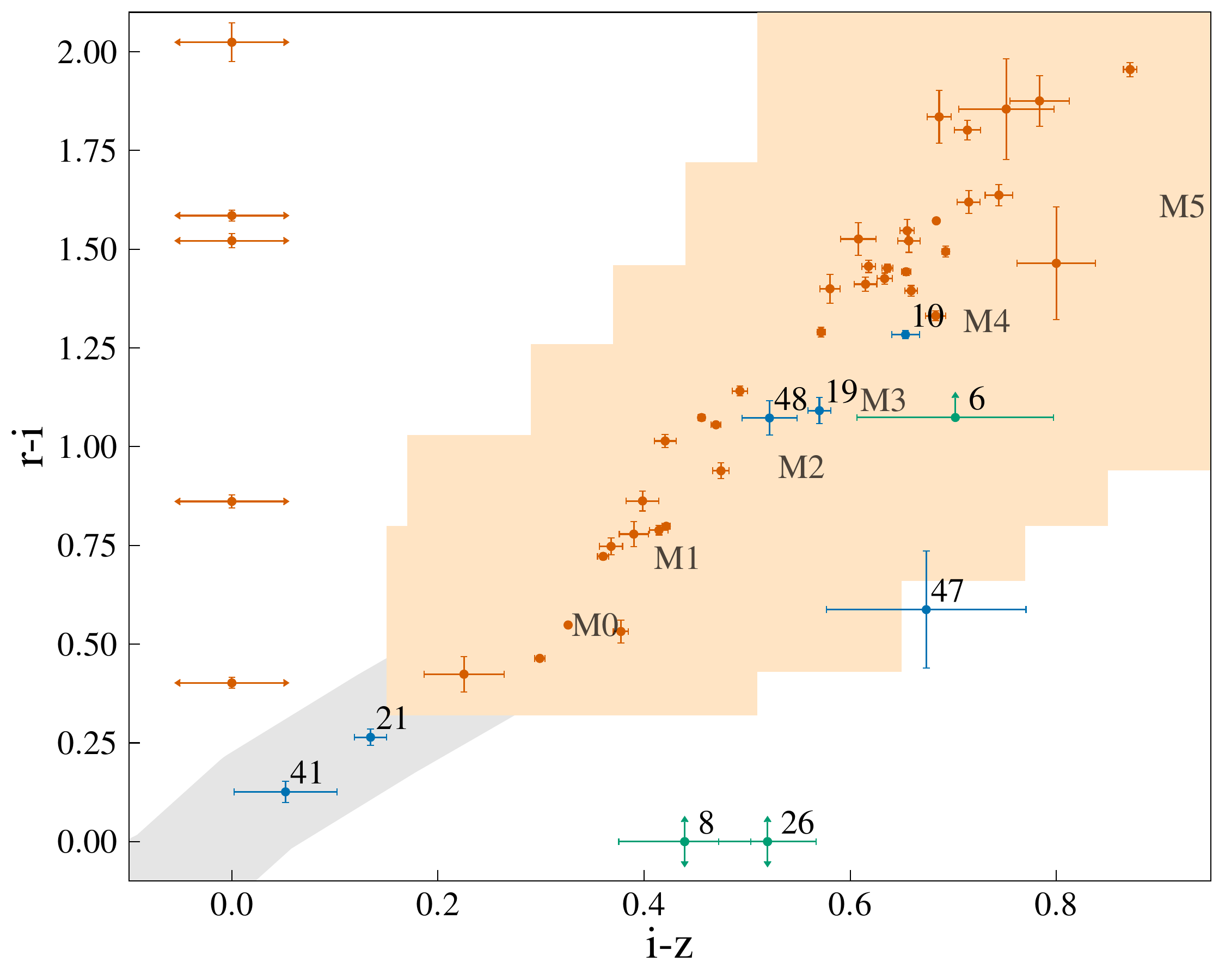}
    \caption{Colour-colour diagrams of the coincident sources, from the \textit{DELVE} catalog, associated with the OA candidates found in this work.  The orange shaded regions are the expected distributions of M-star classes 1-5 \citep{SDSS_Mdwarfs}.  The grey regions denote the main-sequence from DES colour transformations \citep{DES_DR1} to the spectral flux library described in \citet{MS}. An OA candidate coincident source falling in this region indicates that the observed transient is likely a stellar flare.  Orange points are candidates that have a coincident source consistent with a given M-star class to within 3-$\sigma$, Green points are OA candidates without enough coincident source colour information to make a determination of their nature and blue points are OA candidates with coincident source colours inconsistent with M-star with 3-$\sigma$ confidence.}
    \label{fig:colour-colour}    
\end{figure*}

\begin{figure*}
    \includegraphics[width=0.33\textwidth]{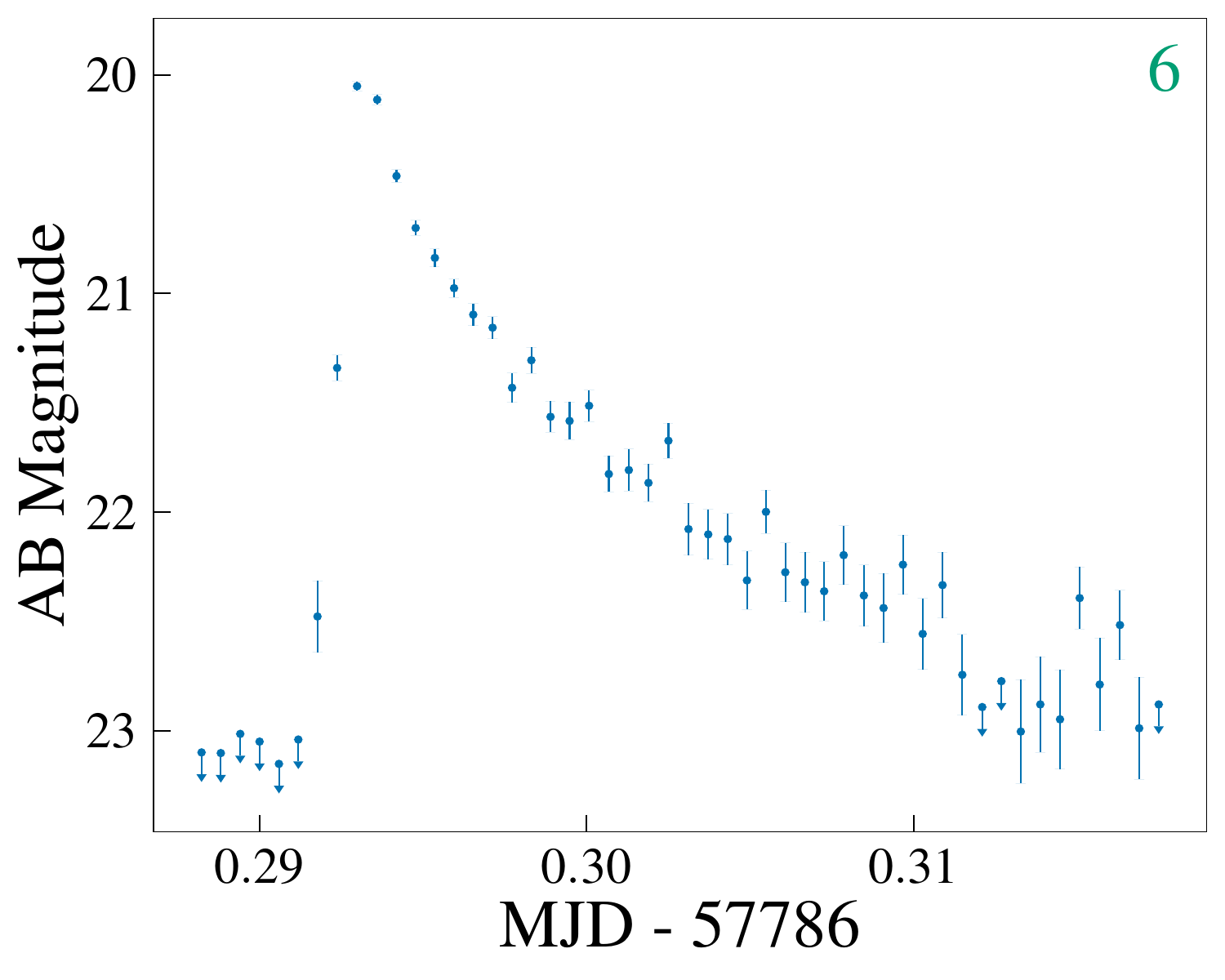}
    \includegraphics[width=0.33\textwidth]{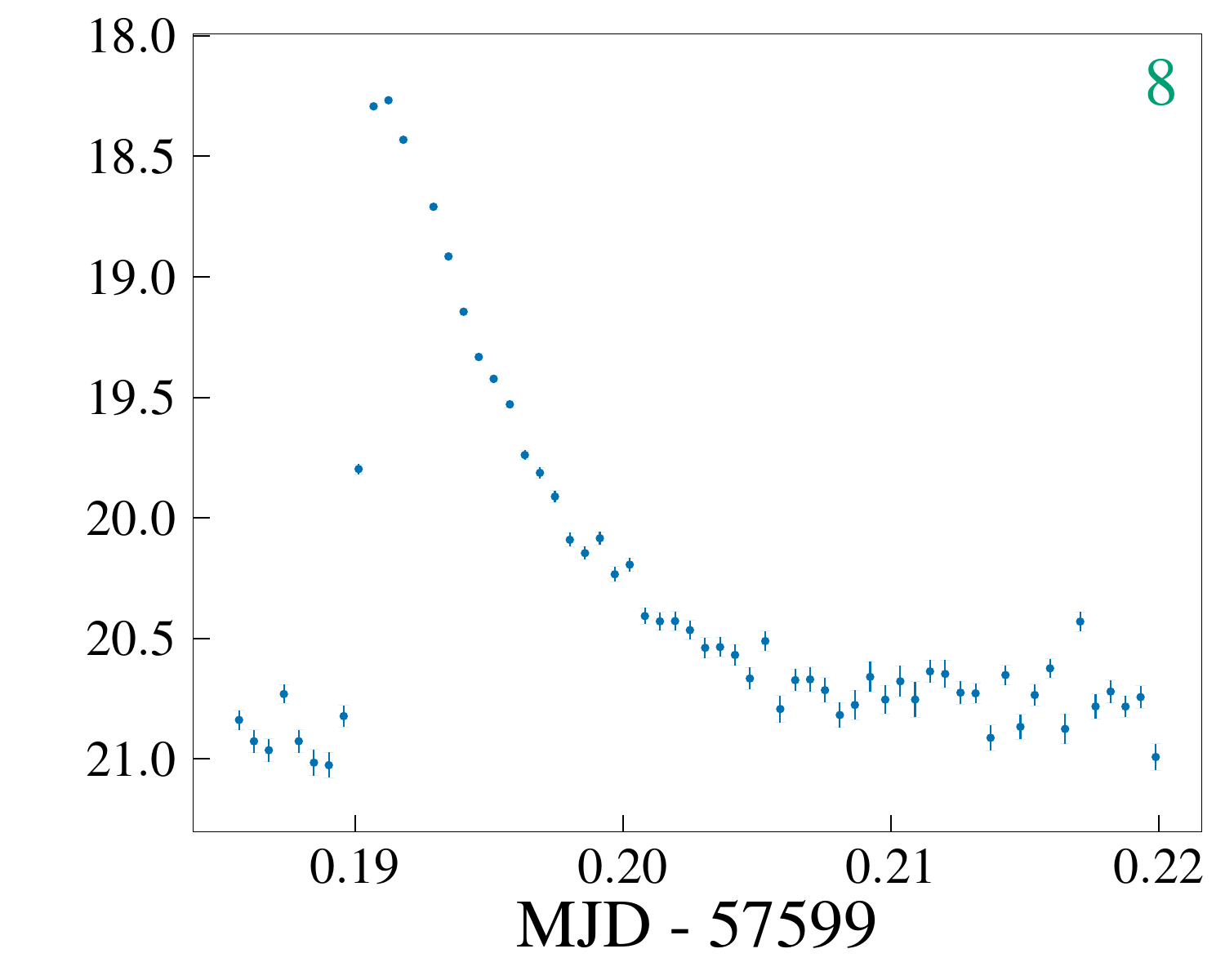}
    \includegraphics[width=0.33\textwidth]{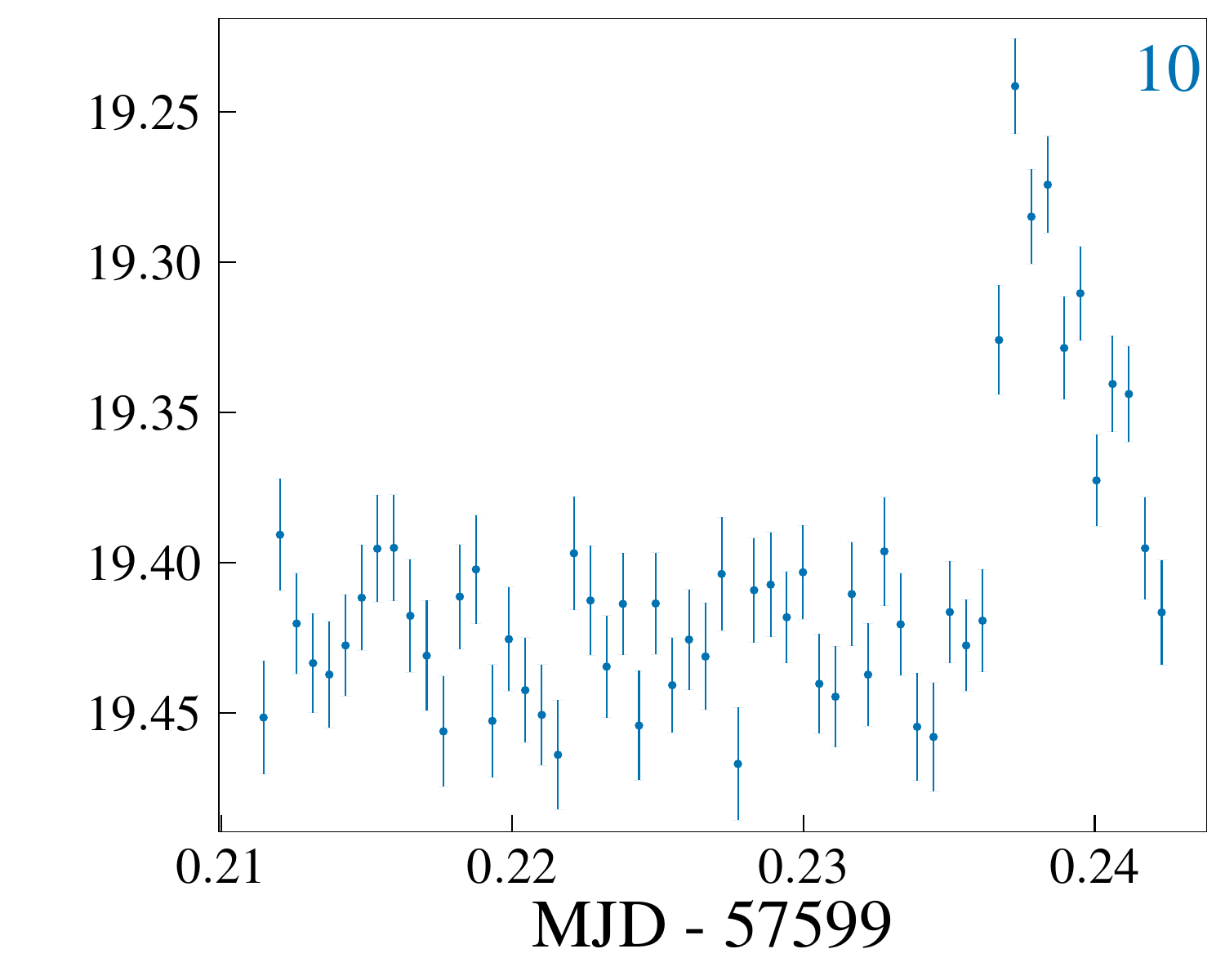}
    \includegraphics[width=0.33\textwidth]{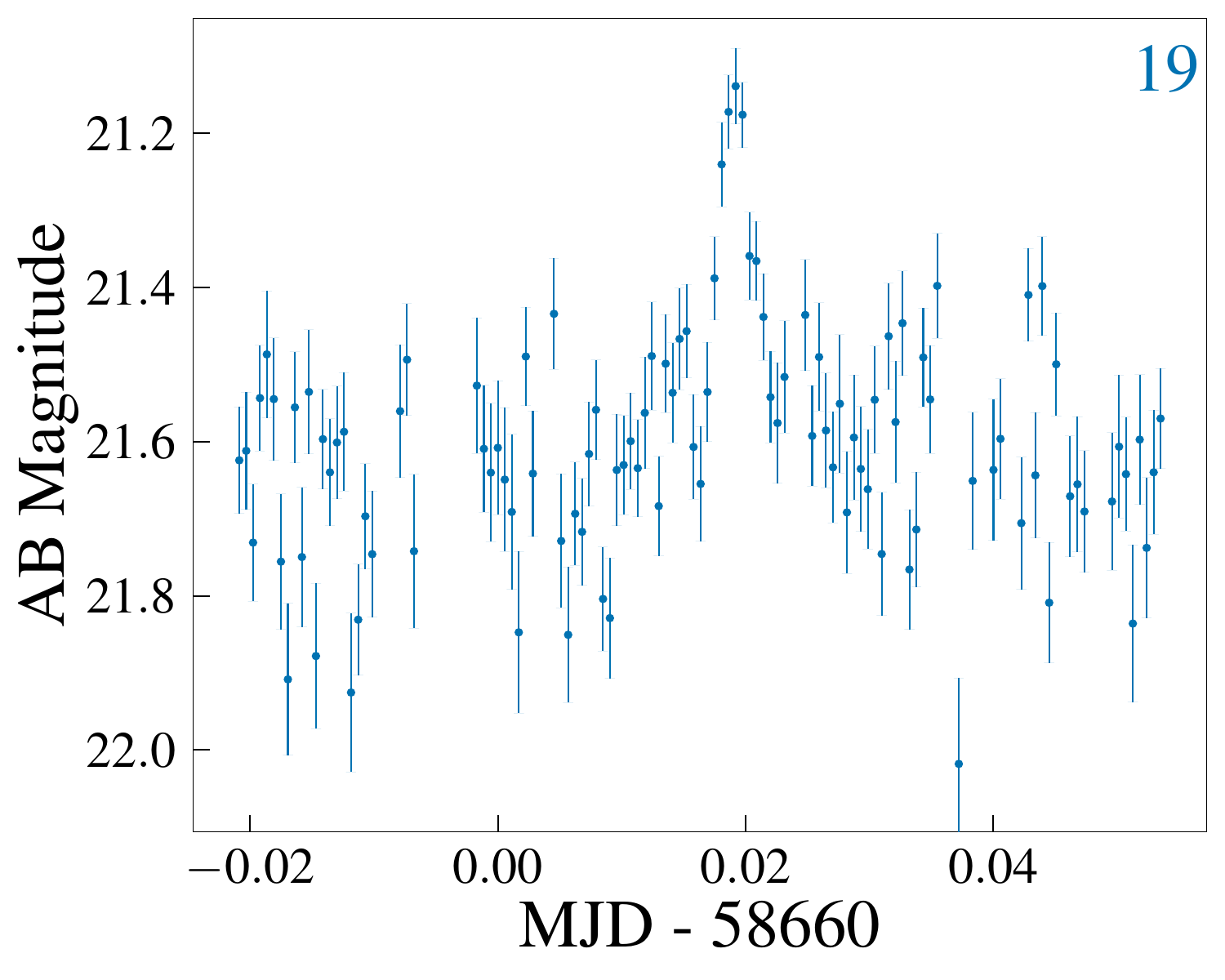}
    \includegraphics[width=0.33\textwidth]{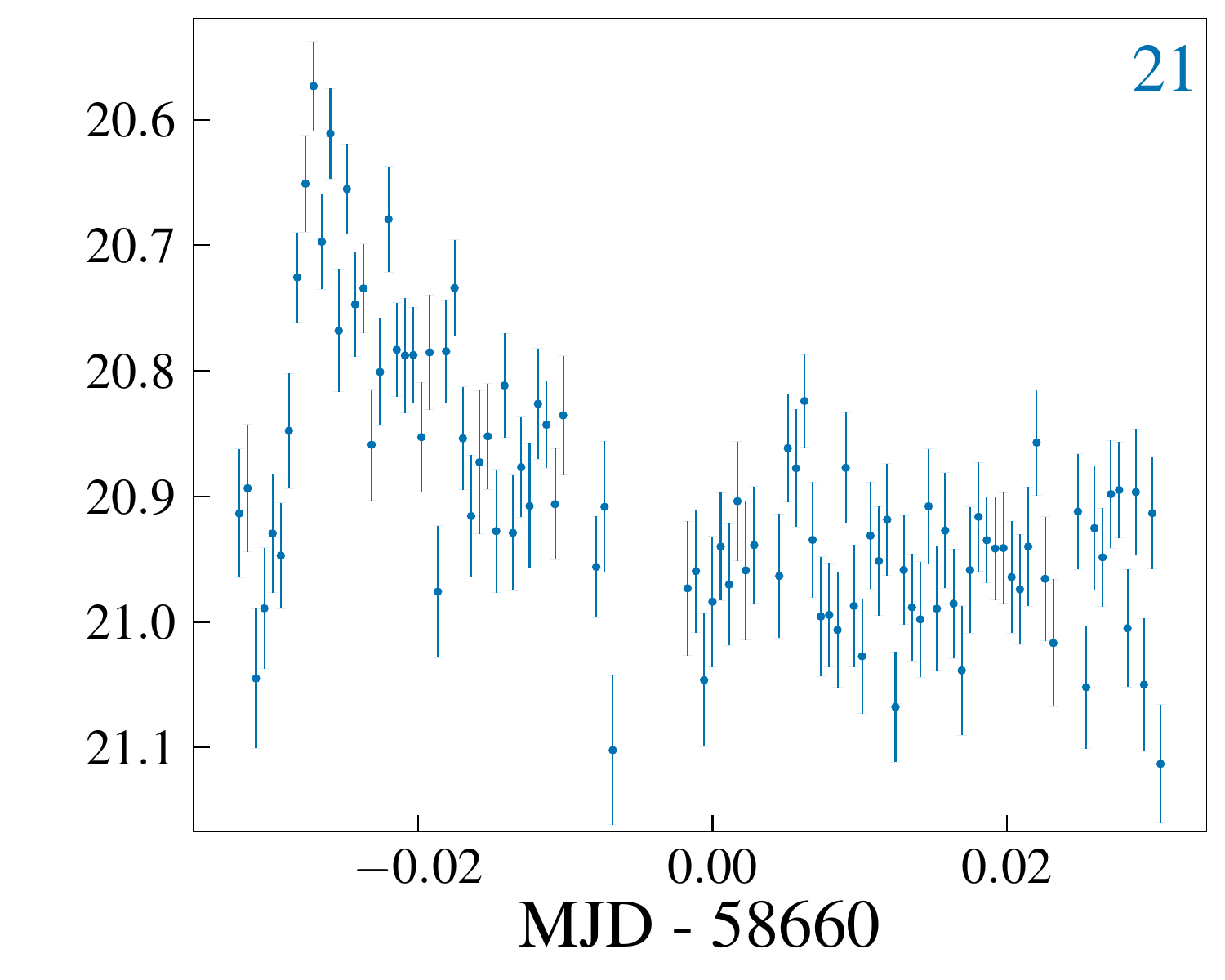}
    \includegraphics[width=0.33\textwidth]{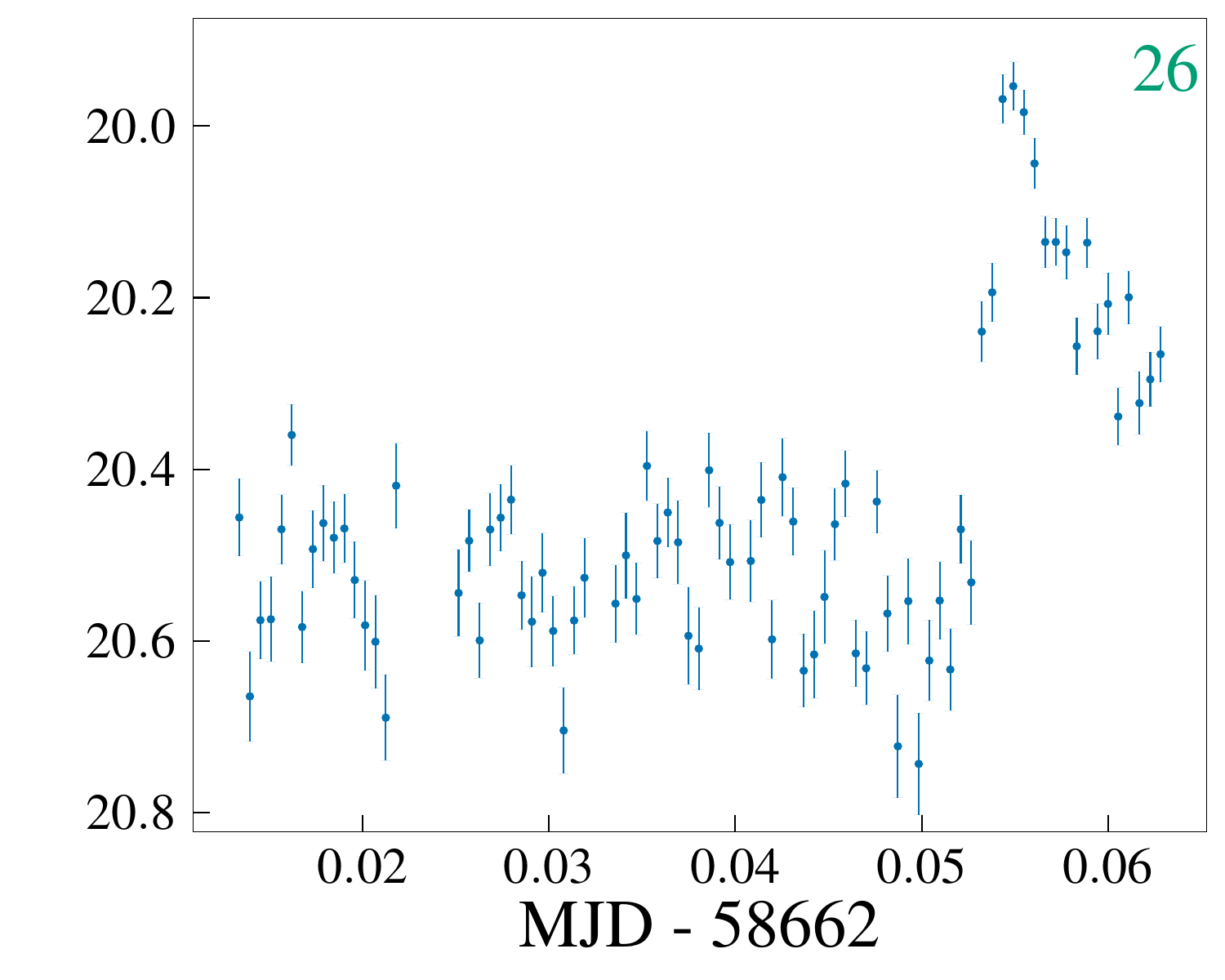}
    \includegraphics[width=0.33\textwidth]{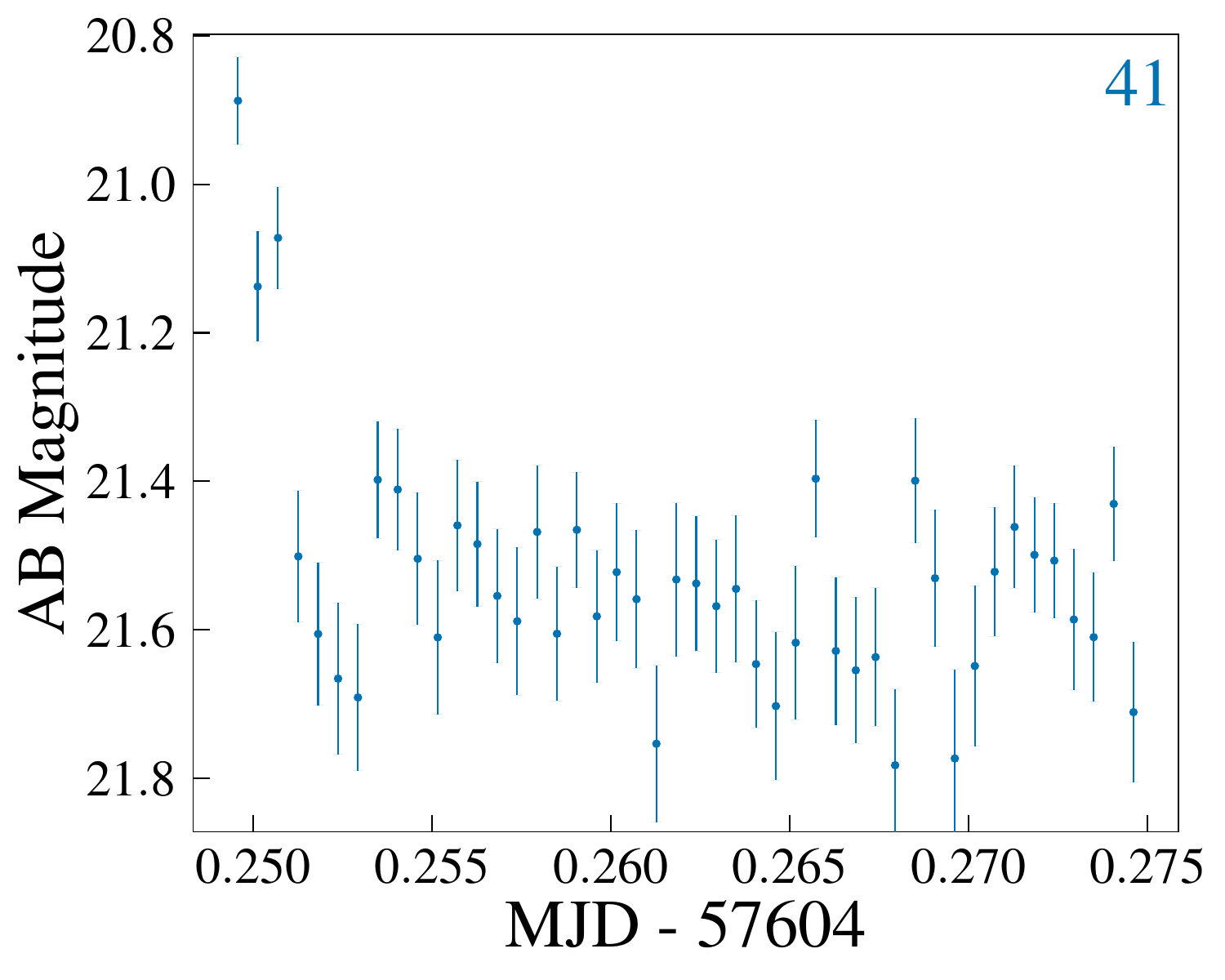}
    \includegraphics[width=0.33\textwidth]{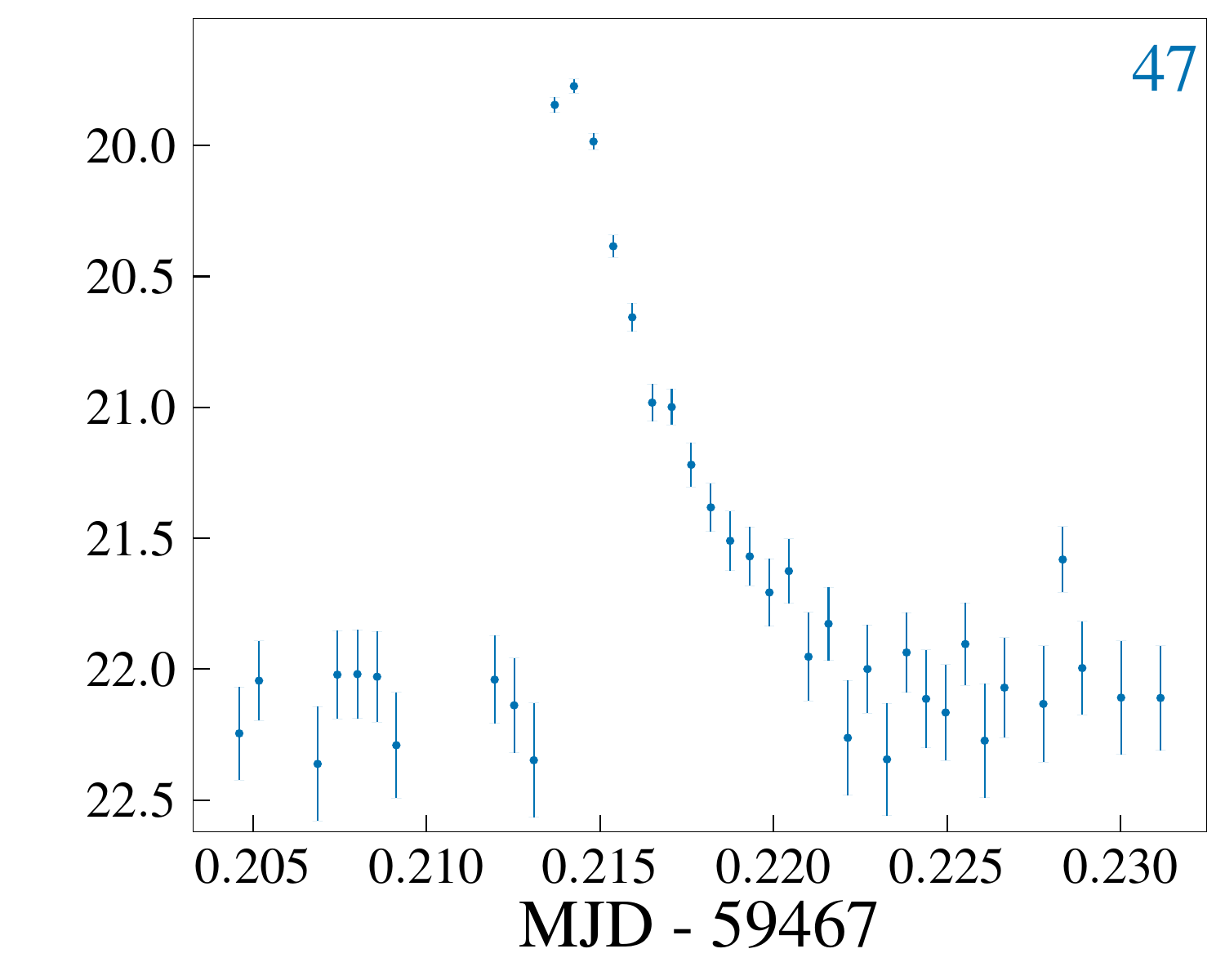}
    \includegraphics[width=0.33\textwidth]{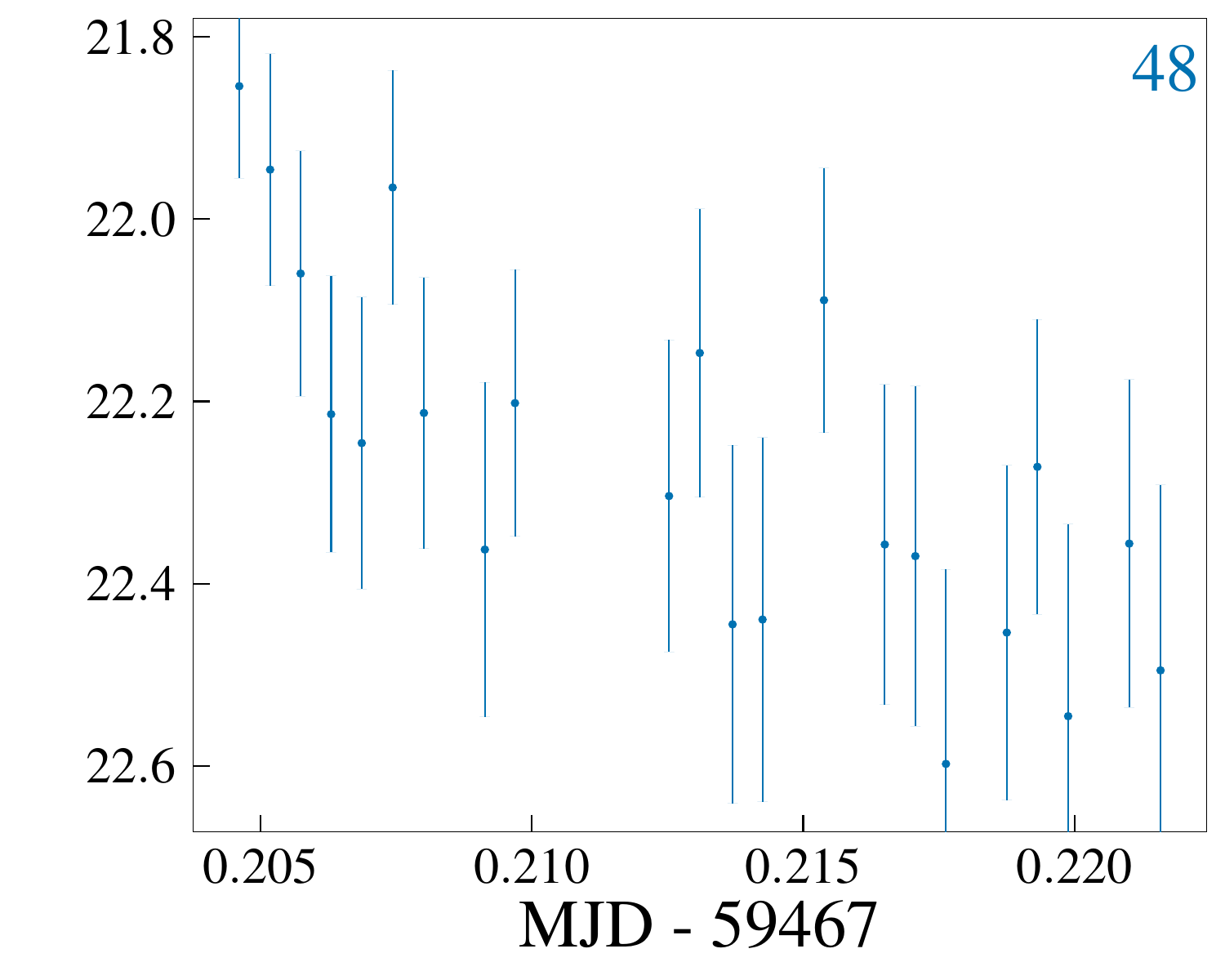}
    \caption{The light curves of each of the nine candidates that do not fall within the M-star regions in Figure \ref{fig:colour-colour}.  We plot the $g$-band AB magnitude against MJD across the observing window in which the transient was detected.  In the top right of each panel, we show the candidate number which is consistent across Table \ref{tab:candidates}, Figure \ref{fig:colour-colour} and Figure \ref{fig:spread_model}.  Candidate 6 was reported in \citet{DWF2}, DWF17x.}
    \label{fig:lcs}    
\end{figure*}

In the absence of a conclusive detection of an extragalactic host for any of the candidates, we analyse the coincident sources' colours to determine whether they are consistent with M-stars or other main sequence stars, as OA and M-dwarf flare light curves can look very similar. To achieve this, we used \textit{DELVE} PSF photometry, supplemented with our measurements in the case of highly blended sources.  In Figure \ref{fig:colour-colour}, we plot the coincident sources' colours along with the observed 3-$\sigma$ distribution of spectroscopically confirmed M-stars reported in \citep{SDSS_Mdwarfs}.  We find that 42 of the coincident sources have colours that are placed within the M-star regions, shown as orange points in Figure \ref{fig:colour-colour}.  Their colours are a strong indication that the candidates may be stellar flares and we classify them here as such.  The light curves of the remaining nine candidates are plotted in Figure \ref{fig:lcs} and are denoted by green or blue points in Figure \ref{fig:colour-colour}.

The three green points in Figure \ref{fig:colour-colour} mark coincident sources where non-detection in two or more of the filters have prevented an M-star classification test.  

The six blue points in Figure \ref{fig:colour-colour} mark coincident sources that are inconsistent with a single M-star.  Two of these coincident sources lie on the main-sequence region in colour-colour space.  While flares from stars bluer than M-stars are comparatively rare, it is likely that some subset of the flares we observe would originate from higher mass stars \citep{HR_flares}.  The other four candidates (10, 19, 47, and 48) have conicident sources that possess colours inconsistent with a main-sequence star.  Candidate 10 is resolved as two distinct sources with an angular separation of less than one arcsecond in \textit{Gaia} DR3.  This explanation could apply to candidates 19 and 48, with smaller angular separations.  This hypothesis is supported by the fact that they have a shallower colour evolution between $g$ and $r$-bands but otherwise have photometry in $r$,$i$ and $z$-band that are consistent with the M-star population.

We also note that candidates 8, 10, 19, 21 and 26 were identified in NGC\,6101 field, which has Galactic latitude close to zero compared to other fields (see Table~\ref{tab:dwf_fields}).  At these galactic latitudes, the stellar density is higher which favours these candidates to be Galactic events.

Candidate 47's coincident source is unusual in its colour evolution compared to the rest of the candidates.  It falls within the M-star region of colour-colour space in Figure \ref{fig:colour-colour} but varies in colour between an M0 and M5 class star.  Despite being a point source, the nature of candidate 47 is unknown.

We fit the light curves in Figure \ref{fig:lcs} to \textsc{afterglowpy} models with varying results in Figure \ref{fig:lc_fits}.  However, we note in Appendix \ref{sec:modelfitting} and Figure \ref{fig:lc_fits_flares} that these fits are not sufficient to rule out a stellar flare explanation.

In Section \ref{sec:fakes} we note that the host galaxies of the theoretical population of OAs explored in this work are expected to predominantly exhibit apparent magnitudes fainter than 23 AB mag.  In Figure \ref{fig:spread_model} we see that none of our candidates satisfy this criteria.  Thus, despite a coincident source having colours that are unexpected for a stellar flare, they are inconsistent with the expected properties of LGRB host galaxies.

\section{Discussion}\label{sec:discussion}

\subsection{Constraints on Jet Structure}

For all of the candidates found in this search, we find that all have coincident sources that are consistent with a point source (Figure \ref{fig:spread_model}).  A subset do not possess colours that are consistent with the M-star population and cannot be confidently associated with a stellar flare.  However, there is a lack of evidence for any of them being associated with an extragalactic host.  We therefore conclude that we have not found any OAs using the procedure in this work.

We calculate the rate at which we would expect to detect a single OA in the data searched using:
\begin{equation}
    \mathcal{R}_{\mathrm{OA}} =  (\Omega \times t_{\mathrm{search}} \times e_{\mathrm{OA}})^{-1}
\end{equation}

Where $\mathcal{R}_{\mathrm{OA}}$ is the OA rate, the average sky coverage of a single pointing is $\Omega=2.14$\,deg$^2$, the efficiency for OAs with a peak detection brighter than $g=22$, drawn from a uniform distribution of $0 < b < 3$, is $e_{\mathrm{OA}} = 0.68$.  Since we distribute the the burst times of our synthetic afterglows up to one day before the start of each observing window, the effects of OA burst times on their detectability is absorbed into the efficiency. We therefore consider the length of time searched to be one day per field per night, $t_{\mathrm{search}}=100$\,d. As a result, with no convincing OA detected in this work, we place an upper limit on the rate of OAs to $g<22$ AB mag of $\mathcal{R}_{\mathrm{OA}}<7.46$\,deg$^{-2}$yr$^{-1}$ at the 95\,\% confidence level and $\mathcal{R}_{\mathrm{OA}}<2.49$\,deg$^{-2}$yr$^{-1}$ at 63.2\,\% confidence.  

This is a novel search, probing an unexplored parameter space.  It is therefore difficult to make a direct comparison to previous work. \citet{DWF2} placed a similar upper limit for extragalactic fast transients of 1.63\,deg$^{-2}$d$^{-1}$.  In this work, however, the authors restricted their search to transients rising and fading within a single observing window.  As our search is sensitive to OAs with burst times up to one day before each observing window, the rate constraints in this work and \citet{DWF2} are not directly comparable. Previous rate constraints on OAs, such as \citet{ZTF_OA_search} place the OA rate from dirty fireballs to be not significantly larger than the LGRB population.  However, the work here probes down to minute-timescales and assumes a luminosity function and light curves from OAs originating from misaligned structured jets rather than dirty fireballs.  Thus, we consider our upper limit on $\mathcal{R}_{\mathrm{OA}}$ to be independent of previous work.

We show the predicted number of OAs with respect to the jet parameters in Figure \ref{fig:expected_afterglows}. We expect more than 1 OA in the DWF if the jet has low power-law index, $b$, and the difference between the angular extent of the wings and the core is large. If GRB\,221009A-like events occurred in the DWF data, with large values of $\theta_w$, we would have expected at least one detection as \citet{221009A_shallow} measure $b=0.8$ for GRB\,221009A.  Thus, our non-detection of OA constrains the possible angles and power-law indices. Assuming a constant, $b=0.8$, we place upper limits on $\theta_w - \theta_c$ of $58.3^{\circ}$ and $56.6^{\circ}$ for smooth power-law and power-law with core jet models respectively. For a steeper angular profile, $b=1.2$, we find upper limits of $\theta_w - \theta_c$ of $75.3^{\circ}$ and $76.8^{\circ}$ for smooth power-law and power-law with core jet models respectively.  These values are calculated for an expected value of 1 OA in the data, corresponding to a confidence of $\sim63.2$\,\%.

\begin{figure*}
    \includegraphics[width=\textwidth]{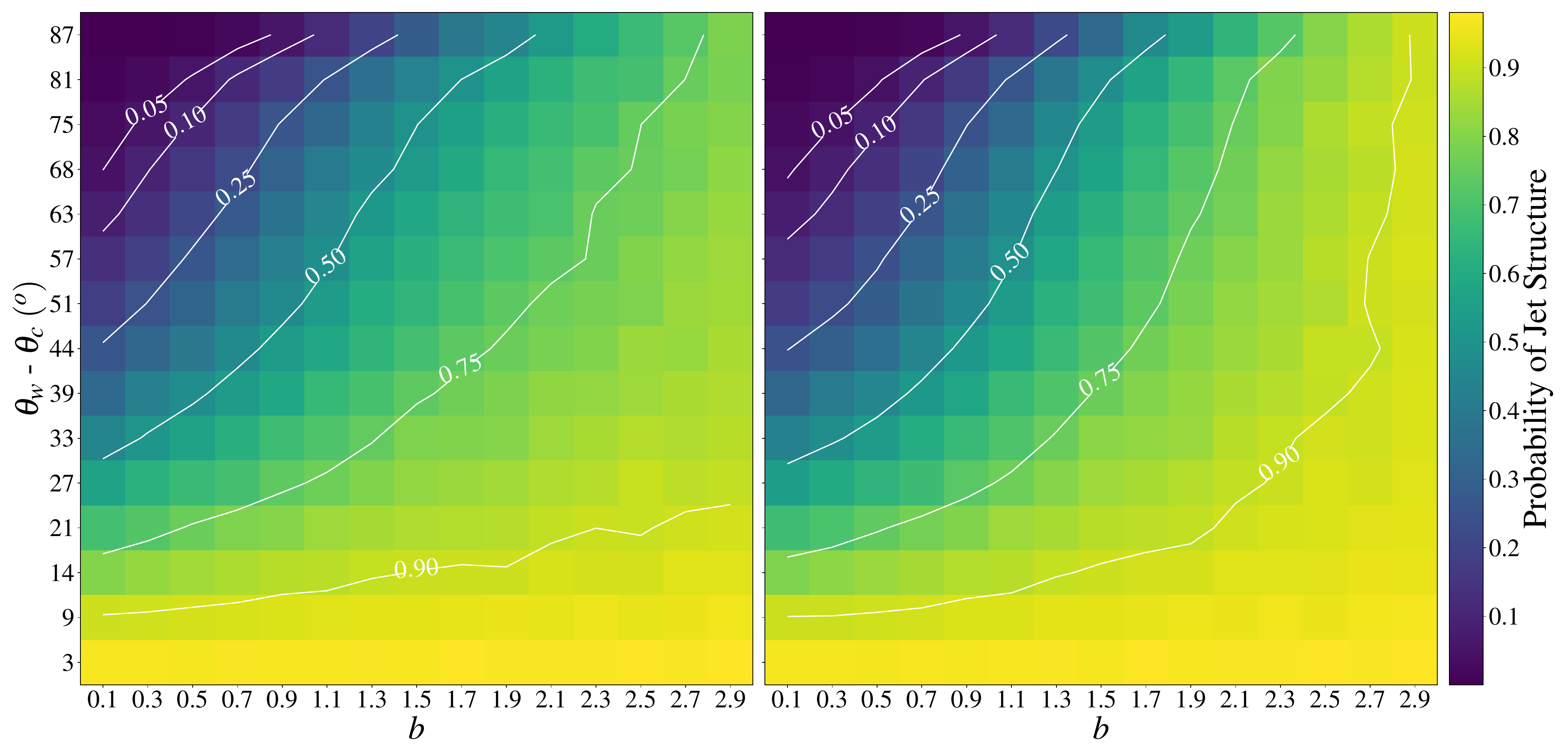}
    \caption{Probability of achieving a non-detection in the DWF data for a range of jet profiles.  Axes are as Figure \ref{fig:expected_afterglows}.}
    \label{fig:probabilities}
\end{figure*}

Using the expected number of afterglows in the data, shown in Figure \ref{fig:expected_afterglows}, we can use Poisson statistics to calculate the probability of our non-detection with a given values of $b$ and $\theta_w - \theta_c$ (see Figure \ref{fig:probabilities}). We find that the non-detection of an OA in this paper disfavours shallow angular jet profiles with a large angular extent outside the $\gamma$-ray emitting region.  Our results favour a scenario where the wings of the jet are small or steeply drop-off in energy outside the $\gamma$-ray emitting region. This is consistent with the results of hydrodynamical simulations of LGRB jets which predict $b>1$ for most of the LGRB population \citep{jetstructure_sim2}.

\subsection{Prospects for Detection with Other Current and Future Surveys}\label{sec:survey_constraints}

\begin{figure}
    \includegraphics[width=\columnwidth]{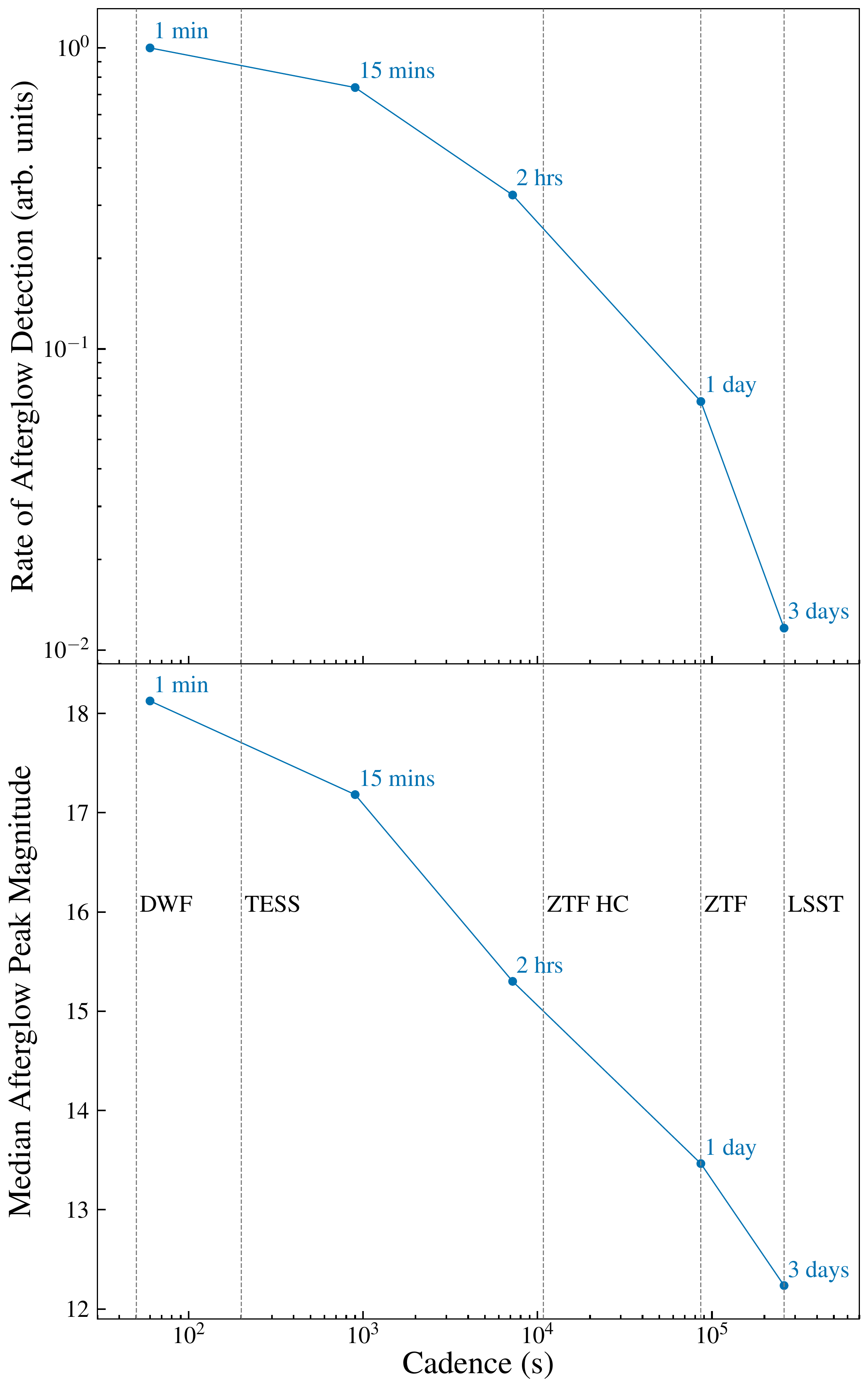}
    \caption{The rate and peak magnitude of OAs, in the top and bottom panel respectively, detected with a theoretical survey possessing a limiting magnitude of $g=23$ AB mag at different observing cadences.  We have simulated the same number of events for each cadence, distributed between a day before the first observation and the last observation.  The OAs were simulated with a jet structure of $b=0.8$ and $\theta_w = 57^{\circ}$. We also require three observations where an OA is $g<23$ AB mag before an event is detected.  We find that at a day cadence, $5.5\,\%$ of the OAs are found compared to minute cadence.  At a day cadence, we find that the median intrinsic peak magnitude for detected OAs are required to be 4.7 magnitudes brighter.}
    \label{fig:fast}
\end{figure}

In Figure \ref{fig:fast}, we see the importance of cadence with searches for OAs.  We assume the same sky coverage and depth for each cadence to enable a direct comparison.  Generally, a lower cadence allows for deeper observations with more sky coverage, which maximises the likelihood of achieving a single detection.  However, to understand and classify a light curve, more detections are required.  This is highlighted in Figure \ref{fig:fast}; a high cadence can also significantly boost OA detection rate by probing the OA population deeper.  DWF's $\sim50$\,s cadence, therefore, has a high detection rate per night and square degree observed, making uniquely positioned amongst transient surveys to search for OAs.

The Vera C. Rubin Observatory's (Rubin) Legacy Survey of Space and Time has an unprecedented combination of depth and a large field-of-view that make it an extremely powerful facility for discovering transients.  However, with a typical cadence of three days, Rubin will be inefficient for a study similar to this one.  At this cadence, the survey will be sensitive to only the brightest and slowest evolving afterglows, detecting a small fraction of those detectable at high cadence, as shown in Figure \ref{fig:fast}.  Fink, a broker for Rubin/LSST, provides alerts and classification in real-time which promises to provide the capability for fast cadenced follow-up to alerts \citep{fink}. Supplementing Rubin alerts with other facilities to achieve a faster cadence could provide a promising avenue for OA detection.  

The Transiting Exoplanet Survey Satellite (TESS) possesses an observing strategy similar to DWF, well suited towards searching for OAs.  Since 2022, TESS has adopted a 200 second cadence, observing a given sector continuously for 27 days at a time.  While TESS's typical 5-$\sigma$ limiting magnitude of 16 AB mag is substantially shallower than the depths that Rubin and DECam are capable of, its 2300\,deg$^2$ field-of-view, cadence and temporal coverage make it a promising instrument for transient detection \citep{TESS}.  TESS imaging has a comparatively large pixel scale at 21 arcseconds compared to DECam's 0.27 arcseconds.  This will present challenges in OA searches, particularly in identifying host galaxies and disentangling them from M-stars.

Difficulties with TESS background subtraction for conducting image subtraction analysis has prevented comprehensive searches for extragalactic fast transients like OAs.  However, recently \textsc{TESSreduce} has made searches like this possible \citep{TESSreduce} and a number of optical afterglows coincident with GRB triggers, serendipitously detected by TESS, have been identified \citep{TESS_afterglows,JayaramanTESSGRBs}.

Evryscope \citep{evryscope} is a ground based facility which utilises a similar, high cadence, observing strategy.  With a depth of $V\sim16$.  it continuously observes an 18400 square degrees at a two minute cadence for six hours per night.  While its sky and temporal coverage fall short of TESS, it is still a powerful facility for searching for OAs and its smaller comparable pixel scale of 13 arcseconds will allow for more effective identification of OA host galaxies.

In the structured jet regime, the OAs luminosity function shifts to fainter peak luminosities as $b$ becomes large.  Due to TESS and Evryscope's comparative shallowness, they are sensitive to probing an OA population with shallower angular jet structure.  It is useful, therefore, to use a two-pronged approach when probing the OA population, with both large field-of-view, shallow surveys like TESS and Evryscope and deep surveys with smaller fields-of-view like Rubin.

\section{Conclusions}\label{sec:conclusion}

Orphan afterglows (OAs) provide a powerful probe into the geometry of the relativistic outflows that give rise to long-duration gamma-ray bursts (LGRBs).  Understanding LGRB jet geometry will help constrain the true, beaming corrected LGRB rate and energy release.  Recent observations, such as the follow-up to GRB\,221009A \citep{oconnor_221009A,221009A_williams}, have supported the possibility of a structured jet where $\gamma$-ray emission is restricted to the jet's core, with OAs detectable at wider viewing angles.

In this work, we conduct a search for OAs in 100 nights of observations from The Deeper, Wider, Faster programme (DWF).  DWF's deep ($g\sim23$), minute-cadence observations provide a unique opportunity to probe this theoretical population of OAs.  We use a machine learning classifier, trained on afterglow models generated with \textsc{afterglowpy} to extract OA candidates in the archival DWF data.

We found 51 OA candidates.  Of these 42 were found to likely originate from M-stars, suggesting they are Galactic, stellar flares.  While their nature is not obvious, the other nine candidates are likely Galactic transients.

We find no strong OA candidates with, or without, apparent host galaxies, in 100 nights of the DWF data, comprising 9033 images and 145 hours of observing time. We measure an upper limit on the rate of OAs $g<22$ AB mag of 7.46\,deg$^{-2}$yr$^{-1}$ (95\,\% confidence).  We also place constraints on GRB jet structure with a structured jet where prompt, $\gamma$-ray emission is restricted to the jet's core.  Setting the power-law index of the structured jet, $b=0.8$ we measure upper limits on the difference between half opening angle of the $\gamma$-ray emitting core and the half opening angle of the jet $\theta_c - \theta_w$.  These values are  $58.3^{\circ}$ and $56.6^{\circ}$ ($75.3^{\circ}$ and $76.8^{\circ}$ for $b=1.2$) for smooth power-law and power-law with core jet models respectively with $\sim63.2$\,\% confidence. We encourage further searches for OAs with other fast-cadenced, wide-field surveys such as TESS to better constrain this parameter space.

The unique deep, fast-cadenced data from DWF allows for a search for OAs at their fastest timescales.  We present the first observational constraints on GRB jet structures with a search for OAs. This work highlights the importance of untargeted, multi-wavelength searches in understanding LGRB jet structure. These efforts not only reveal OA rates but can also provide insights into the jet launching mechanism and the intrinsic properties of LGRBs.

\section*{Acknowledgements}

We thank the anonymous reviewer for their swift and valuable comments and Giancarlo Ghirlanda for his helpful insights.

Parts of this research were conducted by the Australian Research Council Centre of Excellence for Gravitational Wave Discovery (OzGrav), through project numbers CE170100004 and CE230100016.

JC acknowledges funding by the Australian Research Council Discovery Project, DP200102102.

AM is supported by the ARC Discovery Early Career Researcher Award (DECRA) project number DE230100055.

OS acknowledges funding by the Istituto Nazionale di Astrofisica (INAF) `Finanziamento della Ricerca Fondamentale' 2023 projects `POEMS' (1.05.23.06.04) and `MEGA' (1.05.23.04.04). This work has been funded by the European Union-Next Generation EU, PRIN 2022 RFF M4C21.1 (202298J7KT - PEACE)

This project used data obtained with the Dark Energy Camera (DECam), which was constructed by the Dark Energy Survey (DES) collaboration. Funding for the DES Projects has been provided by the US Department of Energy, the U.S. National Science Foundation, the Ministry of Science and Education of Spain, the Science and Technology Facilities Council of the United Kingdom, the Higher Education Funding Council for England, the National Center for Supercomputing Applications at the University of Illinois at Urbana-Champaign, the Kavli Institute for Cosmological Physics at the University of Chicago, Center for Cosmology and Astro-Particle Physics at the Ohio State University, the Mitchell Institute for Fundamental Physics and Astronomy at Texas A\&M University, Financiadora de Estudos e Projetos, Fundação Carlos Chagas Filho de Amparo à Pesquisa do Estado do Rio de Janeiro, Conselho Nacional de Desenvolvimento Científico e Tecnológico and the Ministério da Ciência, Tecnologia e Inovação, the Deutsche Forschungsgemeinschaft and the Collaborating Institutions in the Dark Energy Survey.

The Collaborating Institutions are Argonne National Laboratory, the University of California at Santa Cruz, the University of Cambridge, Centro de Investigaciones Enérgeticas, Medioambientales y Tecnológicas–Madrid, the University of Chicago, University College London, the DES-Brazil Consortium, the University of Edinburgh, the Eidgenössische Technische Hochschule (ETH) Zürich, Fermi National Accelerator Laboratory, the University of Illinois at Urbana-Champaign, the Institut de Ciències de l’Espai (IEEC/CSIC), the Institut de Física d’Altes Energies, Lawrence Berkeley National Laboratory, the Ludwig-Maximilians Universität München and the associated Excellence Cluster Universe, the University of Michigan, NSF NOIRLab, the University of Nottingham, the Ohio State University, the OzDES Membership Consortium, the University of Pennsylvania, the University of Portsmouth, SLAC National Accelerator Laboratory, Stanford University, the University of Sussex, and Texas A\&M University.

Based on observations at NSF Cerro Tololo Inter-American Observatory, NSF NOIRLab (NOIRLab Prop. ID 2015B-0607, 2016A-0095, 2017A-0909, 2019A-0911, 2018A-0137, 2019B-1012 and 2020B-0253; PI: J, Cooke), which is managed by the Association of Universities for Research in Astronomy (AURA) under a cooperative agreement with the U.S. National Science Foundation.

Parts of this work was performed on the OzSTAR national facility at Swinburne University of Technology. The OzSTAR program receives funding in part from the Astronomy National Collaborative Research Infrastructure Strategy (NCRIS) allocation provided by the Australian Government, and from the Victorian Higher Education State Investment Fund (VHESIF) provided by the Victorian Government.

This research made use of \textsc{matplotlib}, a Python library for publication quality graphics \citep{matplotlib}, \textsc{SciPy} \citep{scipy} and \textsc{Astropy}, a community-developed core Python package for Astronomy \citep{astropy1,astropy2}.  We also used XGBoost and \textsc{scikit-learn} \citep{sklearn}.

%%%%%%%%%%%%%%%%%%%%%%%%%%%%%%%%%%%%%%%%%%%%%%%%%%
\section*{Data Availability}

Calibrated images for most observations are publicly available on the NOIRLab Astro Data Archive under program numbers 2015B-0607, 2016A-0095, 2017A-0909, 2019A-0911, 2018A-0137, 2019B-1012 and 2020B-0253. Observations that are still within the 18 month proprietary period will be shared upon reasonable request to the authors. All other code and data underlying this work will be shared upon reasonable request to the authors.



\bibliographystyle{mnras}
\bibliography{ref} % if your bibtex file is called example.bib

%\begin{thebibliography}{99}
%\end{thebibliography}

%%%%%%%%%%%%%%%%%%%%%%%%%%%%%%%%%%%%%%%%%%%%%%%%%%

%%%%%%%%%%%%%%%%% APPENDICES %%%%%%%%%%%%%%%%%%%%%

\appendix

\section{Fitting Afterglow Models to Candidates}\label{sec:modelfitting}

We use \textsc{redback} \citep{redback}, a Bayesian inference software package for fitting electromagnetic transients, to fit our candidate light curves shown in Figure \ref{fig:lcs}.  We use \textsc{afterglowpy}'s power-law with core afterglow model.  Without multi-band data, a prompt GRB detection and a spectroscopic redshift associated with an event, deriving physical parameters from a fit to an afterglow model is difficult.  This is due to the degeneracy between their parameters. As a result, strict priors are required to get reasonable results.  We explore parameter space with the nested sampler, \textsc{dynesty} \citep{dynesty}.

We perform four separate fits, setting static priors for an afterglow at $z=0.5,1,2$ and $3$ based on the parameters used for our synthetic population described in Section \ref{sec:afterglow_sample} with a jet structure consistent with GRB\,221009A as calculated by \citet{221009A_shallow}.  Specifically, we set $p = 2.3$, $\epsilon_e = 0.02$, $\epsilon_B = 0.008$, $\theta_w = 57^{\circ}$, $b=0.8$ and $\theta_c = 5.7^{\circ}$ and we fit for $\log E_{\mathrm{iso}}$, $\log n$ and $\theta_v$.  Informed by the distributions from our synthetic population, described in Section \ref{sec:afterglow_sample}, for $\log E_{\mathrm{iso}}$, we use Gaussian priors with $\mu=53.69$ and $\sigma=1.1$ in units of $\log \mathrm{erg}\,\mathrm{s}^{-1}$. Similarly, we use Gaussian priors and $\mu=0$ and $\sigma=1$ for $\log n$ in units of $\log \mathrm{cm}^{-3}$.  We sample $\theta_v < \theta_w$ and burst time, denoted $t_0$, uniformly.

By default, \textsc{afterglowpy} does not model the deceleration phase of the jet and is therefore unaffected by the initial Lorentz factor of the jet \citep{afterglowpy}. This results in inaccuracies in the \textsc{afterglowpy}'s predicted light curves.  For an on-axis optical afterglow, that is $\theta_v<\theta_w$, from a structured jet, the light curve is most discrepant from numerical models at early times, during the rise.  We therefore fit only to fade phase of the light curve.

\begin{figure*}
    \includegraphics[width=0.33\textwidth]{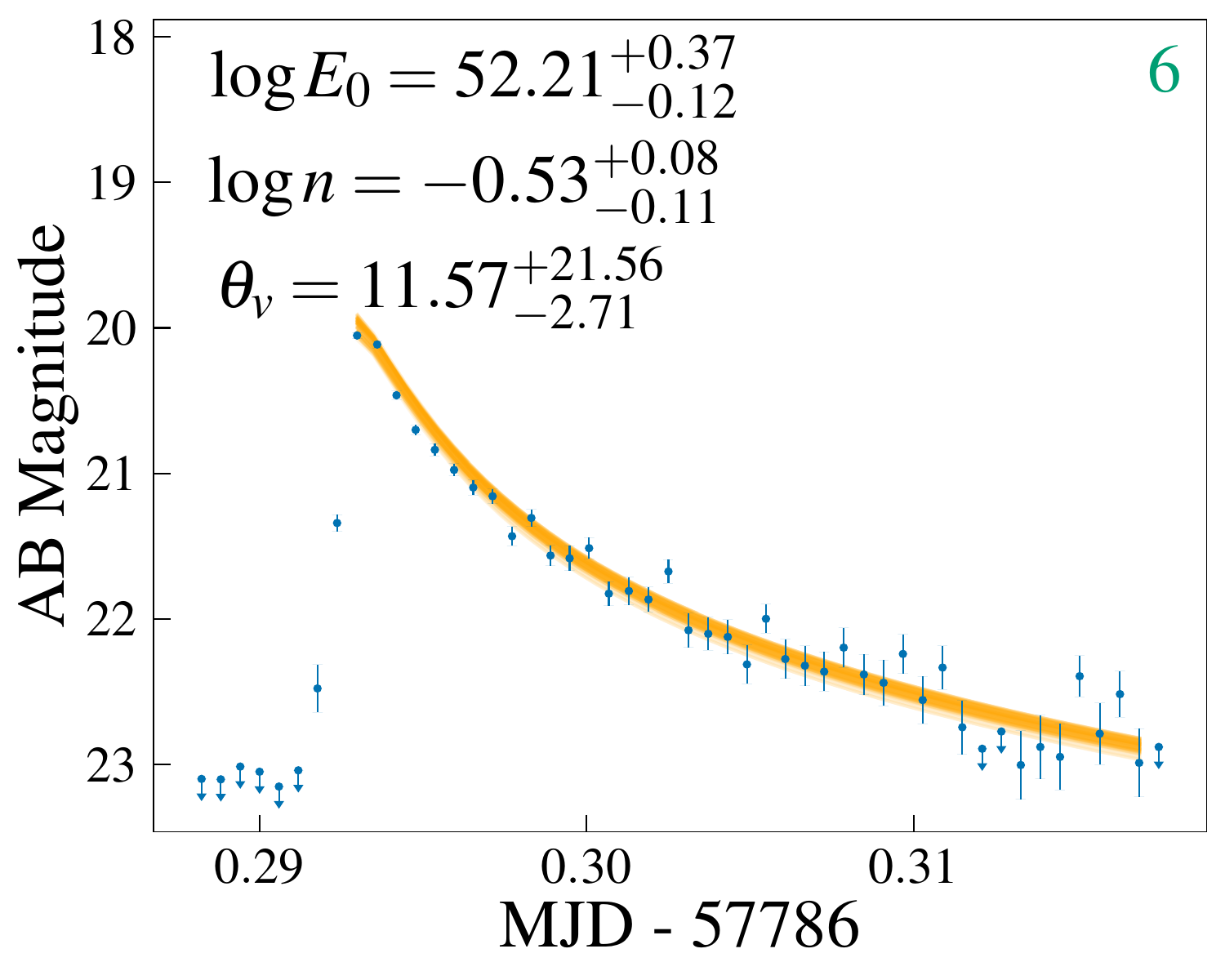}
    \includegraphics[width=0.33\textwidth]{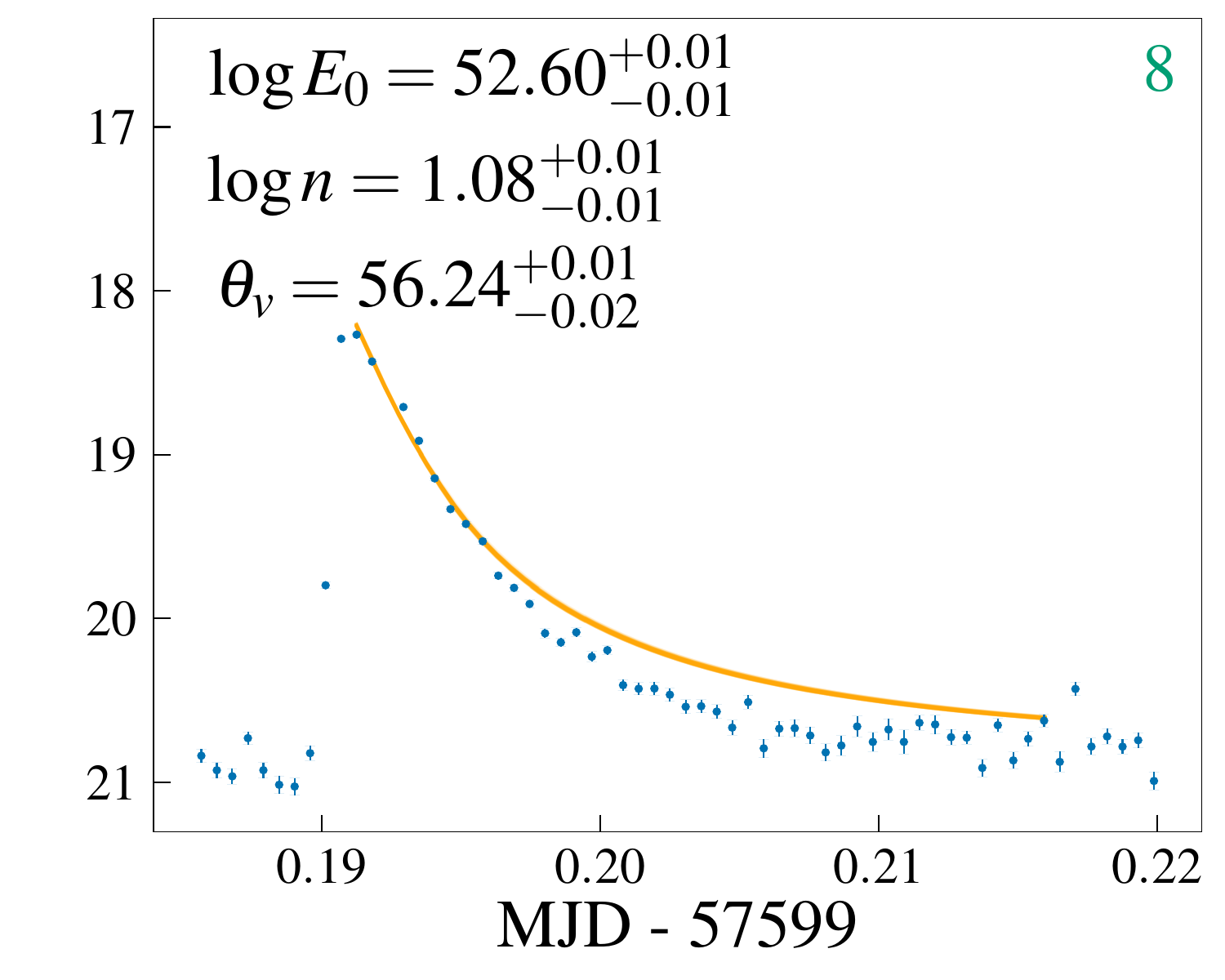}
    \includegraphics[width=0.33\textwidth]{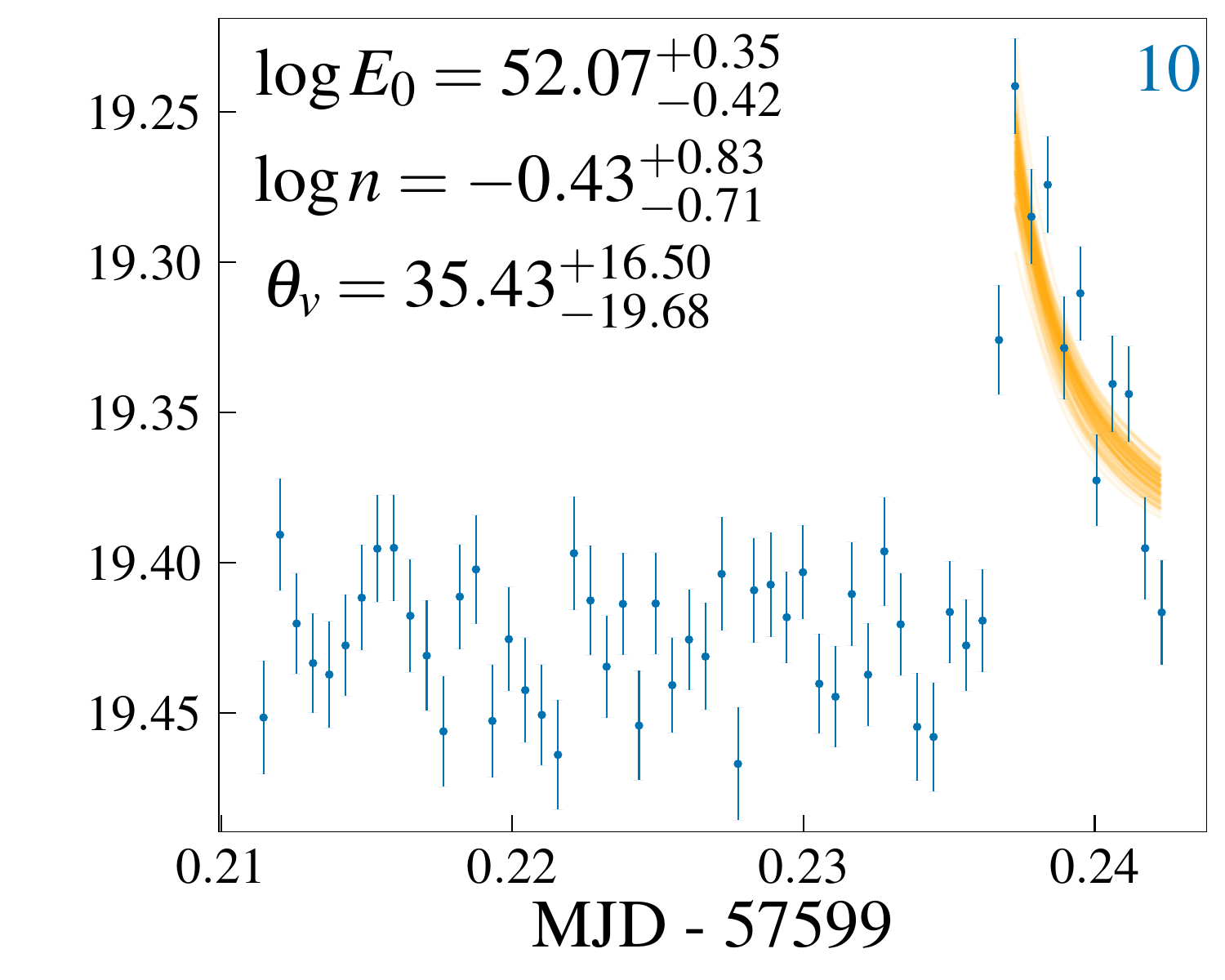}
    \includegraphics[width=0.33\textwidth]{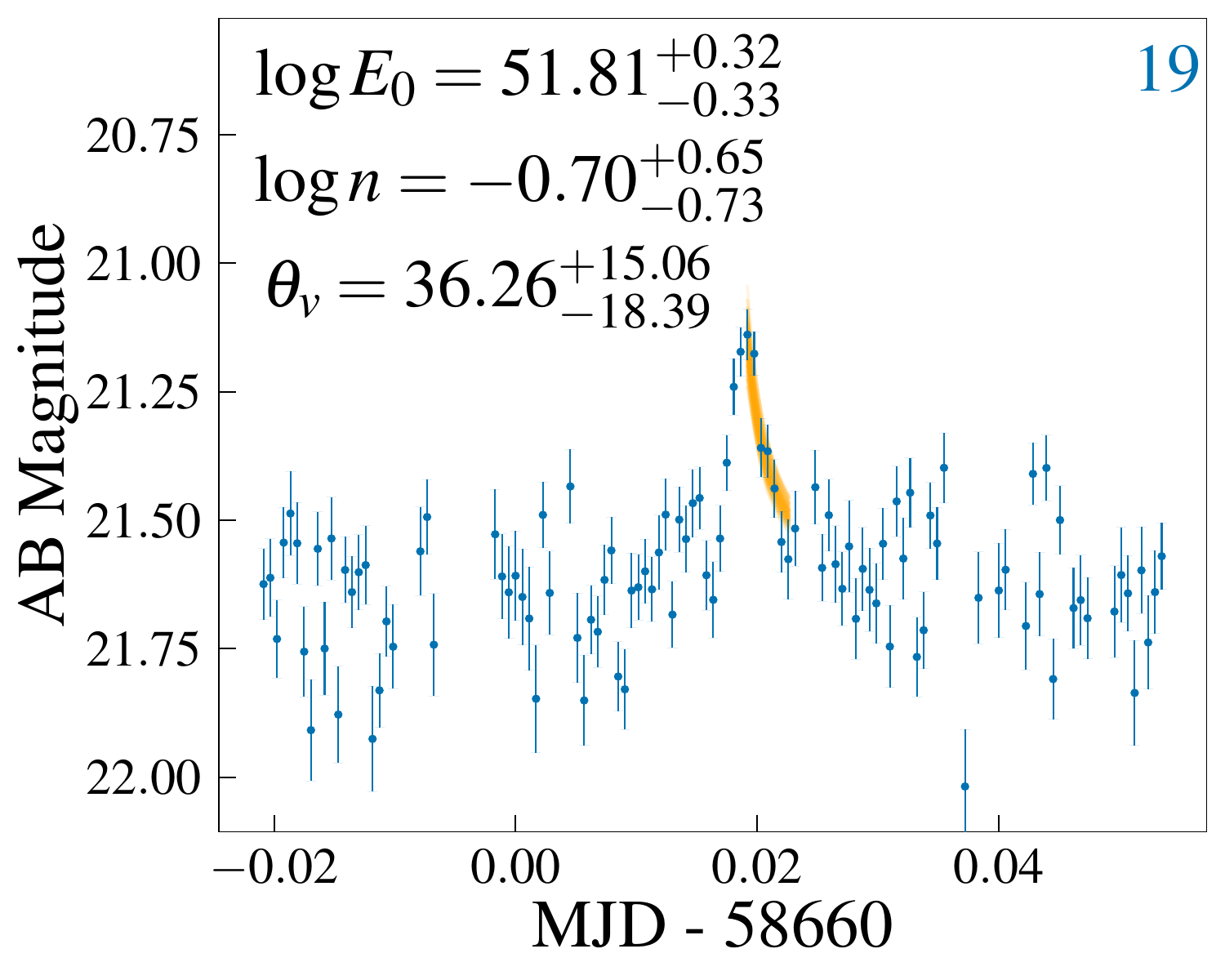}
    \includegraphics[width=0.33\textwidth]{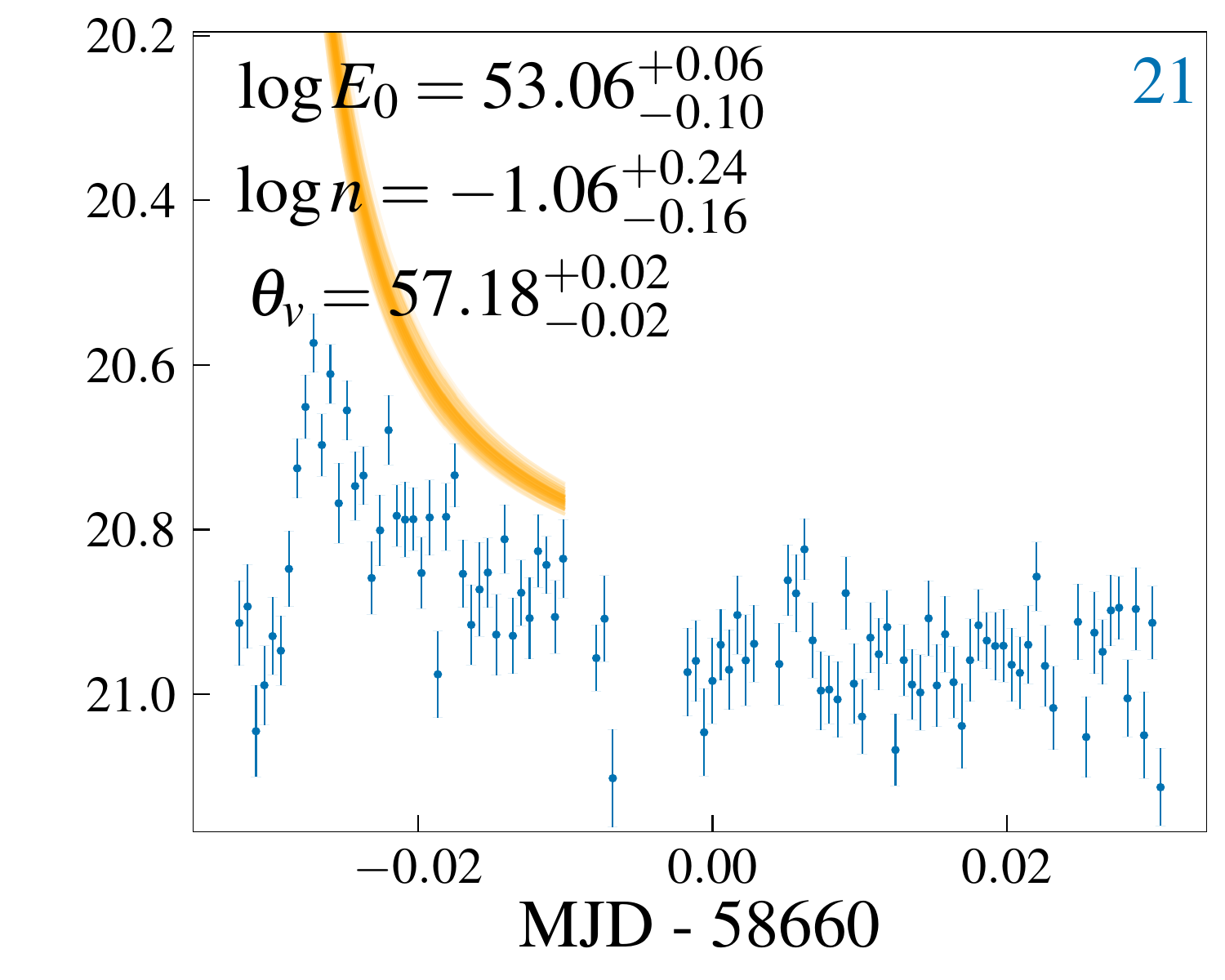}
    \includegraphics[width=0.33\textwidth]{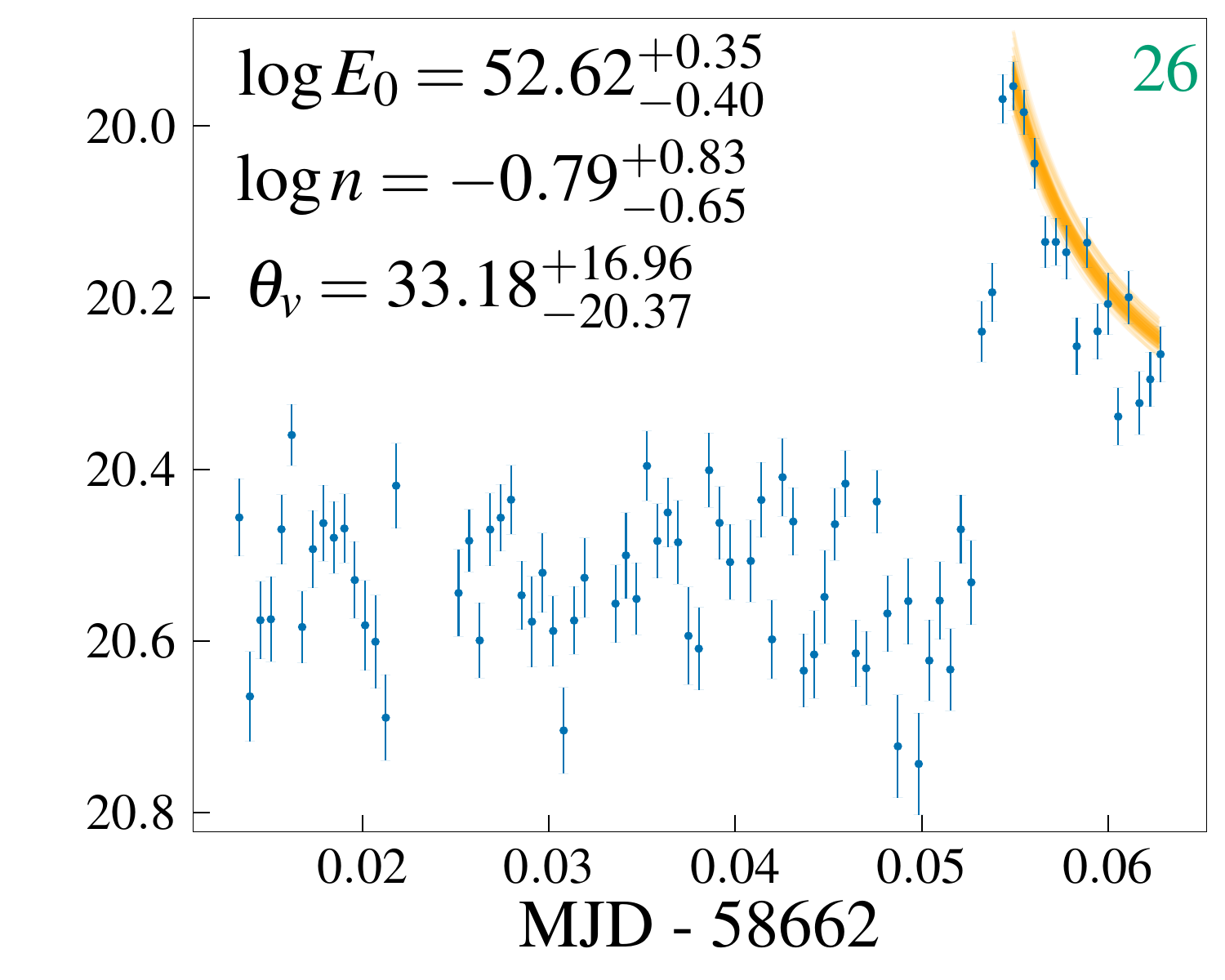}
    \includegraphics[width=0.33\textwidth]{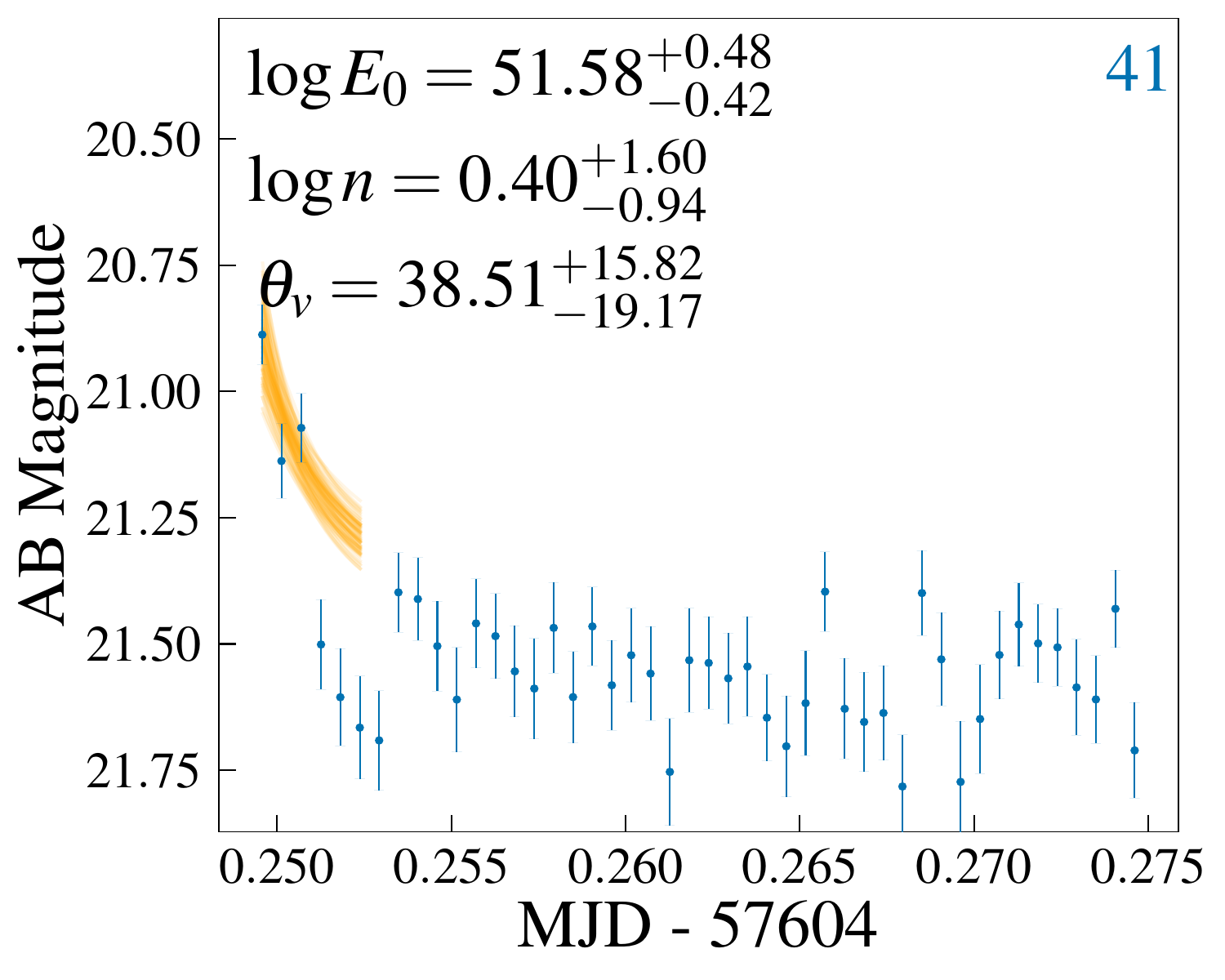}
    \includegraphics[width=0.33\textwidth]{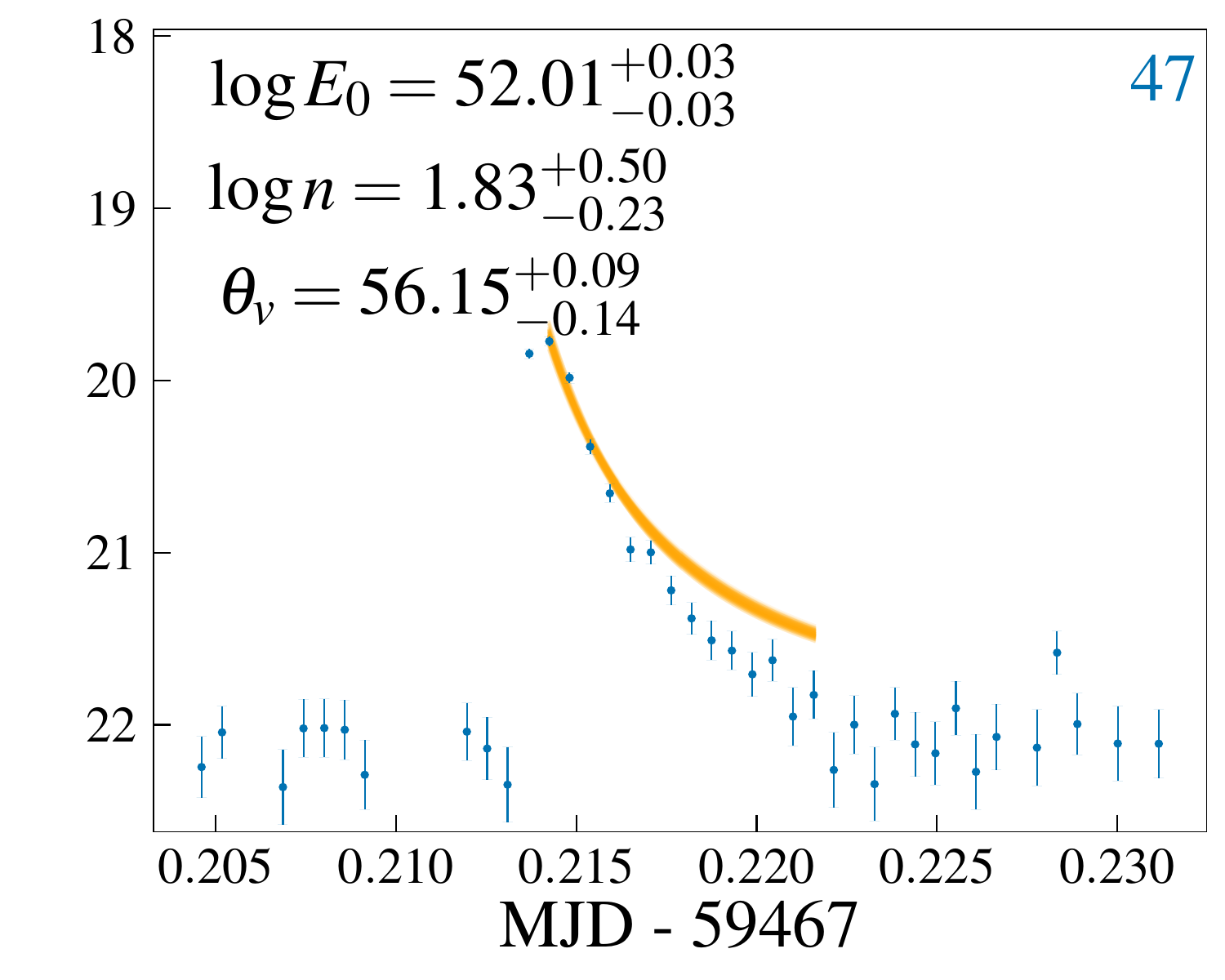}
    \includegraphics[width=0.33\textwidth]{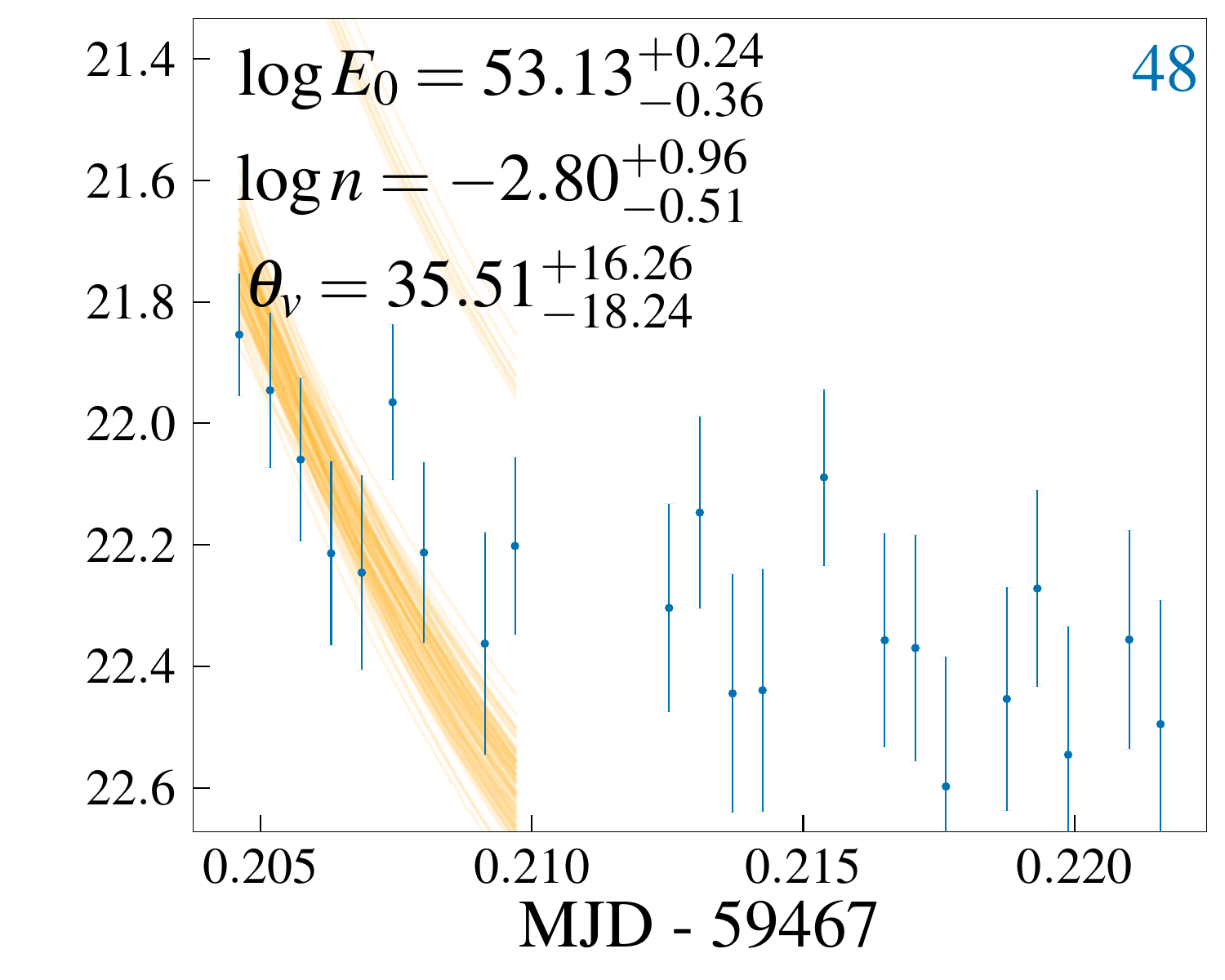}
    \caption{The light curves in Figure \ref{fig:lcs} with 100 \textsc{afterglowpy} models sampled from the posteriors, assuming $z=3$.  The best-fit parameters are shown in the top-left of each light curve.  The units of $\log E_0$, $\log n$ and $\theta_v$ are $\log \mathrm{erg}\,s^{-1}$, $\log \mathrm{cm}^{-3}$ and degrees respectively.  Each light curve is labelled with its candidate number.}
    \label{fig:lc_fits}    
\end{figure*}

In Figure \ref{fig:lc_fits}, we show the results of these fits on the nine candidates without clear M-star colour evolution.  We see qualitative agreement with models in candidates 6, 8, 10, 19, 26, 47 and 48.  However, we have already noted the limitations of \textsc{afterglowpy} models at early times.  In addition, the reverse-shock emission may result in departures from a typical, forward shock afterglow light curve \citep{990123_flash,080319B_flash,130427A_flash,210619B_flash}.  We therefore, cannot rule out any of our candidates purely based off a poor fit.

\begin{figure*}
    \includegraphics[width=0.33\textwidth]{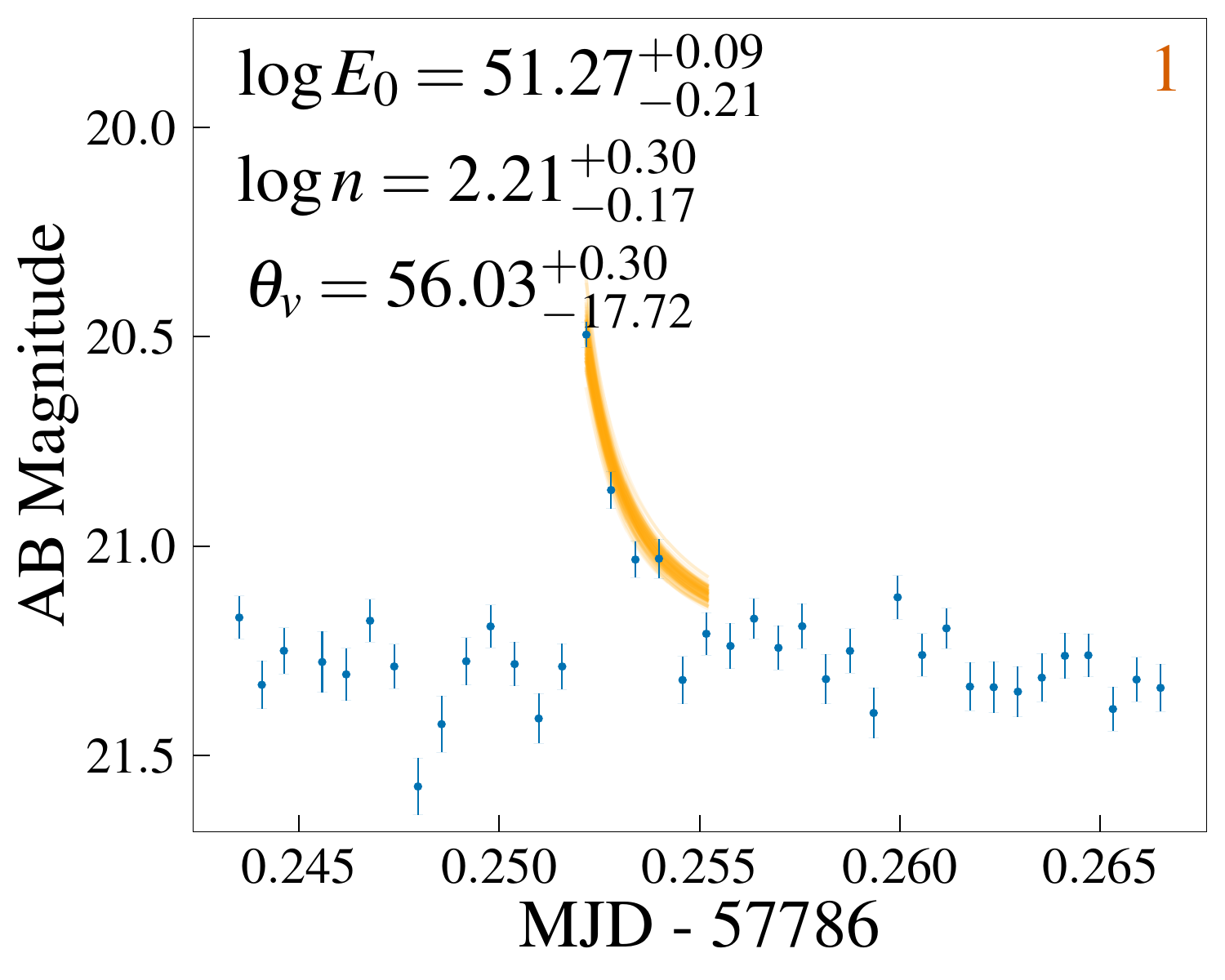}
    \includegraphics[width=0.33\textwidth]{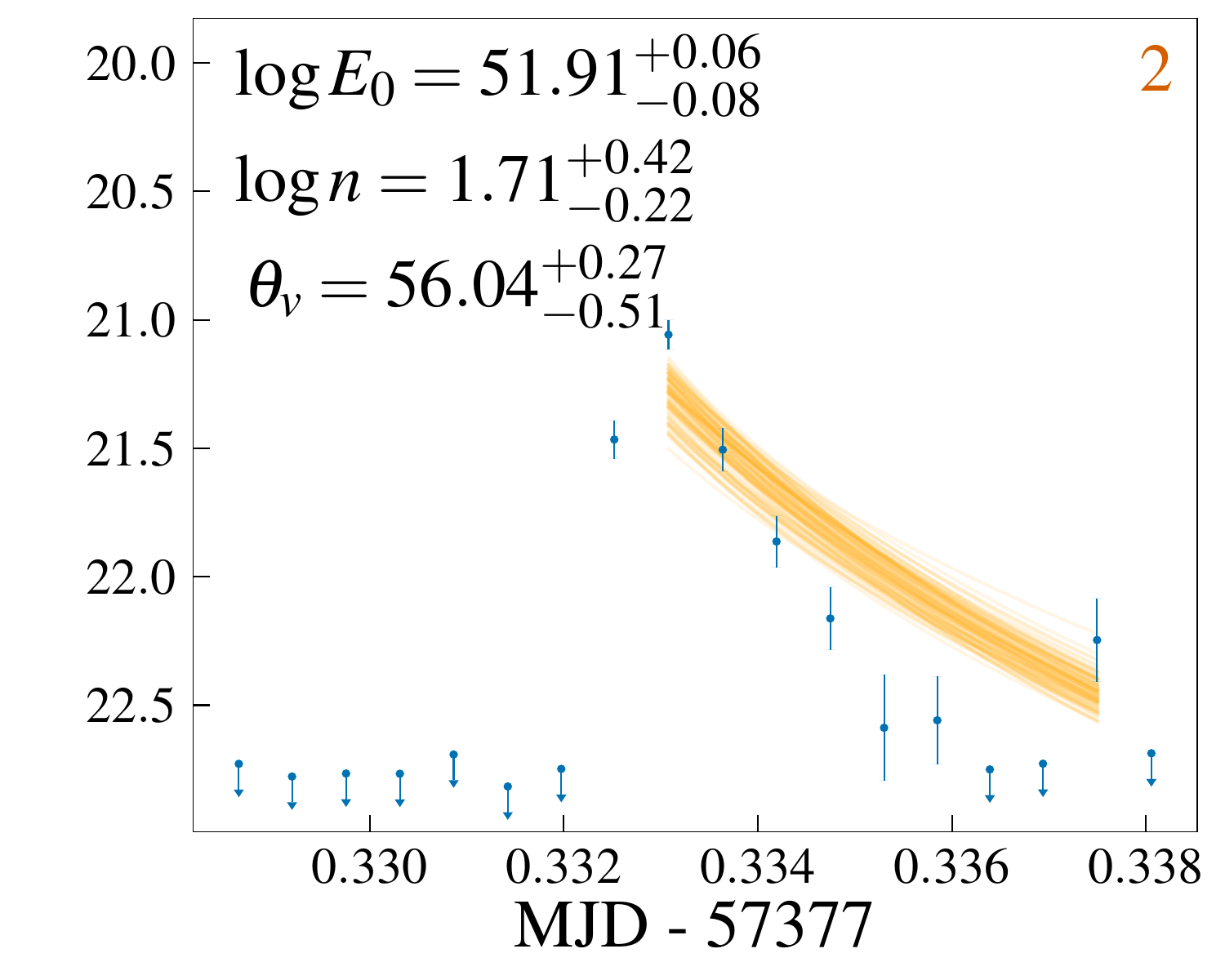}
    \includegraphics[width=0.33\textwidth]{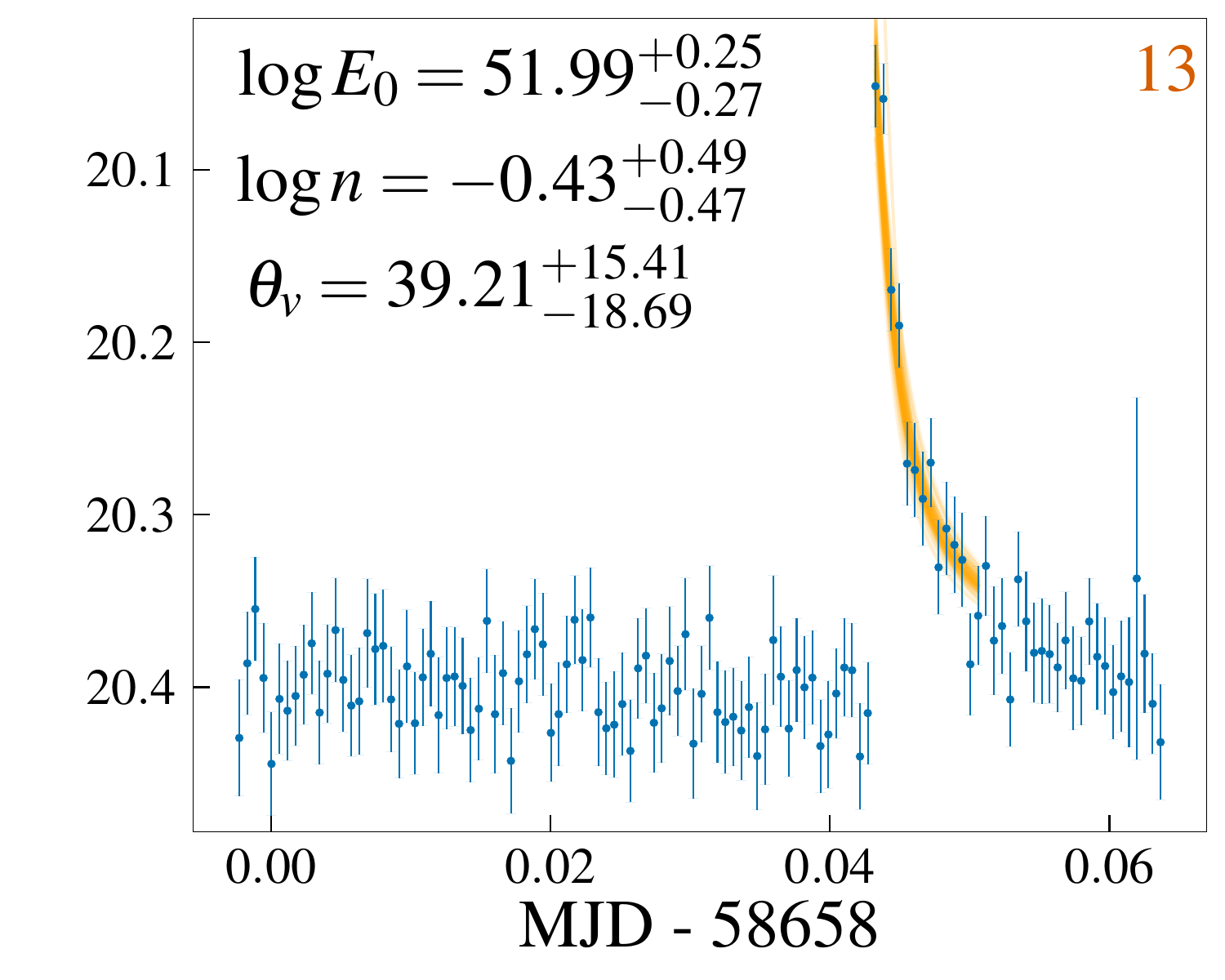}
    \includegraphics[width=0.33\textwidth]{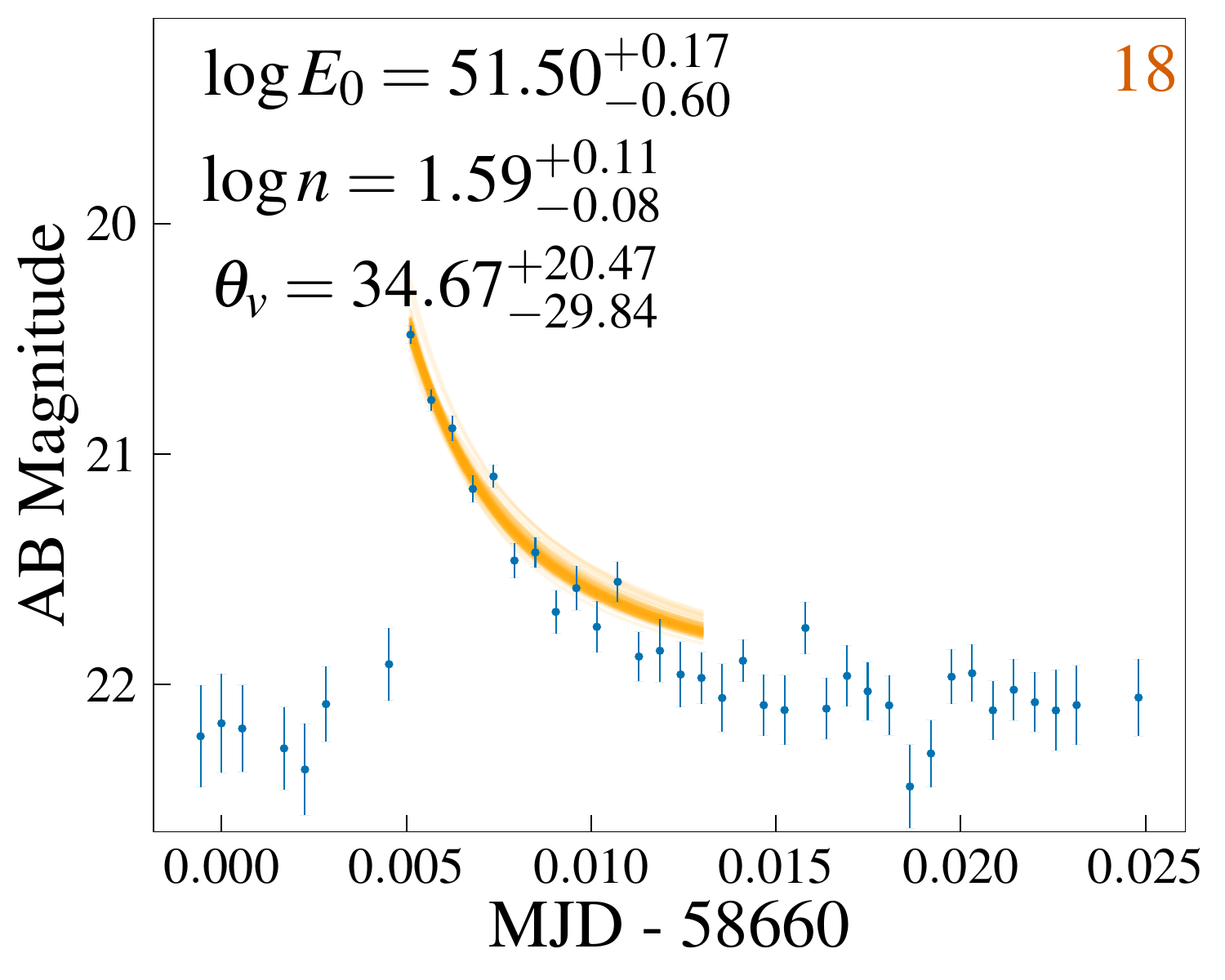}
    \includegraphics[width=0.33\textwidth]{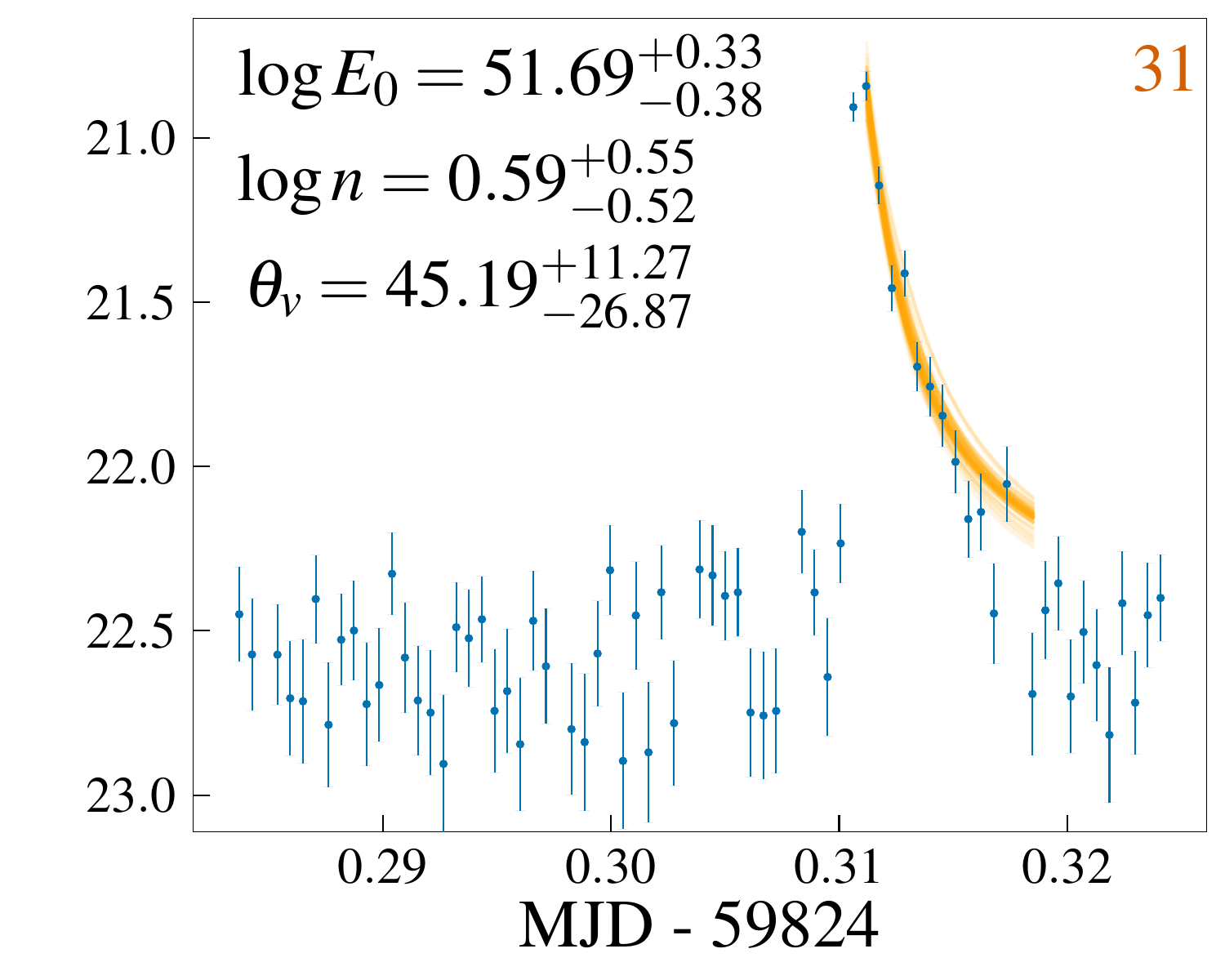}
    \includegraphics[width=0.33\textwidth]{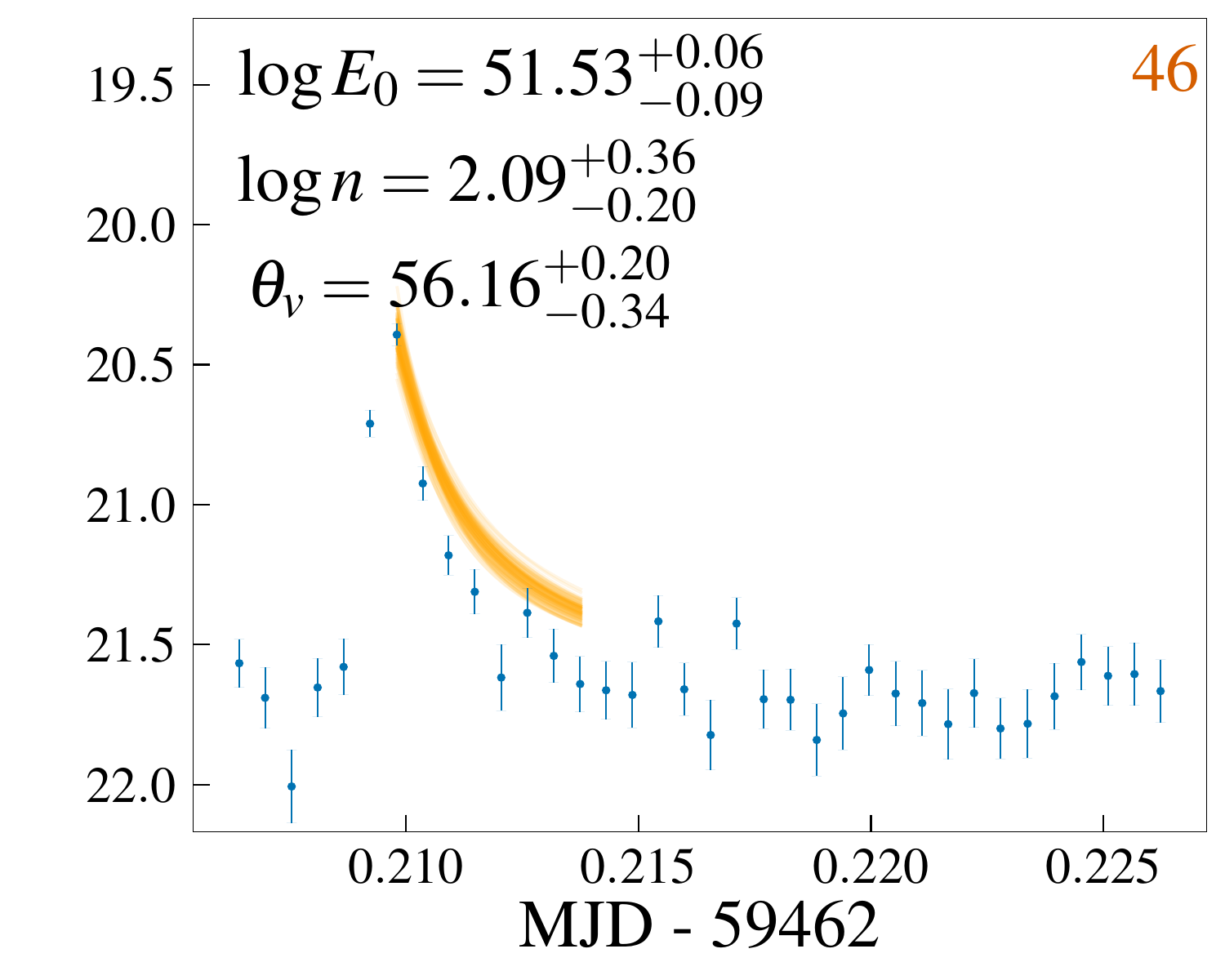}
    \caption{Same as Figure \ref{fig:lc_fits} but with a selection of six candidates that had coincident sources possessing colours consistent with an M-star.  These light curves. therefore, likely originate from stellar flares.  Despite this, our fits yield good results in some of these light curves.}
    \label{fig:lc_fits_flares}    
\end{figure*}

We also fit six of the candidates with M-star colour evolution in Figure \ref{fig:lc_fits_flares}, which are likely to be stellar flares.  We also find qualitative agreement in some of these fits, particularly candidates 13 and 31.  We conclude from this that stellar flares may have an evolution consistent with a GRB afterglow in $g$-band, minutes to hours post-burst.

\begin{table}
 \caption{Parameter estimations from the \textsc{afterglowpy} fits to the candidates in Fig. \ref{fig:lc_fits} and \ref{fig:lc_fits_flares}.  $t_{\mathrm{peak}}$ denotes the time of the brightest detection.  The units are $\log \mathrm{erg}\,s^{-1}$ for $\log E_0$, $\log \mathrm{cm}^{-3}$ for $\log n$, degrees for $\theta_v$ and minutes for $t_0$.}
 \label{tab:param_estimates}
 \begin{tabular*}{\columnwidth}{@{}l@{\hspace*{8pt}}l@{\hspace*{8pt}}l@{\hspace*{8pt}}l@{\hspace*{8pt}}l@{\hspace*{8pt}}}
  \hline
  Candidate No. & $\log E_0$ & $\log_n$ & $\theta_v$ & $t_{\mathrm{peak}} - t_0$\\
  \hline
  Candidates & & & & \\
  \vspace{0.02\columnwidth}
  $6$ & $52.21^{+0.37}_{-0.12}$ & $-0.53^{+0.08}_{-0.11}$ & $11.57^{+0.01}_{-0.02}$ & $1.0013^{+0.0020}_{-0.0009}$\\
  \vspace{0.02\columnwidth}
  $8$ & $52.60^{+0.01}_{-0.12}$ & $1.08^{+0.01}_{-0.01}$ & $56.24^{+0.01}_{-0.02}$ & $1.09^{+0.10}_{-0.06}$\\
  \vspace{0.02\columnwidth}
  $10$ & $52.07^{+0.35}_{-0.42}$ & $-0.43^{+0.83}_{-0.71}$ & $35.43^{+16.50}_{-19.68}$ & $2.2^{+0.4}_{-0.4}$\\
  \vspace{0.02\columnwidth}
  $19$ & $51.81^{+0.32}_{-0.33}$ & $-0.70^{+0.65}_{-0.73}$ & $36.26^{+15.06}_{-18.39}$ & $1.12^{+0.15}_{-0.09}$\\
  \vspace{0.02\columnwidth}
  $21$ & $53.06^{+0.06}_{-0.10}$ & $-1.06^{+0.24}_{-0.16}$ & $57.18^{+0.02}_{-0.02}$ & $3.0008^{+0.0020}_{-0.0006}$\\
  \vspace{0.02\columnwidth}
  $26$ & $52.26^{+0.35}_{-0.40}$ & $-0.79^{+0.83}_{-0.65}$ & $33.18^{+16.96}_{-20.37}$ & $6.5^{+0.3}_{-0.3}$\\
  \vspace{0.02\columnwidth}
  $41$ & $51.58^{+0.48}_{-0.42}$ & $-0.40^{+1.60}_{-0.94}$ & $38.51^{+15.82}_{-19.17}$ & $2.0^{+0.5}_{-0.4}$\\
  \vspace{0.02\columnwidth}
  $47$ & $52.01^{+0.03}_{-0.03}$ & $1.83^{+0.50}_{-0.23}$ & $56.15^{+0.09}_{-0.14}$ & $2.2^{+0.1}_{-0.1}$\\
  \vspace{0.02\columnwidth}
  $48$ & $53.13^{+0.24}_{-0.36}$ & $-2.80^{+0.96}_{-0.51}$ & $35.51^{+16.26}_{-18.24}$ & $4.3^{+0.3}_{-0.3}$\\
  Suspected Flares & & & & \\
  \vspace{0.02\columnwidth}
  $1$ & $51.27^{+0.09}_{-0.21}$ & $2.21^{+0.30}_{-0.17}$ & $56.03^{+0.30}_{-17.72}$ & $1.4^{+0.2}_{-0.2}$\\
  \vspace{0.02\columnwidth}
  $2$ & $51.91^{+0.06}_{-0.08}$ & $1.71^{+0.42}_{-0.22}$ & $56.04^{+0.27}_{-0.51}$ & $5.3^{+0.4}_{-0.4}$\\
  \vspace{0.02\columnwidth}
  $13$ & $51.99^{+0.25}_{-0.27}$ & $-0.43^{+0.49}_{-0.47}$ & $39.21^{+15.41}_{-18.69}$ & $1.03^{+0.05}_{-0.02}$\\
  \vspace{0.02\columnwidth}
  $18$ & $51.50^{+0.17}_{-0.60}$ & $1.59^{+0.11}_{-0.08}$ & $34.67^{+20.47}_{-29.84}$ & $1.8^{+0.2}_{-0.2}$\\
  \vspace{0.02\columnwidth}
  $31$ & $51.69^{+0.33}_{-0.38}$ & $0.59^{+0.55}_{-0.52}$ & $45.19^{+11.27}_{-26.87}$ & $1.03^{+0.06}_{-0.02}$\\
  \vspace{0.02\columnwidth}
  $46$ & $51.53^{+0.06}_{-0.09}$ & $-2.09^{+0.36}_{-0.20}$ & $56.16^{+0.20}_{-0.34}$ & $1.7^{+0.3}_{-0.3}$\\
  \hline
 \end{tabular*}
\end{table}

When using fast-cadenced imaging from a single filter, fitting light curves with currently available models like \textsc{afterglowpy}, is not effective in rejecting contaminants.  A poor fit does not necessarily rule out a GRB afterglow due to limitations of models minutes to hours post-burst.  In addition, a good fit does not rule out a stellar flare as a stellar flare light curve may provide good fits to afterglow models in a single filter.  

The parameter estimations from our fits are shown in Figures \ref{fig:lc_fits} and \ref{fig:lc_fits_flares} as well as Table \ref{tab:param_estimates}. We find that the values of $E_0$ and $n$ extracted from the candidates are within the expected range for LGRBs \citep{Fermi_afterglows}.  

Simultaneous coverage in multiple optical filters may provide an avenue to more effectively use models to disentangle OAs from contaminants like stellar flares.  GROND utilizes a similar approach when following up GRB triggers with simultaneous imaging in seven filters \citep{GROND}.

%%%%%%%%%%%%%%%%%%%%%%%%%%%%%%%%%%%%%%%%%%%%%%%%%%

% Don't change these lines
\bsp	% typesetting comment
\label{lastpage}
\end{document}